 \documentclass[12pt]{article}

\usepackage[margin=2.5cm]{geometry}
\usepackage{times}
\usepackage{setspace}

\RequirePackage{graphicx}

\RequirePackage{natbib}
\usepackage{url}

\usepackage[english]{babel}
\usepackage{array}
\usepackage{rotating}
\usepackage{color}

\usepackage{multirow}
\usepackage{enumerate}
\usepackage{epstopdf}
\usepackage{float}

\usepackage{bbm}
\usepackage{subcaption}
\RequirePackage{amsthm,amsmath,amsfonts,amssymb}
\usepackage{mathrsfs}
\RequirePackage[colorlinks,citecolor=blue,urlcolor=blue]{hyperref}

\usepackage[bottom]{footmisc}
\usepackage{enumitem}
\usepackage{subcaption}

\newcounter{magicrownumbers}

\RequirePackage{geometry}
\RequirePackage{t1enc}
\RequirePackage{enumerate}
\RequirePackage{epsfig}
\RequirePackage{makeidx}
\RequirePackage{graphicx}
\RequirePackage{listings}
\RequirePackage{gensymb}

\usepackage{csvsimple}

\usepackage[utf8]{inputenc}
\usepackage[utf8]{luainputenc}
\usepackage[english]{babel}

\usepackage{xcolor}

\usepackage{algorithm}
\usepackage{algpseudocode}

\newtheorem{remark}{\bf Remark}[section]

\theoremstyle{remark}
\newtheorem{example}{\bf Example}[section]

\def\x{{\boldsymbol x}}

\newcommand{\HD}{\textcolor{black}}

\usepackage{myCmdShortcut}

\title{Efficient subsampling for exponential family models }
\author{
{\normalsize Subhadra Dasgupta, Holger Dette} \\
{\normalsize  Ruhr-Universit\"at Bochum} \\
{\normalsize  Fakult\"at f\"ur Mathematik} \\
{\normalsize  44780 Bochum, Germany}
}

\begin{document}

\doublespacing

        \maketitle

\begin{center}
    \textbf{Abstract}
\end{center}

{We propose a novel two-stage subsampling algorithm based on optimal design principles. In the first stage, we use a density-based clustering algorithm to identify an approximating design space for the predictors from an initial subsample. Next, we determine an optimal approximate design on this design space.  Finally,  we use matrix distances such as the Procrustes, Frobenius, and square-root distance to define the remaining subsample, such that its points are “closest” to the support points of the optimal design. Our approach reflects the specific nature of the information matrix as a  weighted sum of non-negative definite Fisher information matrices evaluated at the design points and applies to a large class of regression models including models where the Fisher information is of rank larger than $1$. 
 }

\noindent%


\section{Introduction} \label{sec1}

Nowadays, with the easy accessibility to data collecting frameworks and computing devices, a large amount of data is encountered in various fields ranging from terrestrial data, manufacturing sector, and e-commerce to name a few. Training statistical models to draw inferences
with such large volumes of data can be very time-consuming. 
A popular method for dealing with large-scale data is sampling, where one performs statistical inference on a subsample, which (hopefully) represents the most informative part of the data and is used as a surrogate. 

Various subsampling algorithms have been proposed in recent times for this purpose using concepts of optimal design theory. Some of the algorithms perform probability-based sampling based on minimizing variances of the predicted response or of the estimator \citep[see, for example][]{ma2015leveragingJMLR, LogisticWang2018optimal}. The others are deterministic \citep[see, for example][]{wang2019information,wang2021orthogonal,ren2021subdata}.

 The IBOSS algorithm \citep{wang2019information} applies to (homoscedastic) linear regression and determines the most informative points by component-wise selecting extreme covariate values. Simulation studies show that it is an improvement over the leverage-based algorithm. \cite{wang2021orthogonal} and \cite{ren2021subdata} consider also the linear regression model and 
use the concept of orthogonal arrays to construct subsampling procedures. An orthogonal array ensures that the covariates are as far apart as possible and also supports the selection of extreme values. As a result, these authors demonstrate by means of a simulation study that orthogonal array-based subsampling is an improvement over IBOSS. 
\cite{LogisticCheng2020information} extend the IBOSS approach to define a deterministic subsampling procedure for the logistic regression. 

Several non-deterministic subsampling procedures have been developed for these regression models as well. For example, \cite{ma2015leveragingJMLR} give a leverage-based algorithm for linear regression, that performs a probability-based subsample with replacement based on the leverage scores of each data point. \cite{LogisticWang2018optimal} consider two-stage subsampling algorithms for the logistic regression model, where the first stage consists in taking a random sample, which is used to obtain a preliminary estimate of the parameter. They then propose a probability score (depending on the initial estimate of the parameter in stage I) based subsampling with replacement based on the A-optimality criterion.  It has been improved by \cite{wang2019more} and extended to several other models and estimation techniques \citep[see][among others]{wangma2021,AiWangYuZhang2021,YuWangAiZhang2022}. 
In principle, these non-deterministic methods apply to a broad class of regression models and we refer to the recent review paper by  \cite{yao_wang_2021} for the current state of the art.

\HD{In this paper, we propose a new deterministic subsampling algorithm that can be easily applied to a wide class of regression models and utilizes different optimality criteria depending on the goals of the experiment.}
Our general idea is to interpret sampling as an optimal design problem. For this purpose, the proposed algorithm first tries to understand the nature/shape of the design space which, roughly speaking, is the domain containing {most of} the predictors of the sample. Thereafter, we determine a subsample approximating the optimal design on the identified design space.  More precisely, in the first step, we obtain an initial parameter estimate (if necessary) and determine an approximation of the design space using a density-based clustering algorithm. Next, we determine the optimal design on the ``estimated design space''. When the number of support points of the optimal design is large, we use the concept of efficiency to eliminate the most unimportant points from the optimal design, which ensures much less execution time for the subsequent subsampling step. 
{To identify the points in the subsample, which are ``closest'' to the remaining support points of the optimal design, we use matrix distances such as the Procrustes, Frobenius, and square-root distance \citep[see, for example,][for a definition]{DrydenKoloydenkoZhou2009,pigoli2014distances}.} 

The algorithm, which is most similar in spirit is the subsampling procedure proposed recently by  \cite{deltom2022} and also accounts for the shape of the design space.  Our approach differs from this work with respect to at least three perspectives.   First, these authors propose to determine an optimal design (more precisely its weights) using the full sample as design space, which is computationally expensive for large-scale data. In contrast, our approach applies clustering techniques to determine a (discrete) design space from an initial subsample (which is also used to obtain initial estimates of the parameter). It is therefore computationally much cheaper because the cardinality of the estimated design space is substantially smaller than the size of the total sample.  Second, it is robust with respect to outliers due to density based clusters identifying the design space. 
Third, we propose to utilize the geometry of the space induced by the Fisher information matrix and thus use matrix distances on the space of non-negative definite matrices \citep[see][]{DrydenKoloydenkoZhou2009,pigoli2014distances} to identify the remaining points for the subsample, such that they are close to the support points of the optimal design.
This reflects the specific nature of the information matrix which is a weighted sum of the non-negative definite Fisher information matrices evaluated at the design points. As a consequence,  our approach is applicable for regression models where the Fisher information is of rank larger than $1$, as they appear, for example, in heteroskedastic regression models \citep[see][]{atkinsoncook1995,dette2009geometric}

The rest of the paper is organized in the following way: Section \ref{sec2} discusses the models under consideration and the idea of approximate optimal design. The subsampling algorithm is introduced in Section \ref{sec3}. 
Finally,  we demonstrate in Section \ref{sec4} through simulation studies that the new approach has better accuracy compared to existing subsampling approaches proposed in \cite{LogisticCheng2020information} and \cite{LogisticWang2018optimal} in the case of a logistic regression model. Compared to the optimal design-based subsampling approach proposed in \cite{deltom2022}, our approach has improved time complexities. In this section, we also illustrate the applicability of our subsampling approach to a heteroskedastic regression model, where the Fisher information has rank larger than $1$.


\section{Preliminaries} \label{sec2} 

  \def\theequation{2.\arabic{equation}}	
  \setcounter{equation}{0}
  
We assume that the full data sample 
$\mathcal{D}  = \{ (  \VectorXi, y_{i}): i= 1, \ldots, n\}$  are realization of 
i.i.d. $\mathbb{R}^{p+1}$-dimensional 
random variables with (conditional) distribution  from an exponential family with density (with respect to a dominating measure, say $\nu$  defined on an appropriate sigma field)
\begin{align} \label{hd21a}
\mathcal{L}(y| \VectorX, \VectorBeta) &= h(y)\; \exp \{ \eta^\top(\VectorX, \VectorBeta) T(y) - A(\pmb x, \VectorBeta)\} ,
\end{align} 
where ($\VectorX$,$y$) is defined on ${\cal X} \times {\cal Y} \subset \mathbb{R}^{p+1}$,
$\VectorBeta = (\beta_0, \beta_1,\ldots, \beta_{p})^\top  \in \Theta \subset \mathbb{R}^{p+1}$ denotes the unknown vector of parameters, $\VectorX$ is a $p$-dimensional predictor and $y$ a univariate response. Here, $ h(y)$ is assumed to be a positive measurable function, $\eta: {\cal X} \times \Theta \to \mathbb{R}^l $, $A: {\cal X} \times \Theta \to \mathbb{R}$, and $T $ denote a $l$-dimensional statistic.

We denote by $ \hat{ \VectorBeta} = ( \hat\beta_0,  \hat\beta_1,\ldots,  \hat\beta_{p})^\top \in \mathbb{R}^{p+1} $ the maximum likelihood estimate of the parameter $\pmb \beta$ from the full sample $\mathcal{D}$  and recall that under standard regularity assumptions 
\citep[see,  for example, Theorem 5.1 in][]{lehmann2006theory}
the statistic $(\mathcal{I}(\VectorBeta, \MatrixX) )^{1/2} \big  ( \hat{ \VectorBeta} - \VectorBeta \big )  $ is asymptotically normal distributed with mean vector $0$ and covariance matrix  $ I_{p+1}$,
where  $\mathcal{I}(\VectorBeta, \MatrixX)$ denotes
the information matrix defined by
\begin{align}  
\mathcal{I}(\VectorBeta, \MatrixX) &= 
\sum_{i=1}^n \mathcal{I}(\VectorBeta, \VectorXi), \label{Information_Matrix_Eqution_2} 
\end{align}
$\MatrixX = [\VectorxOneSample, \ldots,  \VectorxnSample]^\top$ and   
\begin{align}
\mathcal{I}(\VectorBeta, \VectorX ) &= \mathbb{E}\Big[ \Big\{  \dfrac{\partial \log f(y| \VectorX, \VectorBeta)  }{\partial \VectorBeta}    \Big\}^\top \;\; \Big\{  \dfrac{\partial \log f(y| \VectorX, \VectorBeta)   }{\partial \VectorBeta}  \Big\} 
\Big]  \label{Information_Matrix_Eqution_1}
\end{align}

is the Fisher information matrix at the point $\VectorX$.
We are interested in identifying a most informative subsample from ${\cal D}$ of size $k$, say 
\begin{align*}
\mathcal{D}_{k} &= \{ (  \VectorXsi, y_{s_i}): i= 1, \ldots, k\} ~\subset ~\mathcal{D}.
\end{align*}and denote by  
$\hat\VectorBeta_{\mathcal{D}_k} $   the parameter estimate based on the {subsample} $\mathcal{D}_{k}$. 
Before we continue, we present three examples, which will be in the focus of this paper. \\

\begin{example} \label{ex1} 
{\rm
  Assume that ${\cal Y} = \mathbb{R}$, 
    $\nu$ is the Lebesgue measure, and  that $g({\pmb x, \pmb \beta})$ is a sufficiently smooth function. Then, for 
    $$
    h(y) = \dfrac{\exp{(- {y}^2/2 \sigma^2)}}{\sqrt{2 \pi \sigma^2}}, ~ \eta(\VectorX, \VectorBeta ) = \dfrac{g({\pmb x, \VectorBeta})}{\sigma^2 } ,~ T(y) = y, ~ A(\pmb x, \VectorBeta)= \dfrac{ g({\pmb x, \VectorBeta})^2}{2  \sigma^2} 
    $$ 
    we obtain the normal distribution (with known variance $\sigma^2$) and the common nonlinear regression model with $$
    \mathbb{E} [y | \VectorX, \VectorBeta ] = g({\pmb x, \VectorBeta}) ~.
    $$
Here the  Fisher-information matrix 
    for the parameter $\VectorBeta$ at the point $\VectorX$ is given by 
    \begin{align}
    \mathcal{I}(\VectorBeta , \VectorX ) &= \dfrac{\partial g(\VectorX, \VectorBeta )^\top   }{\partial\VectorBeta }  \;   \dfrac{\partial  g(\VectorX,\VectorBeta ) }{\partial \VectorBeta }  \label{Information_Matrix_Eqution_NonLinearRegression_1} .
    \end{align}
    In particular, if $
    g(\VectorX, \VectorBeta )= f(\VectorX)^\top \VectorBeta 
    $ for a $(p+1)$-dimensional vector of regression functions $f$,  this model reduces to the common linear regression model.
In this case, the Fisher-information at the point $\VectorX$ is given by 
$    \mathcal{I}(\VectorBeta, \VectorX ) =  f(\VectorX) f^\top(\VectorX) $.

}
\end{example}

\medskip

\begin{example} \label{ex2} 
{\rm
    If ${\cal Y} = \{0, 1\} $, $\nu $ is the counting measure
    and 
    \begin{align*}
    h(y) & = 1, ~ \eta (\VectorX,\VectorBeta ) =
    \log \big [ \pi (\VectorX, \VectorBeta ) / (1- \pi (\VectorX, \VectorBeta ) \big ], \\
    T(y) & = y, ~A(\pmb x,\VectorBeta )= -\log(1-\pi (\VectorX, \VectorBeta )) 
    \end{align*}
    for some function $\VectorX \to \pi (\VectorX, \VectorBeta ) \in (0,1)$,
    we obtain the Binomial response model, that is
\begin{align}
    \label{hd1}
   \HD{  \mathcal{L}(y_i| \VectorXi, \VectorBeta) = \pi (\VectorX_i, \pmb \beta ) ^{y_i}  \big (1-\pi (\VectorX_i, \pmb \beta )  \big )^{1-y_i}~},
\end{align}
and 
\begin{align} \label{hd2} 
       \mathbb{P}(y =1 | \VectorX ,\pmb \beta )
       = \pi (\VectorX, \VectorBeta ).
\end{align}
For the special choice 
$$
\pi (\VectorX, \VectorBeta )        = \frac{\exp( \pmb z^\top \pmb \beta     )}{1+\exp( \pmb z^\top \pmb \beta   )}
$$
where $\pmb z  = (1 , \pmb x^\top)^\top$
 we get the frequently used logistic regression model with parameter $\VectorBeta$. 
In this case, the Fisher-information at the point $\VectorX$ is given by 
  \begin{align}
    \mathcal{I}(\VectorBeta, \VectorX ) &=  \big(\phi(\VectorZ,\VectorBeta) \VectorZ \big)  \;\; \big(\phi(\VectorZ,\VectorBeta) \VectorZ \big)^\top \label{Information_Matrix_Eqution_LogisticRegression_1},
    \end{align}
    where 
    $$
    \phi(\VectorZ,\VectorBeta)= \displaystyle{\dfrac{\exp(\VectorZ^\top\VectorBeta/2)}{1+\exp(\VectorZ^\top\VectorBeta)}}.
    $$

}

\end{example}

\medskip 

\begin{example} \label{ex3} 
{\rm
 Assume that ${\cal Y} = \mathbb{R}$, $\nu$ is the Lebesgue measure, and  that $g({\pmb x, \VectorBeta})$ is a sufficiently smooth function. Then, for $T(y) = \big[y^2 ~~ y \big]^\top, ~$
    $$
    h(y) = \dfrac{1}{\sqrt{2 \pi}}, ~ \eta(\VectorX, \VectorBeta ) = \Big[ \dfrac{-1}{2 \sigma^2({\pmb x, \VectorBeta}) } ~~ \dfrac{g({\pmb x, \VectorBeta})}{\sigma^2({\pmb x, \VectorBeta}) }   \Big]^\top ,~  A(\pmb x, \VectorBeta)= \dfrac{g^2({\pmb x, \VectorBeta})}{2 \sigma^2({\pmb x, \VectorBeta}) } - \dfrac{1}{2} \log (\sigma^2({\pmb x, \VectorBeta}))
    $$ 
    we obtain the normal distribution and the common heteroskedastic nonlinear regression model with
    \begin{align}
    \mathbb{E} [y | \VectorX, \VectorBeta ] & = g({\pmb x, \VectorBeta}) ~,~~
     {\rm Var} [y | \VectorX, \VectorBeta ] = \sigma^2({\pmb x, \VectorBeta}).
    \end{align} In this case, Fisher-information matrix for the parameter $\VectorBeta$ at the point $\VectorX$ is given by 
    \begin{align}
    \mathcal{I}(\VectorBeta , \VectorX ) &= \dfrac{1}{\sigma^2({\pmb x, \VectorBeta}) } \dfrac{\partial g(\VectorX, \VectorBeta)^\top}{\partial\VectorBeta}  \;   \dfrac{\partial  g(\VectorX,\VectorBeta )}{\partial \VectorBeta }   +  \dfrac{1}{2 \sigma^4({\pmb x, \VectorBeta})} \dfrac{\partial \sigma^2({\pmb x, \VectorBeta})^\top }{\partial\VectorBeta }  \;   \dfrac{\partial {\sigma^2({\pmb x, \VectorBeta})} }{\partial \VectorBeta }  \label{Information_Matrix_Eqution_NonLinearRegression_3}
    \end{align} \cite[see][]{dette2009geometric}. Note that in general the matrix in \eqref{Information_Matrix_Eqution_NonLinearRegression_3} has rank $2$ (in contrast to Example \ref{ex1} and \ref{ex2}).

}
\end{example}

\medskip



In the following, we will develop an algorithm for selecting a subsample $\mathcal{D}_k$ from $\cal D$ 
 such that the resulting maximum likelihood estimator $\hat{\VectorBeta}_{\mathcal{D}_k}$
 based on ${{\mathcal{D}_k}}$  is most efficient.
 Similar to \cite{deltom2022} the subsampling algorithm proposed in this article is based on optimal design principles. For this purpose, we briefly describe some basic facts from optimal design theory 
 \citep[see, for example, the monographs of][for more details]{silvey1980optimal,Puke1993,Randall2007}
and present some tools, which will be useful for our approach.
Following \cite{kiefer1974general} we define an  approximate design $\xi$ on  a given  design space ${\cal X} $ (which will be  determined as described in Section \ref{sec32} below) as a probability measure with weights  $w_1, \ldots , w_b$ 
at the points $\x_1, \ldots , \x_b  \in {\cal X}$.  If
$N$ observations can be taken, the quantities ${w}_{\ell} N$ are rounded to non-negative integers, say $N_{\ell}$, such that $\sum_{\ell=1}^{b} N_{\ell} =N$ and the experimenter takes $N_{\ell}$ (independent) observations at each $\x_{\ell}$ ($\ell =1,  \ldots , b$). In this case, under standard assumptions,  the covariance matrix of the maximum likelihood estimator
$ \sqrt{N} \hat{ \VectorBeta} $  for the parameter $ { \VectorBeta}  $ in model \eqref{hd21a}
converges to the matrix $M^{-1}( \xi , \VectorBeta )$, where 
$$
M( \xi , \VectorBeta ) := 
\sum_{i=1}^b  w_i \mathcal{I}(\VectorBeta, \VectorXi), 
$$
which is the analog of the matrix \eqref{Information_Matrix_Eqution_2} and 
used to measure the accuracy of the estimator $ \hat{\VectorBeta}$.
An optimal design maximizes an appropriate functional, say $\Psi$,  of the matrix $M(\xi, \VectorBeta)$  
with respect to the design $\xi$. Here $\Psi$ is an information function in the sense of \cite{Puke1993}, that is a positively homogeneous, concave, non-negative, non-constant, and upper semi-continuous function on the space of non-negative definite matrices and the optimal design is called $\Psi$-optimal design. Numerous criteria have been proposed in the literature to discriminate between competing designs \citep[see][]{Puke1993} and we exemplary mention  
Kiefer's $\Psi_q$-criteria, which are defined for $-\infty \leq q < 1$ as
\begin{align}  
\Psi_q(\xi) = (\text{tr}  \big  \{  \big  ( M^{q }(\xi , \VectorBeta)   \big  \} \big  )^{1/q} 
&= \Big( \text{tr}  \Big  \{  \Big  (  \sum_{i=1}^b  w_i \mathcal{I}(\VectorBeta, \VectorXi) \Big)^q \Big \} \Big)^{1/q}, \label{2.4}
\end{align}
and contain the famous  $E$-optimality $(q=-\infty )$, $A$-optimality $(q=-1)$ and  $D$-optimality   $(q=0)$ criterion as special cases.
Under some continuity assumptions a $\Psi$-optimal design, say  
\begin{align*}
    \xi^\ast(\VectorBeta, {\pmb \chi}) = \biggl\{   \genfrac{}{}{0pt}{}{\VectorxOneOptimal}{w^\ast_1} \; \genfrac{}{}{0pt}{}{\VectorxTwoOptimal}{w^\ast_2} \ldots \genfrac{}{}{0pt}{}{\pmb x^\ast_b}{w^\ast_b} \biggl\},
\end{align*}
maximizing \eqref{2.4} exists, where $\VectorxOneOptimal , \VectorxTwoOptimal,\ldots ,\pmb x^\ast_b \in {\pmb \chi}$ and $ w^\ast_1+ w^\ast_2+\ldots + w^\ast_b =1 $ are the support points and weights of the optimal design. Note that in general, this design depends on the unknown parameter $\VectorBeta$ and on the design space ${\cal X}$, which is reflected in our notation. Of course, the design also depends on the optimality criterion $\Psi$. However, as the criterion does not play an important role in the following discussion (in fact the proposed method is basically applicable for any information function in the sense of \cite{Puke1993}), this dependence will not be reflected in the following discussion.

In most cases of practical interest, a $\Psi$-optimal design is unknown and has to be found numerically, and for a given design $\xi$ 
its $\Psi$-efficiency is defined by 
\begin{align}
\label{det1}
\mbox{eff} (\xi , \VectorBeta)  = {\Psi  (M (\xi , \VectorBeta) ) \over  \Psi  ( M(  \xi^\ast(\VectorBeta, {\pmb \chi}),  \VectorBeta )) } 
\in [0,1]
\end{align}

\section{Optimal Design Based Sub-sampling } \label{sec3}
  \def\theequation{3.\arabic{equation}}	
  \setcounter{equation}{0}
  
In this section, we explain the basic structure of the proposed algorithm to obtain a subsample of size $k$, which will be called ODBSS and summarized in Algorithm \ref{Algorithm1}: first, we use an initial sample to obtain an {\it estimate of a design space} and of the parameter $\pmb  \beta$. Second, based on the estimated design space and the parameter estimate, we determine the (locally) optimal design
with respect to some optimality criterion. Third, we determine a subsample by choosing the data points, which are close to the support points of the optimal design. For this purpose, we will define an appropriate ``metric'' not on the set $\pmb \chi $ but on the set of Fisher informations $\{ \mathcal{I}(\pmb \beta, \pmb x) | \pmb x \in \pmb \chi$\}.

\begin{figure}[H]
	\centering
	\begin{subfigure}{.5\textwidth}
		\centering
		\includegraphics[width=.9\linewidth]{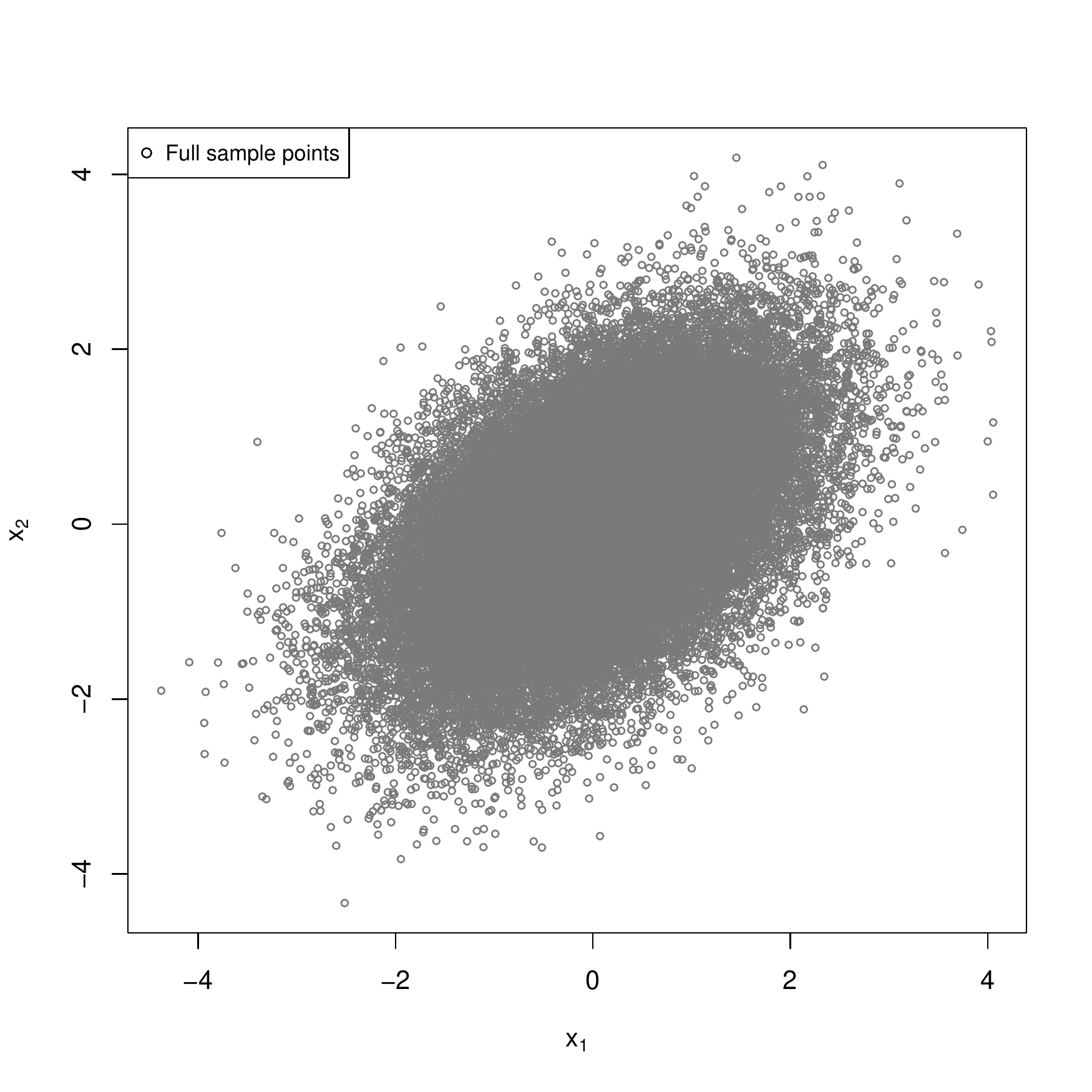}
		\caption{\scriptsize Covariates of full data $(n=50000)$.    }
		\label{fig1_1}
	\end{subfigure}%
	\begin{subfigure}{.5\textwidth}
		\centering
		\includegraphics[width=.9\linewidth]{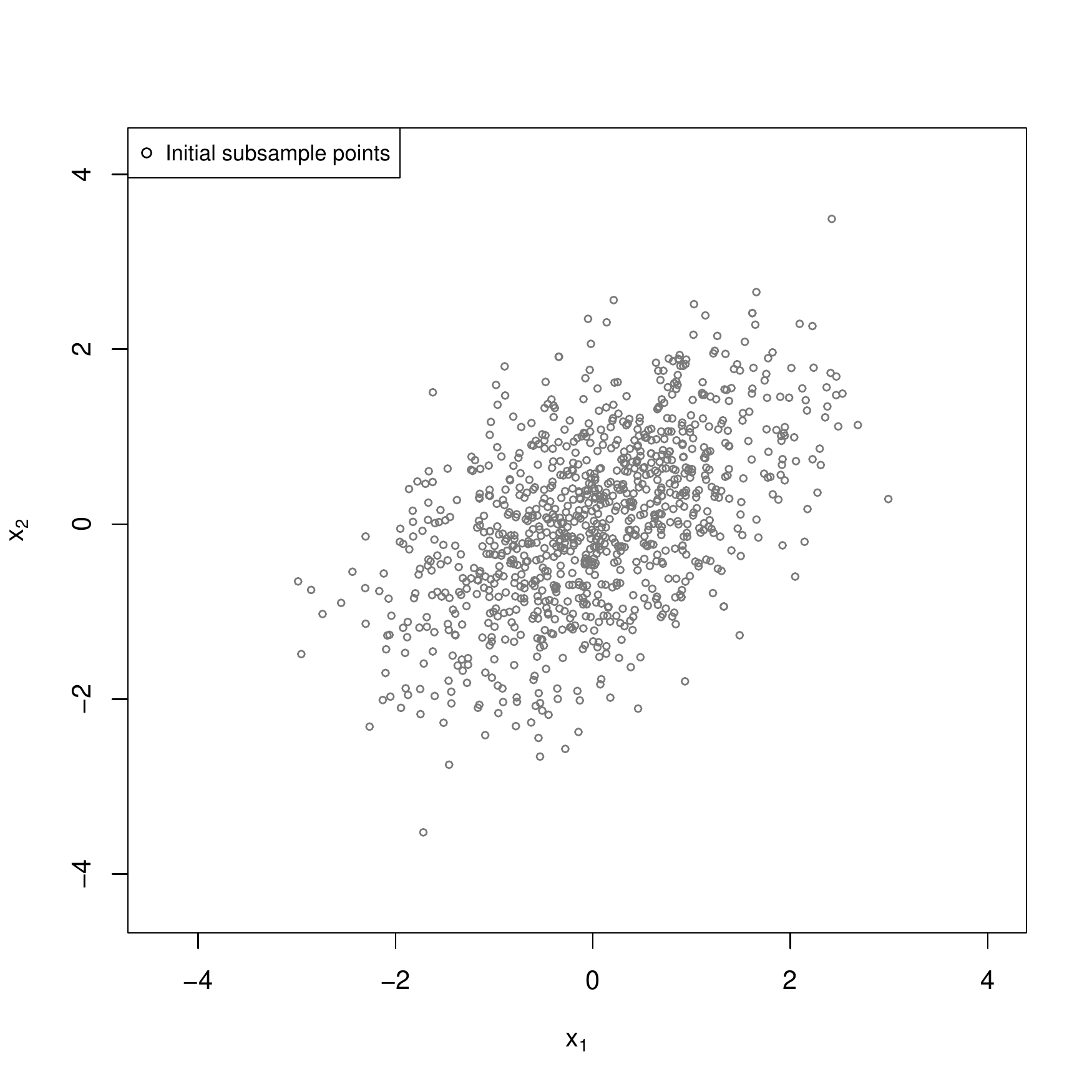}
		\caption{\scriptsize Initial uniform  subsample ($k_0 = 1000$). 
  }
		\label{fig1_2}
	\end{subfigure}
	\vspace{-1em}
	\centering
	\begin{subfigure}{.5\textwidth}
		\centering
		\includegraphics[width=.9\linewidth]{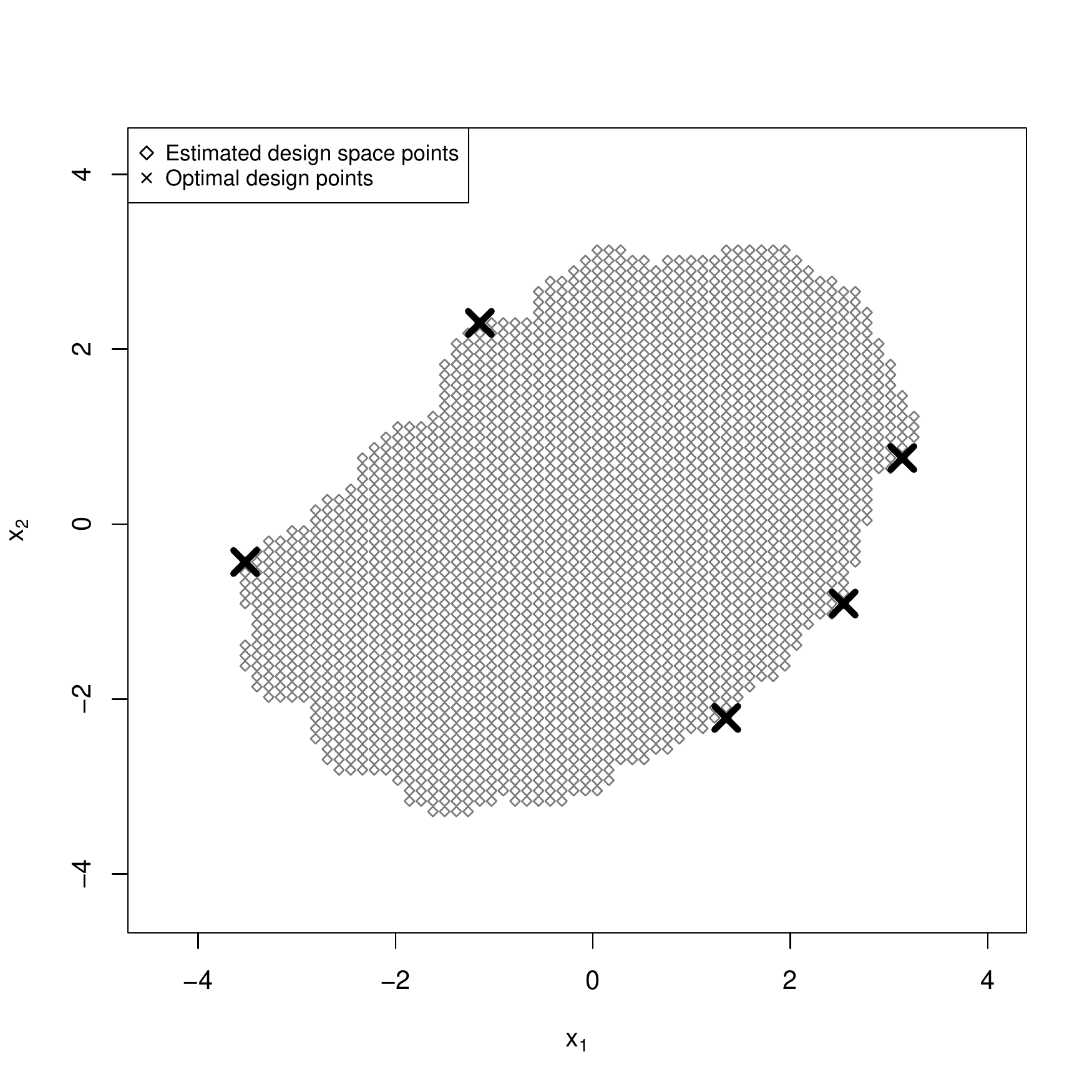}
		\caption{ \scriptsize Estimated design space and optimal approximate design.  }
		\label{fig1_3}
	\end{subfigure}%
	\begin{subfigure}{.5\textwidth}
		\centering		\includegraphics[width=.9\linewidth]{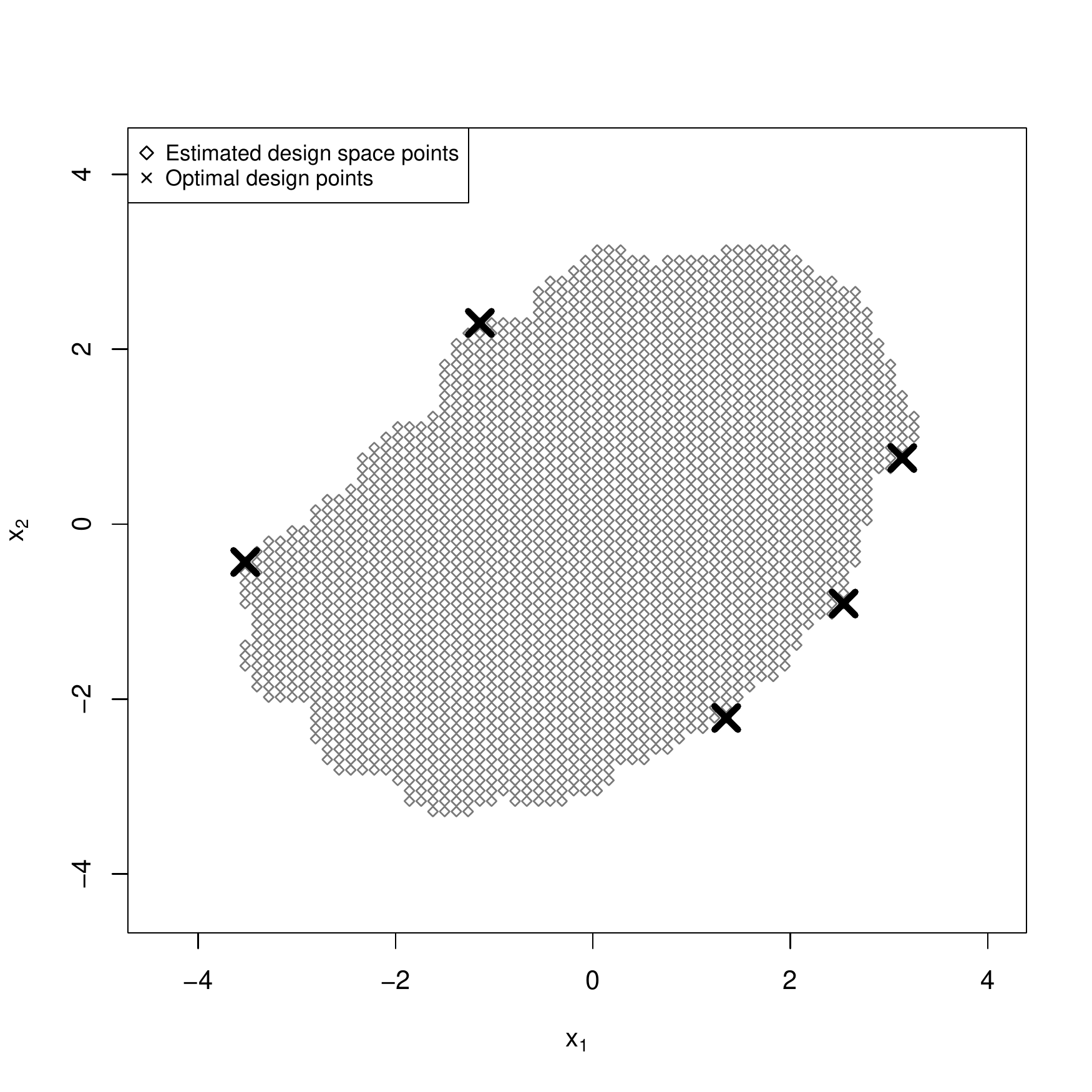}
		\caption{ \scriptsize Final ODBSS subsample($k = 5000$)   }
		\label{fig1_4}
	\end{subfigure}
	\vspace{1em}
	\centering	\caption{ \it Illustration of key steps in ODBSS algorithm for a logistic regression model. }
	\label{fig1} 
\end{figure}

Before we give details for each step in Section \ref{sec31} - \ref{sec33}, we illustrate our approach in a concrete example.  In Figure \ref{fig1}, we display a typical situation, where we use ODBSS to determine a subsample for estimating the parameter of a logistic regression model with $p=2$ covariates, and where the true parameter is given by $\pmb \beta = (0.1,0.5,0.5)$. The full sample size is $n=50000$ and we want to find the most informative subsample of size $k=5000$. Figure \ref{fig1_1} displays the corresponding predictors simulated from a $2$-dimensional normal distribution centered at $(0,0)^\top $ with 
unit variances and correlation $0.5$. Figure \ref{fig1_2} displays predictors corresponding to the initial subsample of size $k_{0}=1000$, which is chosen by uniform random sampling. This initial subsample  $\mathcal{D}_{k_0}$  is used for two purposes: first, to obtain an initial parameter estimate (say $\hat{\VectorBeta}_{\mathcal{D}_{k_0}}$) for the parameter $\VectorBeta$, and second, to get an estimate of the  ``design space''  (say ${{\pmb\chi}}_{k_0}$), which is used in the calculations for obtaining the optimal design. Figure \ref{fig1_3} shows the estimated design space ${{\pmb\chi}}_{k_0}$ and  the support points of the  $A$-optimal design $ \xi^\ast(\hat{\VectorBeta}_{\mathcal{D}_{k_0}},  {{\pmb\chi}}_{k_0}) $ on the design space ${{\pmb\chi}}_{k_0}$. The details of design space estimation and optimal design determination can be found in Section \ref{sec31} and \ref{sec32}, respectively. Finally, Figure \ref{fig1_4} shows the $k-k_{0}$ (= 4000) optimal subsample points along with the initial subsample, such that they are in some sense close to the support points of the optimal design $   \xi^\ast(\hat{\VectorBeta}_{\mathcal{D}_{k_0}}, {{\pmb\chi}}_{k_0}) $ (see Section
\ref{sec33} for more details). 

\begin{algorithm}[H] 
\caption{ODBSS}
\label{AlgorithmMain}  \label{Algorithm1}
\smallskip
{\textbf{Input:}} The sample  $\mathcal{D}$   of size $n$
\begin{itemize}
\item[(1)] 
\textbf{Initial sampling {and estimation of the design space} } 
    \begin{itemize}
    \item[(1.1)]  Take a uniform subsample of size $k_0$ denoted by $\mathcal{D}_{{k_0}}$ 
    \item[(1.2)]   Find an estimate of the  design space ${{\pmb\chi}}_{k_0}$ based on $\mathcal{D}_{{k_0}}$
    \item[(1.3)]  Calculate an initial parameter estimate $\hat{\VectorBeta}_{\mathcal{D}_{k_0}}$ based on $\mathcal{D}_{{k_0}}$
\end{itemize}
    
\item[(2)]  \textbf{Optimal design determination }   
  
  Find a (locally)  approximate optimal design
         $
     \xi^\ast(\hat{\VectorBeta}_{\mathcal{D}_{k_0}}, {{\pmb\chi}}_{k_0}) 
         $
on the design space  ${{\pmb\chi}}_{k_0}$ 
 for the parameter $\hat{\VectorBeta}_{\mathcal{D}_{k_0}}$
\item[(3)] \textbf{Optimal design based sampling} 
    \begin{itemize}
    \item[(3.1)]
        {Determine the remaining subsample $\mathcal{D}_{{k_1}}$ ($k_1 =  k - k_0$), such that,    $\lfloor w^\ast_i ~  k_1 \rfloor$ 
        observations are ``close'' to the support points $\VectorxiOptimal$  of  the optimal design $
        \xi^\ast(\hat{\VectorBeta}_{\mathcal{D}_{k_0}}, {{\pmb\chi}}_{k_0}) $
        ($i=1, \ldots , b$)}
        \item[(3.2)] The final subsample  $\mathcal{D}_{{k}} = \mathcal{D}_{{k_0}} \cup \mathcal{D}_{{k_1}}  $  
     \end{itemize}
\end{itemize}
{\textbf{Output:}} The subsample  $\mathcal{D}_{{k}}$ of size $k$ 
\end{algorithm}

\begin{remark} {\rm
In cases, where the optimal design does not depend on the parameter (for example in linear models) the parameter estimation in step (1) of Algorithm \ref{Algorithm1} is not necessary (but the design space has still to be estimated).
Moreover, we emphasize that the points of the optimal design calculated in step (2) of Algorithm \ref{Algorithm1} are not necessarily part of the original sample $\mathcal{D}$.}
\end{remark}

\subsection{\HD{Step 1: {Initial sampling} and estimation of the design space} }\label{sec31}

To obtain the  required design space ${\pmb \chi}_{k_0}$ from the 
 the initial set $\mathcal{D}_{{k_0}}$
we  view it as   a union of clusters
and use the set of clusters in $\mathcal{D}_{{k_0}}$ to estimate ${\pmb \chi}_{k_0}$. We found that a density-based clustering algorithm popularly referred to as {DBSCAN} \citep[][]{DBSCAN_1996_Paper} is suitable for our problem. In the following we describe 
the two main steps for design space estimation using DBSCAN.
\begin{enumerate}
    \item[(a)]
    \textit{Training DBSCAN from the initial sample $ {\mathcal{D}}_{k_0} $.}  DBSCAN identifies several  clusters labeled by $1,2,\ldots , m $
    and  a  cluster of  ``outliers''  labeled by $ 0$ using two 
 inputs: a constant  $\epsilon >0$  and 
 an integer $ m_p > 0$. 
    Consider first the case where there is only one cluster in $\mathcal{D}_{{k_0}}$. A point $\pmb c \in \mathcal{D}_{{k_0}} $ is said to be a {\it core point}, if its $\epsilon$ neighbourhood $N_\epsilon(\pmb c)$
    contains at least $m_p$ points from $\mathcal{D}_{{k_0}}$, that is $|N_\epsilon(\pmb c) \bigcap \mathcal{D}_{{k_0}}|\geq m_p $. A point $\pmb 
   c^\prime  \in  \mathcal{D}_{{k_0}} $ is said to be a boundary point, if $\pmb c^\prime \in N_\epsilon(\pmb c)$ but $|N_\epsilon(\pmb c^\prime) \bigcap \mathcal{D}_{{k_0}}| < m_p $, where $\pmb c$ is any core point. The cluster is the union of the core points and the boundary points.
If there is more than one cluster present in the data, using a notion of {\it density reachability } and {\it density connectedness}, this algorithm identifies the different clusters  \cite[for more details see][]{DBSCAN_1996_Paper}. 
We denote by  $\mathcal{C} = \{ C_1, C_2,  \ldots , C_m \} $  the set of clusters present in the initial sample  $\mathcal{D}_{{k_0}}$ and by $C_0$ the set of outliers.

\HD{
In our simulation studies, we use the R-implementation in the $dbscan$ package in $R${\it -software} \citep[see][]{R_Software_DBSCAN}. 
The tuning parameters $\epsilon$ and $m_p$ are set on the lines of the recommendation for parameter tuning in \citep[][]{DBSCAN_1996_Paper}. We set $m_p= 5$ to ensure a reasonable definition for the cluster, that is, a cluster point $\pmb c$ must contain at least 5 points in $N_{\epsilon}(\pmb c)$.  \HD{For this purpose setting the parameter $\epsilon$ is more important}. Let $NN(\pmb x,k)$ denote the $ k^{th} \text{ nearest neighbour of } \pmb x$ and let $dist_k(\pmb x) : = \| \pmb x - NN(\pmb x,k) \|_2 $. Based on our numerical studies we recommend  \begin{equation}
    \epsilon =  \min\Bigg\{0.1 (p-1) \Big(\max_{i,j}  {( {\pmb X}_{\mathcal{D}_{k_0}})}_{ij}- \min_{i,j}  {({\pmb X}_{\mathcal{D}_{k_0}})}_{ij}\Big) , \max_{\pmb x \in  \mathcal{D}_{k_0}}(dist_4(\pmb x)) \Bigg\}, \label{eq_3.1}
\end{equation} where $\pmb X_{\mathcal{D}_{k_0}}$ is 
the design matrix corresponding to the initial subsample $\mathcal{D}_{k_0}$. The first term in the minimum in \eqref{eq_3.1} provides a bound for $\epsilon$ in the case where covariates have a heavy-tailed distribution. This choice of $\epsilon$ allows the majority of the points in $\mathcal{D}_{k_0}$ to be included in the cluster except for outliers and also allows the identification of separate clusters. Most importantly, any given (or new) point is included in the cluster if it is reasonably close to the density cloud of $\mathcal{D}_{k_0}$. The original article \citep{DBSCAN_1996_Paper} recommends a graphical approach for parameter tuning. However, this is not practical for our problem as such a graphical parameter tuning step would make ODBSS time consuming. We simply take the maximum of $4^{th}$ nearest neighbor distance to ensure that we make a loose approximation of the design space and leave out as little as possible from the initial sample $\mathcal{D}_{k_0}$ (as $\mathcal{D}_{k_0}$ is relatively sparse compared to the full data). This also allows adjustment for the fact that the design space could be generated from different distributions. }
 
\item[(b)]  \textit{Using DBSCAN for estimating a design space.} 
Once the DBSCAN model is trained we can, in principle, decide for any $\pmb c^{\prime\prime} \in \mathbb{R}^p$ if it belongs to  a cluster from $\mathcal{C}$ or if it is an outlier.  More precisely, if $\pmb c^{\prime\prime} \in  N_\epsilon(\pmb c)$ for some core point $\pmb c\in \mathcal{D}_{{k_0}}$ associated with a  cluster $C_i$, then $\pmb c^{\prime\prime} \in {C}_i$. Thus, after training  DBSCAN formally defines a function,  which assigns each point to a cluster and for a given set $ \mathcal{M} \subset \mathbb{R}^p$,  we obtain a decomposition 
$$
\mathcal{M}  = \mathcal{M}_0 \cup   \mathcal{M}_1 \cup   \mathcal{M}_2 \cup   \ldots ~
\cup   \mathcal{M}_m, 
$$
where $ \mathcal{M}_j   $ contains the points 
in  $\mathcal{M}  $ which are assigned to $C_j$
$(j=1, \ldots  ,m)$. Note that $\mathcal{M}_j$ does not necessarily contain points  from the 
cluster $C_j \subset \mathcal{D}_{k_0}$, it just contains all points from $\mathcal{M}$, which are ``close'' to the  set  $C_j $.
We  define for a set $ \mathcal{M} \subset \mathbb{R}^p$ the function 
$$
\Omega_{\mathcal{D}_{{k_0}}}( \mathcal{M} ) = \mathcal{M}_1  \cup \ldots  \cup  \mathcal{M}_m  ~,
$$
where the index  $\mathcal{D}_{{k_0}} $ reflects the fact that the clusters are defined by the initial sample  $\mathcal{D}_{{k_0}}$. We finally define the estimated design space by 
\begin{align}  \label{31}
{{\pmb\chi}}_{k_0} &= \Omega_{ \mathcal{D}_{{k_0}}}( \mathcal{G} )    ~, 
\end{align}
where  
\begin{align*} 
\mathcal{G}  &=  \Big  \{    ({i_{1} } , \ldots , { i_{j} })^{\top}~\Big |~ i_{j} \in 
\Big  \{ \underline{K}_{j},  \tfrac{L \underline{K}_{j} + (\overline{K}_{j}-   {\underline{K}_{j} })}{ L} , \ldots, \tfrac{L \underline{K}_{j} + (\overline{K}_{j}-   {\underline{K}_{j} })(L-1)} {L}, \overline{K}_j
\Big  \} ; ~j=1, \ldots , p  \Big    \}~
\end{align*}
is a grid 
and the bounds  $\underline{K}_{1}, \ldots ,\underline{K}_{p}$ and $\overline{K}_{1}, \ldots ,\overline{K}_{p}$ are chosen such that the grid covers the covariate space corresponding to the initial sample $\mathcal{D}_{k_0}$ and $L$ corresponds to the partition size of the grid. 

\HD{\item[($\text{b}^\prime$)]  \textit{Using  DBSCAN for estimating a design space in high dimensional settings.} }
 \HD{While the approach in step (b) works well for moderate dimensions $p$, two challenges arise when $p$ is large: first, for moderate values of $L$ the grid $\mathcal{G}$ is sparse in $\mathbb{R}^{p}$ and second, we cannot take high values of $L$ as that makes enumerating $\mathcal{G}$ computationally infeasible due to memory constrains of any software. Therefore, in higher dimensions, we have to adopt a different approach to estimate the design space ${\pmb\chi}_{k_0}$. As before perform the step (a) for training $DBSCAN$ algorithm to identify the clusters $\mathcal{C}$ in $\mathcal{D}_{k_0}$. Instead of step (b), that is, generating the grid $\mathcal{G} $ and approximating the design space, we generate samples from the uniform distribution on the clusters $\mathcal{C}$ using the random walk Metropolis-Hastings algorithm \cite[see ][page 287]{robert1999monte}. To be precise we first consider the case, where $\mathcal{C}$ has only one cluster, and generate a sample of $t$ different points ${\pmb c}^{(1)},\ldots, {\pmb c}^{(t)} $ from the target distribution $\frac{1}{\text{Vol}(\mathcal{C})} \mathbbm{1}_{\mathcal{C}}(\cdot)$ by the recursion
 \begin{align}
     \label{det101}
 {\pmb c}^{(\ell)}=
\begin{cases}
    {\pmb c}^{(\ell-1)} + \Delta_{\ell-1} & \text{with probability } \min\left\{1, \dfrac{\mathbbm{1}_{\mathcal{C}}({\pmb c}^{(\ell-1)} + \Delta_{\ell-1} )}{\mathbbm{1}_{\mathcal{C}}({\pmb c}^{(\ell-1)})}\right\}\;\;\;\; \ell = 1, \ldots, t, \\
    {\pmb c}^{(\ell-1)} & \text{otherwise } 
\end{cases}
 \end{align}
where $\Delta_1, \ldots, \Delta_t $ are independent and random variables with a (centered) $p$-dimensional $t$-distribution with 3 degrees of freedom and ${\pmb c}^{(0)}$ is some random point in the cluster $\mathcal{C}$. We consider the estimate of the design space to be $\pmb \chi_{k_0} = \{{\pmb c}^{(1)},\ldots, {\pmb c}^{(t)} \}$. We keep generating the sample ${\pmb c}^{(\ell)}$ until the number of times where ${\pmb c}^{(\ell-1)} + \Delta_{\ell-1}$ is accepted in \eqref{det101} is equal to $5 \{  \frac{1}{2} p(p+1)+1 \}$. Therefore, $t$ in each case varies and depends upon $\mathcal{C}$. {This approach ensures that $\pmb \chi_{k_0}$ is specified by sufficient (distinct) points and an optimal design can be determined over $\pmb \chi_{k_0}$ without running into any computational challenges   \citep[as there exists an optimal design with at most   $\frac{1}{2} p(p+1)+1$ support points][]{silvey1980optimal}}.\\
Finally,  if $DBSCAN$ detects $m (>1)$ clusters, we generate samples  of sizes $t (= t(C_i)$, $i=1,\ldots, m)$  on each cluster as discussed above.
\begin{remark}
{\rm 
The random variables $\Delta_\ell $ can be generated from any symmetric distribution, in particular from a normal distribution. We use the  $t$-distribution, as this allows an efficient  approximation of   design spaces when the covariates  have  heavy-tailed distributions.
}
\end{remark}
}
 \end{enumerate}


\subsection{\HD{Step 2: Optimal design determination} }\label{sec32}

For most cases of practical interest, the optimal designs on $\pmb \chi_{k_0}$ have to be found numerically and for this purpose, we use the R-package  {\it OptimalDesign} \citep[see][]{R_Software_OptimalDesign}.
 \HD{Note that only the optimal weights corresponding to each element in $\pmb \chi_{k_0}$ have to be determined, where most of the points in ${\pmb \chi}_{k_0} $ will get weight $0$, because there always exists an approximate optimal design with at most $\frac{1}{2} p(p+1)+1$ support points  \citep[see][]{silvey1980optimal}.} 
 The resulting optimal design will be denoted by 
\begin{equation}
    \xi^\ast(\hat{\VectorBeta}_{\mathcal{D}_{k_0}}, {{\pmb\chi}}_{k_0}) 
          = \Big \{   \genfrac{}{}{0pt}{}{\VectorxOneOptimal}{w^\ast_1} \; \genfrac{}{}{0pt}{}{\VectorxTwoOptimal}{w^\ast_2} ~ \ldots ~\genfrac{}{}{0pt}{}{\pmb x^\ast_b}{w^\ast_b} \Big\}, 
\label{hol13}
\end{equation}
     where we assume without loss of generality that the weights are ordered, that is  $w^\ast_1 \geq  {w^\ast_2} \geq   \ldots  \geq  {w^\ast_b} $.
        For a large dimensional parameter, the number of support points of the optimal design is rather large, but in many cases, most of the mass is concentrated at a smaller number of support points. Since, in Step (3) of 
        Algorithm \ref{Algorithm1} we determine 
        points in the sample $\mathcal{D}$ which are close to support points of the optimal design,  we propose to reduce the number of support points by deleting support points with small weights as long as the efficiency of the design is not decreasing substantially. In other words, for a given 
     efficiency bound   $\zeta \in (0,1)$ we consecutively omit the support points with small weights  in $\xi^\ast(\hat{\VectorBeta}_{\mathcal{D}_{k_0}}, {{\pmb\chi}}_{k_0})  $ (and rescale the weights) such that the resulting design has at least efficiency
        $\zeta$.
        The details are given in Algorithm \ref{Algorithm2} and in the following discussion the resulting design will also be denoted $  \xi^\ast(\hat{\VectorBeta}_{\mathcal{D}_{k_0}}, {{\pmb\chi}}_{k_0}) $. 
        In Section \ref{sec413}, we study the impact of the choice of $\zeta$ on the accuracy of the resulting subsampling in a logistic regression model. In particular, we demonstrate that the number of support points can be reduced substantially without losing too much efficiency.

\begin{algorithm}
\caption{Optimal design with reduced support points}\label{algred}  \label{Algorithm2}
\begin{algorithmic}
\State {\bf Input:} $\Psi$-optimal design $\xi^\ast = \xi^\ast(\hat{\VectorBeta}_{\mathcal{D}_{k_0}}, {{\pmb\chi}}_{k_0}) $ defined in \eqref{hol13}; accuracy $\zeta \in (0.5, 1)$
\State \qquad \quad \quad Set ${b^\prime} = {b}$ and $
 \xi^\ast_{b^\prime}= \xi^\ast(\hat{\VectorBeta}_{\mathcal{D}_{k_0}}, {{\pmb\chi}}_{k_0}) 
$ 
\smallskip
\smallskip

\While{$\Psi (\xi^\ast_{{b^\prime}}) > \zeta \Psi (\xi^\ast)  $ and $b^\prime > p$}
\smallskip
\smallskip
\smallskip
\State \qquad   \qquad ${b^\prime} \gets {b^\prime}-1$
\smallskip
\smallskip
\State \qquad   \qquad   $u^\ast_{i} \gets \frac{w^\ast_{i}}{\sum^{b^\prime}_{i=1}w^\ast_{i} }$ (for $i=1,\ldots,b^\prime$)
\smallskip
\smallskip
\State \qquad   \qquad $w^\ast_{i}  \gets u^\ast_{i} $ (for $i=1,\ldots,b^\prime$)
\smallskip
\smallskip
\State \qquad   \qquad  $         \xi^\ast_{b^\prime} \gets \Big \{   \genfrac{}{}{0pt}{}{\VectorxOneOptimal}{w^\ast_{1}} \; \genfrac{}{}{0pt}{}{\VectorxTwoOptimal}{w^\ast_{2}} \ldots \genfrac{}{}{0pt}{}{\pmb x^\ast_{b^\prime} }{w^\ast_{b^\prime}} \Big\}$
\EndWhile
\smallskip
\smallskip

\smallskip
$\xi^\ast(\hat{\VectorBeta}_{\mathcal{D}_{k_0}}, {{\pmb\chi}}_{k_0}) \gets \xi^\ast_{b^\prime+1}$ 

\State {\bf Output:} 
 Design 
 $\xi^\ast(\hat{\VectorBeta}_{\mathcal{D}_{k_0}}, {{\pmb\chi}}_{k_0})$  with a minimal number of support points and   $\Psi$-efficiency larger or equal than $\zeta$.
\end{algorithmic}
\end{algorithm}

\subsection{\HD{Step 3: Optimal design based subsampling}} \label{sec33}
Once the approximate  $\Psi$-optimal design 
$ \xi^\ast(\hat{\VectorBeta}_{\mathcal{D}_{k_0}}, {{\pmb\chi}}_{k_0}) $  
 has been determined (if necessary with a reduced number of support points), the algorithm proceeds to find the remaining $k_1= k-k_0$  points for the subsample such that they are close to the support points 
 $\VectorX_1^*, \ldots  , \VectorX_b^*$ of the approximate $\Psi$-optimal design. For this purpose, we introduce several distances, which will be discussed first.

\HD{In particular, we are not comparing the points $\VectorX,  \VectorX^\prime \in \pmb \chi$ with respect to a distance defined on the design space $\pmb \chi$, }
but we use distances between the Fisher information matrices 
$\mathcal{I}(\VectorBeta, \pmb x)   $ and $\mathcal{I}(\VectorBeta, \pmb x^\prime) $. We concentrate on the Frobenius distance 
\begin{align} 
 d_F(  \VectorX , \VectorX^\prime ) & :=   \| \mathcal{I}(\VectorBeta, \pmb x) - \mathcal{I}(\VectorBeta, \pmb x^\prime)\|_F:=  tr\Big \{ \big(\mathcal{I}(\VectorBeta, \pmb x) - \mathcal{I}(\VectorBeta, \pmb x^\prime)\big)^\top ~\big( \mathcal{I}(\VectorBeta, \pmb x) - \mathcal{I}(\VectorBeta, \pmb x^\prime)\big)
 \Big \}^{1/2}~, \label{d1} 
 \end{align}
the  square root distance 
\begin{align} 
   d_s(  \VectorX , \VectorX^\prime ) & :=   \| \mathcal{I}(\VectorBeta, \pmb x)^{1/2} - \mathcal{I}(\VectorBeta, \pmb x^\prime)^{1/2} \|_F~, \label{d2}  
\end{align}
 and the Procrustes distance 
 \begin{align} 
   d_p(  \VectorX , \VectorX^\prime ) &:= \displaystyle {\inf_{\pmb K \in  O(  \mathbb{R}^{p \times p}) }  \Big  \{  \| \mathcal{I}(\VectorBeta, \pmb x) - \mathcal{I}(\VectorBeta, \pmb x^\prime) \pmb K \|_F \Big \}^{1/2} }  \label{d3} 
\end{align}
 between the information matrices $\mathcal{I}(\VectorBeta, \pmb x)$ and $ \mathcal{I}(\VectorBeta, \pmb x^\prime) $, which reflect the geometry of the space $\{ \mathcal{I}(\VectorBeta, \pmb x)~: ~ \VectorX \in \pmb{\chi } \} $ as a subset of the non-negative definite (symmetric) matrices \citep[see][]{DrydenKoloydenkoZhou2009,pigoli2014distances}. 
Here the infimum in \eqref{d3} is taken over the set of all $p \times p$ orthogonal matrices.
%
Note that
the Procrustes distance can be further simplified if the Fisher information  at $\pmb x$  and 
$ \pmb x^\prime$
can be represented as $\mathcal{I}(\VectorBeta, \pmb x) = \pmb L_1 \pmb L_1^T $ and $ \mathcal{I}(\VectorBeta,\pmb x^\prime) = \pmb L_2 \pmb L_2^T$, respectively. In this case  we have $d_p^2(  \VectorX, \VectorX^\prime ) =  \| \pmb L_1 \|_F^2 + \| \pmb L_2 \|_F^2- 2 \Sigma_{k} \sigma_k $, where $\sigma_1 , \sigma_2 , \ldots ,  $ are  the singular values of  the matrix $\pmb L_2^T \pmb L_1$ \citep[see][]{DrydenKoloydenkoZhou2009,pigoli2014distances}.

We can use any of these distances (and also other distances) in step (3) of  Algorithm \ref{Algorithm1} 
and in the following, we denote this distance by $d$. The points, which are closest to the  support points of the optimal design 
$ \xi^\ast(\hat{\VectorBeta}_{\mathcal{D}_{k_0}}, {{\pmb\chi}}_{k_0}) $ in \eqref{hol13} with respect to $d(\cdot)$ are retained in the subsample. The number of points corresponding to each design point $\VectorX_i^*$ is proportional to the corresponding weight $w_i^*$, which means that only $\lfloor w^\ast_i ~ k_1\rfloor$  points closest to $\VectorXi $ (with respect to the distance  $d(\cdot)$) are retained in the subsample. The details of distance-based subsample allocation are summarized in Algorithm \ref{algsub}.

\begin{algorithm}
\caption{Subsampling of the points closest to the support of the locally  optimal design } \label{algsub}   \label{Algorithm3}
\begin{algorithmic}
\State {\bf Inputs:} Sample $\mathcal{D}$, locally optimal design    $\xi^\ast(\hat{\VectorBeta}_{\mathcal{D}_{k_0}}, {{\pmb\chi}}_{k_0}) $, and one of the distance metric $d(.)$  as defined in \eqref{d1} - \eqref{d3}.

\smallskip
\smallskip

\State	Define $\mathcal{D}_{k_1}= \emptyset$.

    \For{ $i = 1, \ldots, 	b $}
		\For{ $j = 1, \ldots, 	n $}
			\State{ Calculate $d_{ij} = d(\VectorxiOptimal ,  \VectorxjSample  )$, }
		\EndFor
			\State{Define $\pmb{d}_i= (d_{i1},d_{i2}, \ldots,d_{in})$}
			\State{Let $\pmb d_{i_{(\cdot )}} = ( d_{i_{(1)}}, 
   d_{i_{(2)}}, \ldots, d_{i_{(n)}})$, where ${\pmb d_{i_{(\cdot )}}}$ 
   is the vector of ordered  components of ${\pmb d_i}$}
      \State{($d_{i_{(1)}}$ is the smallest value).}
	\EndFor 
\smallskip
\smallskip
\For{ $i = 1, \ldots, 	b $}
    \State From $\pmb d_{i_{(\cdot )}}  $, remove all components corresponding to points in $\mathcal{D}_{k_1}$.
        \smallskip
    \State Take sample points $\pmb x \in \mathcal{D}$ corresponding to the first $\lfloor w^\ast_i~ k_1 \rfloor$ elements in $\pmb d_{i_{(\cdot )}}$ and add
    \State{them to $\mathcal{D}_{k_1}$.}
\EndFor
\State {\bf Output:} $\mathcal{D}_{k} = \mathcal{D}_{k_0} \cup \mathcal{D}_{k_1}$ is the subsample of size $k$
\end{algorithmic}
\end{algorithm}

\subsection{Computational complexity  of  ODBSS} \label{time_complexity} \label{sec34}
In this section we briefly discuss the computational complexity of the ODBSS algorithm, which is constituted of three main parts:
\begin{enumerate}
        \item[(1)]  Area estimation: The complexity of DBSCAN algorithm with  $p$-dimensional $k_0$ points is $\mathcal{O}(k^2_0 p)$ \citep[see][]{DBSCAN_1996_Paper_A_Folloup}. Note that  $k_0$ and $p$ are small compared to $n$.
        
        \item[(2)] Calculation of the $\Psi$-optimal design on the design space ${{\pmb\chi}}_{k_0}$: 
        The R-package  {\it OptimalDesign} 
        determines the optimal weights  $w_i^\ast$ maximizing \eqref{2.4}, but this time the points  $\VectorXi $ of the estimated designs space
         ${{\pmb\chi}}_{k_0}$  are candidates for the support of the design $\xi^\ast$. \cite{sagnol2015computing} formulates the problem of finding approximate optimal design into a mixed integer second-order cone programming and discusses the time complexity for finding approximate optimal designs under various criteria. It can be seen that by using second-order cone programming an approximate A-optimal design could be determined in time $\mathcal{O}(s p r )^3 ~ \sqrt{s p}~ log(1/\delta)$, where $\delta$ is the permissible error, $s$ is the cardinality of ${{\pmb\chi}}_{k_0}$, $r$ is the rank of the information matrix at a point defined in equation \eqref{Information_Matrix_Eqution_1}  \cite[see also][for more details on complexity for solving second order cone programming]{ben2001lectures}. From equation~\eqref{31} we see that in Algorithm~\ref{Algorithm1}, in step (2) where the optimal design is determined, $ s  = |\mathcal{X}_0| \leq |\mathcal{G}| = L^p$. Therefore, the experimenter has control over the computation time for the determination of the optimal design by varying the grid size parameter $L$ as opposed to calculating the optimal design on the full sample.
    \item[(3)] Subsample allocation: For the subsample allocation the distances $ \pmb{d}_i $ for $i=1, \ldots, b$ needs to be computed and then $\lfloor w_ik_1 \rfloor$ smallest elements $ \pmb{d}_i $ are determined (see Algorithm \ref{Algorithm3}). In the case when the information matrix $\mathcal{I} (\VectorBeta,\VectorX )$ at point $\VectorX$ has rank $1$, that is, in the case of usual logistic and linear regression, calculation of $ \pmb{d}_i $ has computational complexity $\mathcal{O}(bnp) $ and then determining $\lfloor w_ik_1 \rfloor$ smallest elements $ \pmb{d}_i $ has complexity $\mathcal{O}( nb  )$  \citep{martinez2004partial}. Therefore, the total computational complexity of this step of the Algorithm in the rank $1$ case is $\mathcal{O}(nb (p+1)) $.  To draw some comparisons we would highlight that, the time-complexity for IBOSS linear regression \cite{wang2019information} is $\mathcal{O}(np)$, IBOSS for logistic regression \cite{LogisticCheng2020information} is approximately $\mathcal{O}(np)$, and OSMAC-1 for logistic regression \cite{LogisticWang2018optimal} is also $\mathcal{O}(n  p)$.  Although this step  is computationally more expensive, we emphasize that the 
    algorithms proposed in these references are specifically constructed for the linear and logistic regression model, while ODBSS is generally applicable.

\end{enumerate}

\HD{
\begin{remark} {\rm 
If the information matrix has rank $ > 1$ then the time complexity the procedure depends the choice of the matrix distance.
In particular we demonstrate in the supplementary material that for a models where rank of Fisher information matrix is $2$, the time complexity of ODBSS with Frobenius distance remains the same as that of the rank $1$ case. { However, for the other two distances used in ODBSS, the time complexity much is higher. Therefore, we recommend using ODBSS with Frobenius norm when Fisher information matrix at a point has rank > 1.}
}
\end{remark}  }

\section{Numerical results} 
\label{sec4}

  \def\theequation{4.\arabic{equation}}	
  \setcounter{equation}{0}

In this section, we investigate the performance of the new algorithm (ODBSS) by means of a simulation study. In Section \ref{Simulation_Study_1} we consider the logistic regression model, which corresponds to a Fisher information of rank $1$.
We provide a comparison of versions of ODBSS with different distance metrics $d_F$, $d_s$ and $d_p$ as defined in \eqref{d1}, \eqref{d2} and \eqref{d3}, respectively, and also compare  ODBSS with alternative subsampling algorithms for logistic regression \citep[see][]{LogisticCheng2020information,LogisticWang2018optimal}.  Moreover, 
we  illustrate the effect of reducing  support points of the optimal design on ODBSS  (using Algorithm  \ref{Algorithm3})
and  investigate the computational time  of an alternative method for design space estimation 
(more precisely, using the full sample as an estimate of design space in step 1 of Algorithm \ref{Algorithm1} rather than a cluster-based area determination). 
Finally, Section \ref{sec43} is devoted to the performance of ODBSS for a model with a Fisher information matrix of rank larger than $1$. 

In the following illustrations, different subsample algorithms will be compared  by the mean squared error ({MSE})
\begin{equation}
\label{det21} 
\mathbb{E} \big [ \|\VectorBeta- \hat{\VectorBeta}_{\mathcal{D}_k} \|^2 \big ] ,
\end{equation}
which is estimated by $100$ simulation runs. \HD{We consider the {MSE} to compare the subsampling algorithms, as the existing subsampling algorithms for logistic regression in \cite{LogisticWang2018optimal, LogisticCheng2020information} are compared by this criterion. The $A$-optimality criterion is equivalent to minimizing the average variance of the estimates of the parameters. For unbiased estimators, this corresponds to the minimization of the MSE. Moreover, in the nonlinear models considered in this paper, this is asymptotically equivalent to minimizing MSE. Therefore, for the simulation studies in this paper, we use $A$-optimality in ODBSS. However, if the experiment had a different objective, the optimality criteria in ODBSS could be set to a different one (like $D$-optimality, $G$-optimality, etc.).} \\
    
\HD{In Section 1 of the supplementary material, we study the impact of the size $k_0$ of the initial uniform sample on the estimation of MSE using ODBSS. Based on these results we recommend and take $k_0= 0.2k$, in this paper. }\\

Note, when the rank of the Fisher information matrix $\mathcal{I} (\VectorBeta, \VectorX )$ at point $\VectorX$ is $1$, then 
\begin{align*}
   \mathcal{I} ( \VectorBeta, \VectorX) =  \Phi(\VectorX, \VectorBeta) \Phi(\VectorX, \VectorBeta)^\top 
\end{align*}
where $\Phi(\VectorX, \VectorBeta) \in \mathbb{R}^{p+1}$. 
In that case, calculating the matrix distances  becomes very easy and we obtain
\begin{align*}
d_F(  \VectorX , \VectorX^\prime ) &=   
\Big \{ 
\|  \Phi(\VectorX, \VectorBeta)  \|^4 + \|  \Phi(\VectorX^\prime  , \VectorBeta) \|^4 - 2 (\Phi(\VectorX^\prime  , \VectorBeta)^\top \Phi(\VectorX, \VectorBeta) )^2
\Big \}^{  1/2},   \\ 
d_s(  \VectorX , \VectorX^\prime ) &=   
\Big \{ 
\|  \Phi(\VectorX, \VectorBeta)  \|^2 + \|  \Phi(\VectorX^\prime  , \VectorBeta) \|^2 - 2 \dfrac{(\Phi(\VectorX^\prime, \VectorBeta)^\top \Phi(\VectorX, \VectorBeta) )^2}{\|  \Phi(\VectorX, \VectorBeta)  \| ~ \|  \Phi(\VectorX^\prime  , \VectorBeta) \|}
\Big \}^{1/2},\\
d_p(  \VectorX , \VectorX^\prime ) &=   \|  \Phi(\VectorX, \VectorBeta)  -  \Phi(\VectorX^\prime  , \VectorBeta) \| = 
\Big \{ 
\|  \Phi(\VectorX, \VectorBeta)  \|^2 + \|  \Phi(\VectorX^\prime  , \VectorBeta) \|^2 - 2 (\Phi(\VectorX^\prime  , \VectorBeta)^\top \Phi(\VectorX, \VectorBeta) )
\Big \}^{  1/2},   
\end{align*}
 where $\| \cdot \|$ is the Euclidean norm. However, no such simple representations exist in the case, where the rank of the information matrix $\mathcal{I} ( \VectorBeta, \VectorX)$ is  larger than $1$.

\subsection{Logistic regression - rank $1$ case} 
\label{Simulation_Study_1} \label{sec41}
\HD{We consider a logistic regression model with no intercept  
\begin{equation}
\mathbb{P}(y =1 | \VectorX ,\pmb  \beta )     = \frac{\exp( \pmb x^\top \pmb \beta     )}{1+\exp( \pmb x^\top \pmb \beta   )} \label{41LogReg_NoIntercept},
\end{equation}
where $\pmb \beta = (\beta_1, \ldots, \beta_p)^\top$. In this section, this parameter is set to ${\pmb \beta} = (0.5,0.5, \ldots, 0.5)^\top $. Assume that $\pmb x$ follows either a $p$-dimensional normal distribution 
 or a $t$-distribution with $\kappa$ degrees of freedom, centered at $\pmb \mu = (\mu_1, \ldots, \mu_{p})$ and covariance $ \pmb \Sigma$ with densities 
 \begin{align}
 \varphi_p(\pmb x; \pmb \mu, \pmb \Sigma) &= \frac{1}{(2\pi)^{p/2} |\boldsymbol{\Sigma}|^{1/2}} \exp\left(-\frac{1}{2} (\mathbf{x}-\pmb \mu)^T \boldsymbol{\Sigma}^{-1} (\mathbf{x}-\pmb \mu) \right)
 \end{align} 
  and 
  \begin{align}
         \mathcal{T}_{p}(\pmb x;\pmb \mu, \pmb \Sigma, \kappa) &= \dfrac{\Gamma[(\kappa+p)/2]}{\Gamma(\kappa/2) (\pi\kappa)^{p/2} { |\pmb \Sigma|}^{1/2}}  \Big[ 1+ \dfrac{1}{\kappa} \pmb (\mathbf{x}-\pmb \mu)^\top \pmb \Sigma^{-1}\pmb (\mathbf{x}-\pmb \mu)\Big]^{-(\kappa+p)/2}~, \label{4.4}
 \end{align} 
respectively. }For our simulation studies, we used $\kappa= 3$ in \eqref{4.4} to study the effect 
of more heavy-tailed covariates.  We consider  three types of covariance structures: 
  \begin{enumerate}
    \item[(1)] $ \pmb \Sigma_1 = ( 0.5^{|i-j|} )_{i,j=1, \ldots , p} $.
    \item[(2)]  In each simulation run we choose randomly 
  $5$ mutually orthogonal dominant directions $\pmb e_1$, $\pmb  e_2$, $\pmb  e_3$, $\pmb  e_4$, and $\pmb  e_5$ $\in $  on $\mathbb{S}_{(p-1)} 
  = \{  \pmb x  \in \mathbb{R}^p  ~ | ~ \| \pmb x \| =1 \} $ using the {\it randortho}-function in $R$. Then, the covariance matrix is defined by 
  $$
  {\pmb \Sigma_2} = 2 \;  \pmb  e_1 \pmb  e_1^\top +  1.8 \;  \pmb  e_2 \pmb  e_2^\top + 1.6 \;  \pmb  e_3 \pmb  e_3^\top + 1.4 \;  \pmb  e_4 \pmb  e_4^\top + 1.2 \;  \pmb  e_5 \pmb  e_5^\top  + 0.1 \; {\pmb \Sigma_1}.
  $$
  where ${\pmb \Sigma_1}$ is defined in (1).
  Note, that this data is concentrated in the neighborhood of a $5$-dimensional plane determined by the vectors  $\pmb  e_1, \pmb  e_2, \pmb  e_3, \pmb  e_4$, and $\pmb  e_5$. 
  \item[(3)] Similarly, to (2) we consider data concentrating in a neighborhood of a $3$-dimensional plane. Thus we randomly choose in each run  mutually orthogonal dominant directions $\pmb e_1$, $\pmb  e_2$, and $\pmb  e_3$ $\in  \mathbb{S}_{(p-1)}  \subset \mathbb{R}^p$, and define the covariance matrix by
  $$
  {\pmb \Sigma_3} = 3 \;  \pmb  e_1 \pmb  e_1^\top +  2 \;  \pmb  e_2 \pmb  e_2^\top + 1 \;  \pmb  e_3 \pmb  e_3^\top  + 0.1 \; {\pmb \Sigma_1}.
  $$
  \end{enumerate}
%

\subsubsection{The impact  of different distances on ODBSS }
\label{sec411} 
\begin{figure}[H]
	\centering
	\begin{subfigure}[b]{.3\textwidth}
		\centering
		\includegraphics[width=\linewidth]{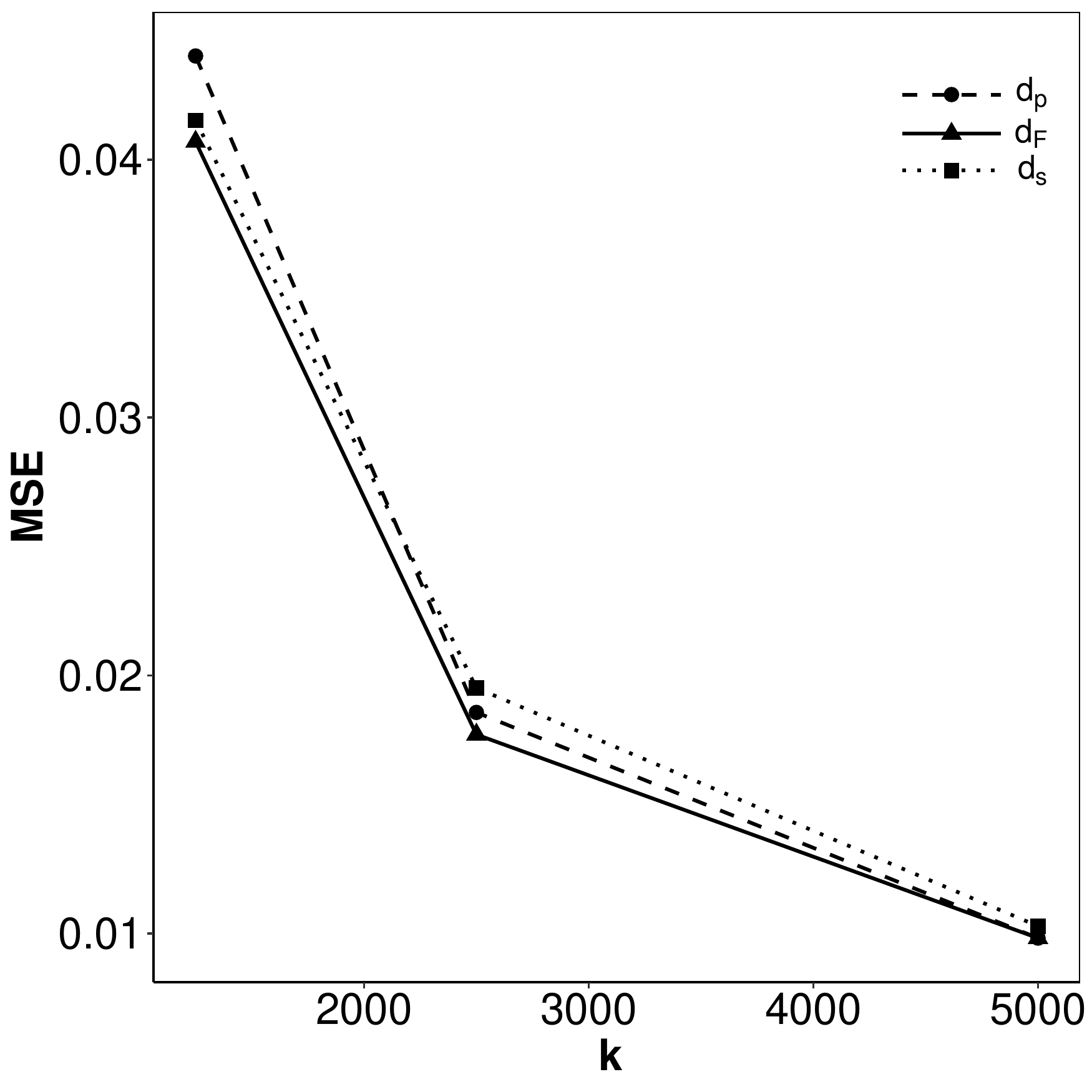}
		\caption{\scriptsize { $  \varphi_p(\pmb x; \pmb 0, \pmb \Sigma_1) $ }}
		\label{fig2_1}
	\end{subfigure}%
 \hfill
   \begin{subfigure}[b]{.3\textwidth}
		\centering
		\includegraphics[width=\linewidth]{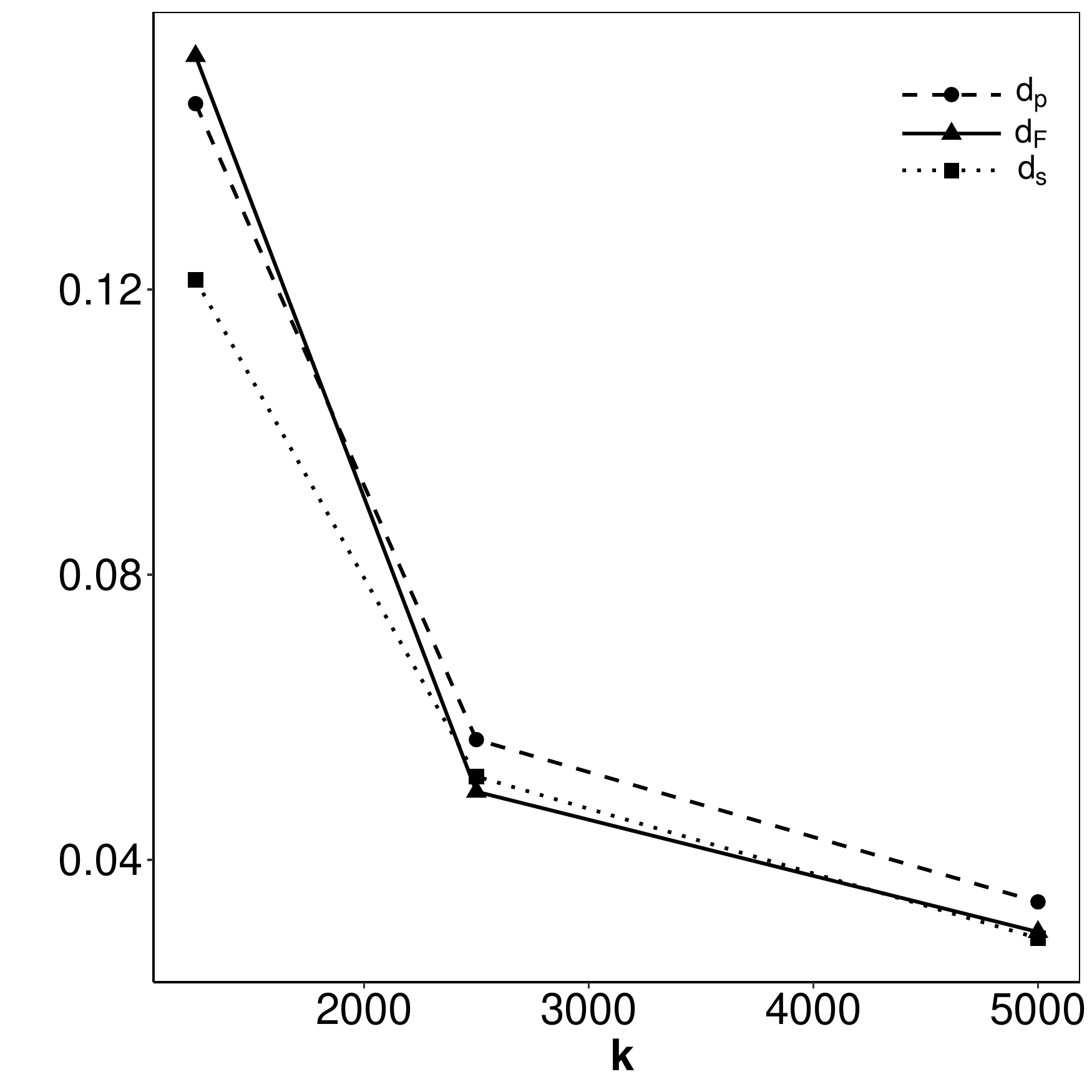}
		\caption{\scriptsize $ \varphi_p(\pmb x; \pmb 0, \pmb \Sigma_2)$ }
		\label{fig2_2}
	\end{subfigure}
 \hfill
  \begin{subfigure}[b]{.3\textwidth}
		\centering
		\includegraphics[width=\linewidth]{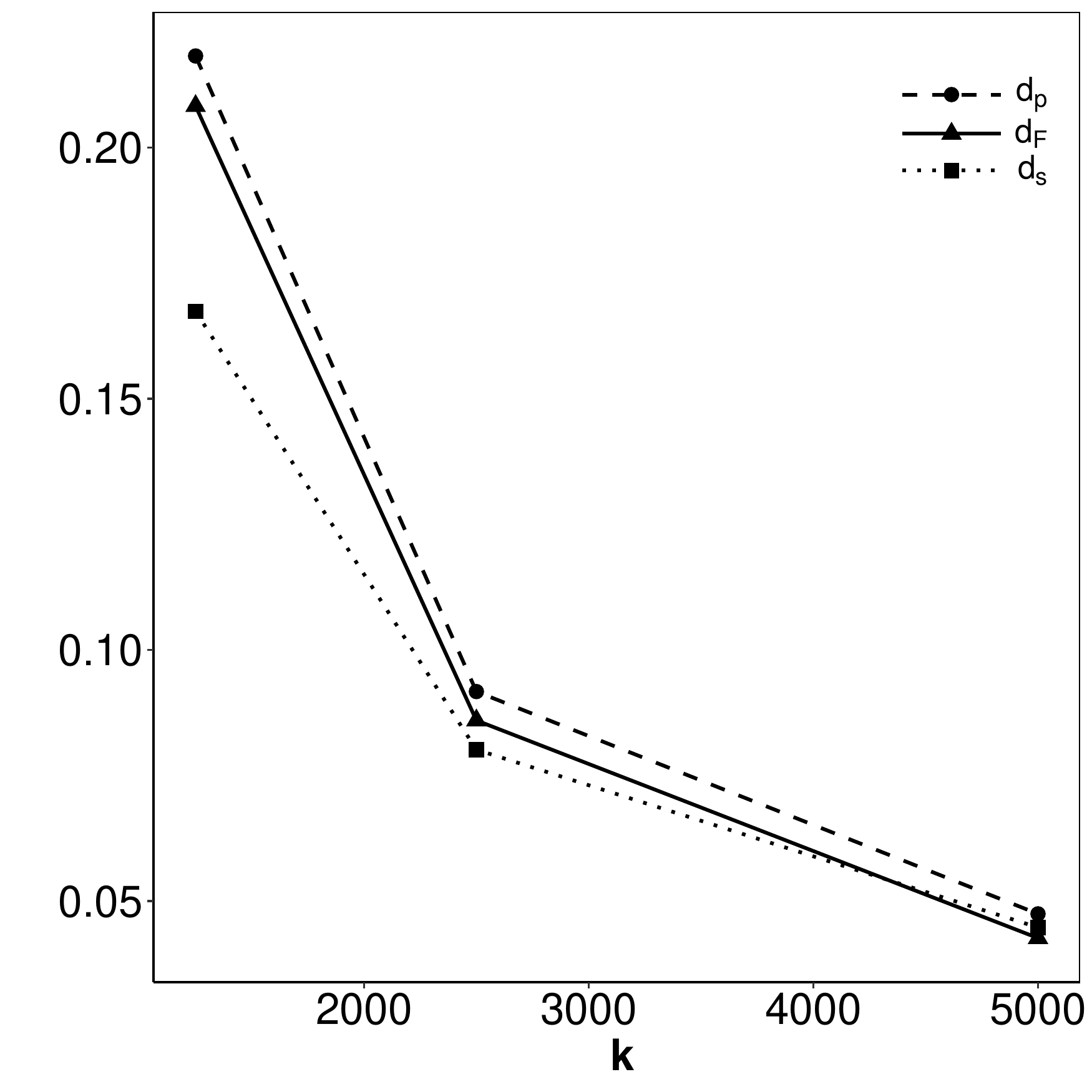}
		\caption{\scriptsize  $ \varphi_p(\pmb x; \pmb 0, \pmb \Sigma_3)$}
		\label{fig2_3}
	\end{subfigure}
	\centering	\caption{ \it  Simulated mean squared error of the parameter estimate in the logistic regression model \eqref{41LogReg_NoIntercept} with p = 7 using 
ODBSS subsampling from {$n = 10^5$} observations with the  metric $d_F(\cdot)$, $d_s(\cdot)$, and $d_p(\cdot)$ defined in \eqref{d1}, \eqref{d2} and \eqref{d3}, respectively. The covariates are normally distributed with different covariance matrices.
 }
	\label{fig2} 
\end{figure}

In this section, we investigate the impact of the metric (step 3 of Algorithm \ref{Algorithm1}) on the performance of ODBSS. More precisely, we display in Figure \ref{fig2}  the simulated mean squared error (MSE) of ODBSS, which is used with the different metrics $d_F$, $d_s$, and $d_P$ defined in \eqref{d1}, \eqref{d2}, and \eqref{d3}, respectively. For the sake of brevity, we restrict ourselves to the case of a normal distribution. The results for the $t$-distribution are very similar and not reported here. We observe that the three metrics do not yield substantial differences in ODBSS. Since there is not much difference in the performance of ODBSS  for the different metrics in the logistic regression model,  we use ODBSS with Frobenius distance $d_F(\cdot)$ in the following sections
to illustrate other aspects of the ODBSS algorithm in the logistic regression model. 

\subsubsection{Comparison with other subsampling procedures}
\label{sec412}

For the logistic regression model \eqref{41LogReg_NoIntercept} there exist several alternative subsampling procedures. {Here we consider the IBOSS procedure introduced by \cite{LogisticCheng2020information} 
and  a  modification  of the score-based subsampling developed by \cite{LogisticWang2018optimal}, which is called OSMAC-1 ($\pi^{mVc}$) and OSMAC-2 ($\pi^{mMSE}$).
This modification is described below and will always yield an improvement of the original procedure. We also include uniform subsampling in all comparisons.}

\begin{figure}[H]
	\centering
	\begin{subfigure}[b]{.3\textwidth}
		\centering
		\includegraphics[width=\linewidth]{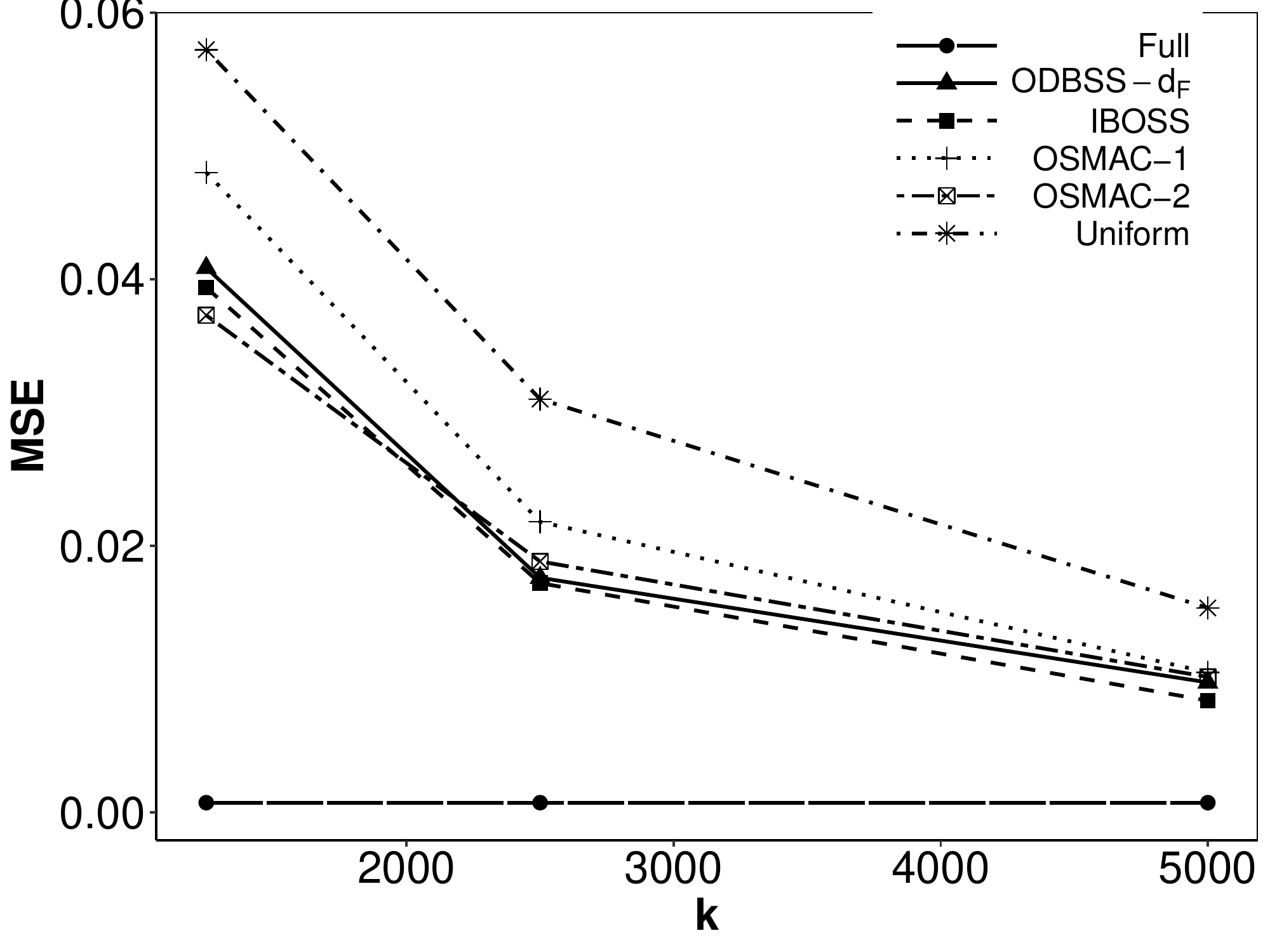}
		\caption{\scriptsize { $  \varphi_p(\pmb x; \pmb 0, \pmb \Sigma_1) $ } } 
		\label{fig3_1}
	\end{subfigure}%
 \hfill
   \begin{subfigure}[b]{.3\textwidth}
		\centering
		\includegraphics[width=\linewidth]{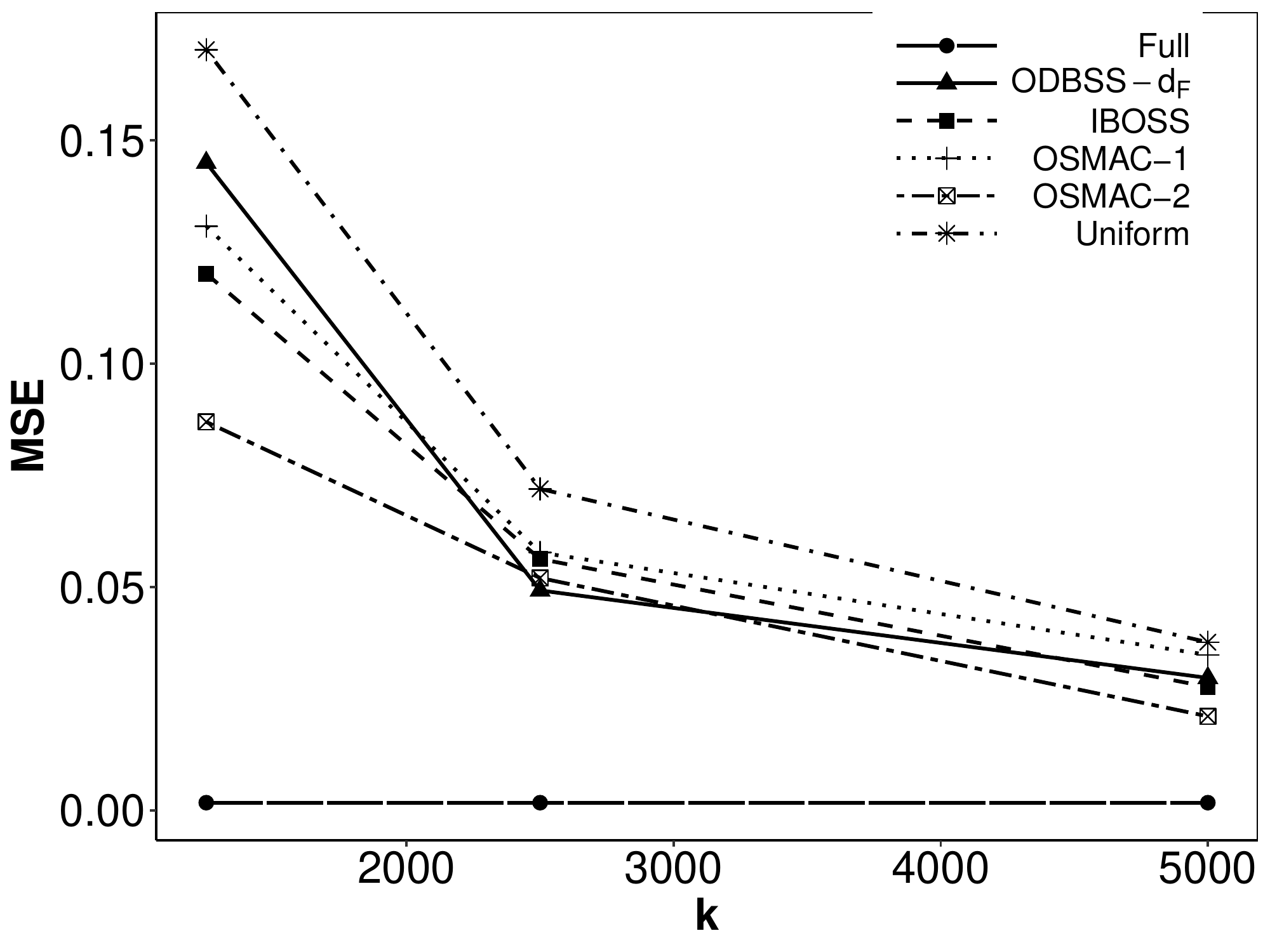}
		\caption{\scriptsize $  { \varphi_p(\pmb x; \pmb 0, \pmb \Sigma_2) }$ }
		\label{fig3_2}
	\end{subfigure}
 \hfill
  \begin{subfigure}[b]{.3\textwidth}
		\centering
		\includegraphics[width=\linewidth]{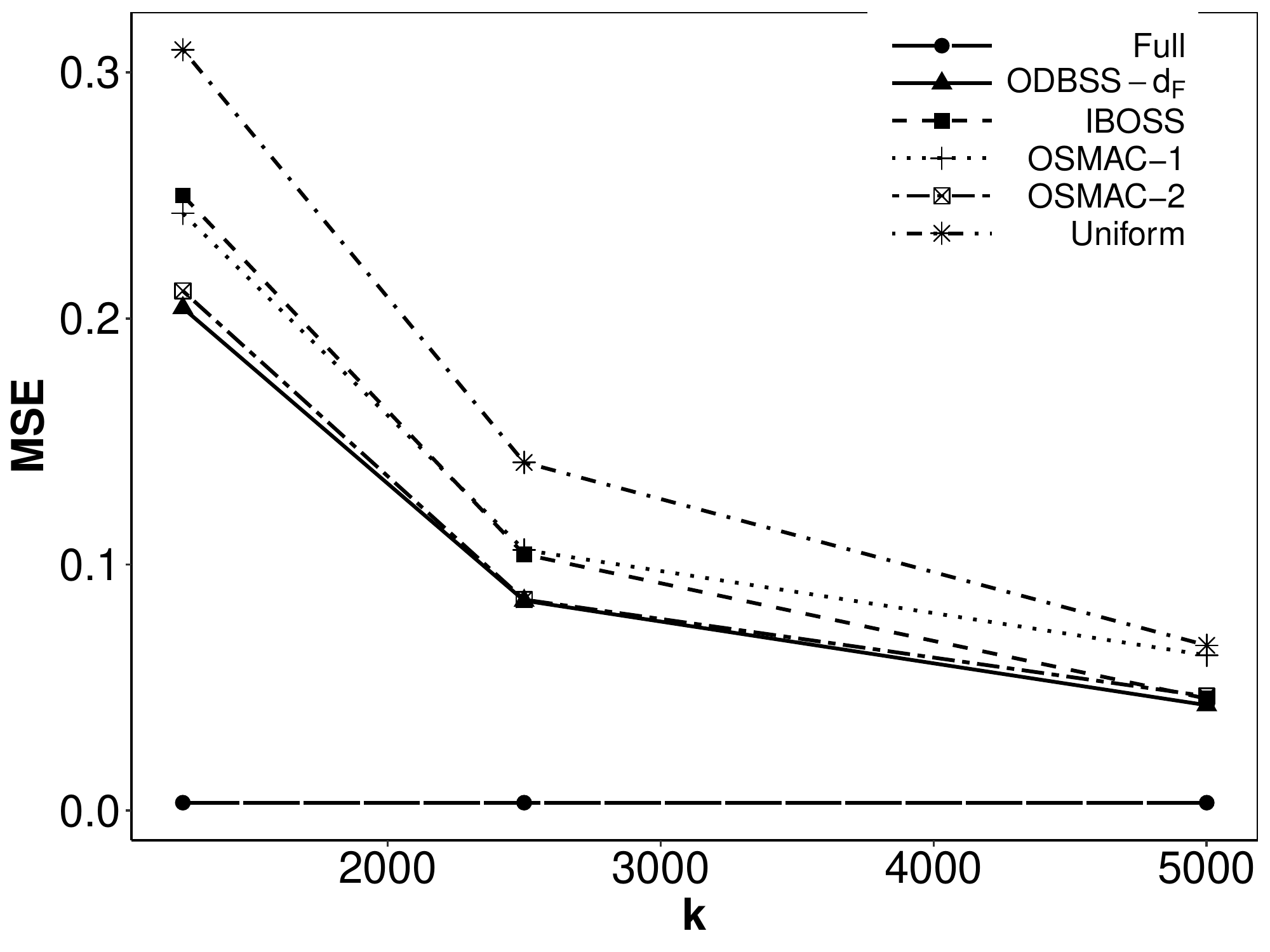}
		\caption{\scriptsize  $ \varphi_p(\pmb x; \pmb 0, \pmb \Sigma_3)$}
		\label{fig3_3}
	\end{subfigure}
 \vspace{1cm}
 
 \begin{subfigure}[b]{.3\textwidth}
		\centering
		\includegraphics[width=\linewidth]{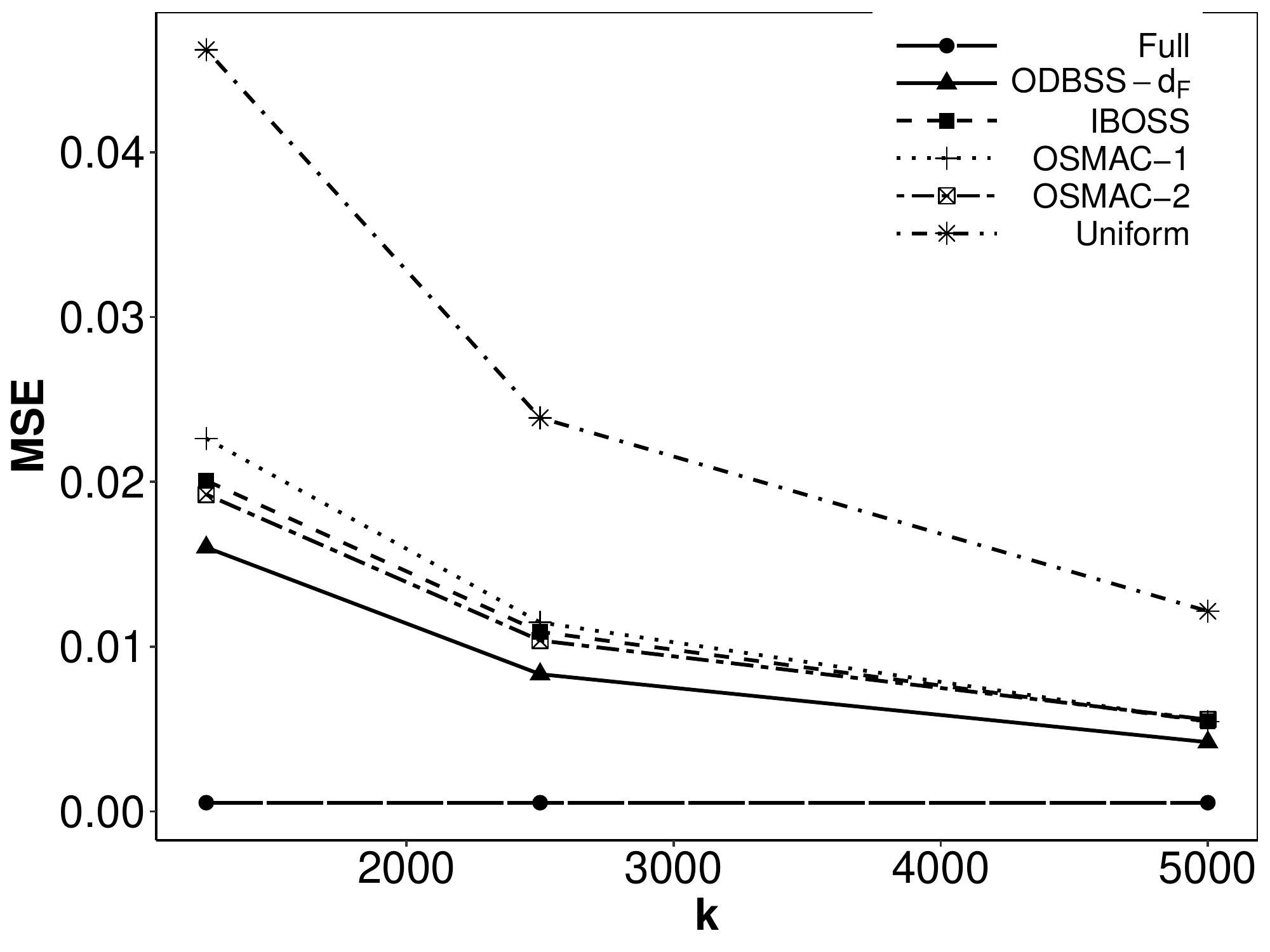}
		\caption{\scriptsize $ \mathcal{T}_{p}(\pmb x; \pmb 0, \pmb \Sigma_1, 3)$ }
		\label{fig3_4}
	\end{subfigure}%
 \hfill
   \begin{subfigure}[b]{.3\textwidth}
		\centering
		\includegraphics[width=\linewidth]{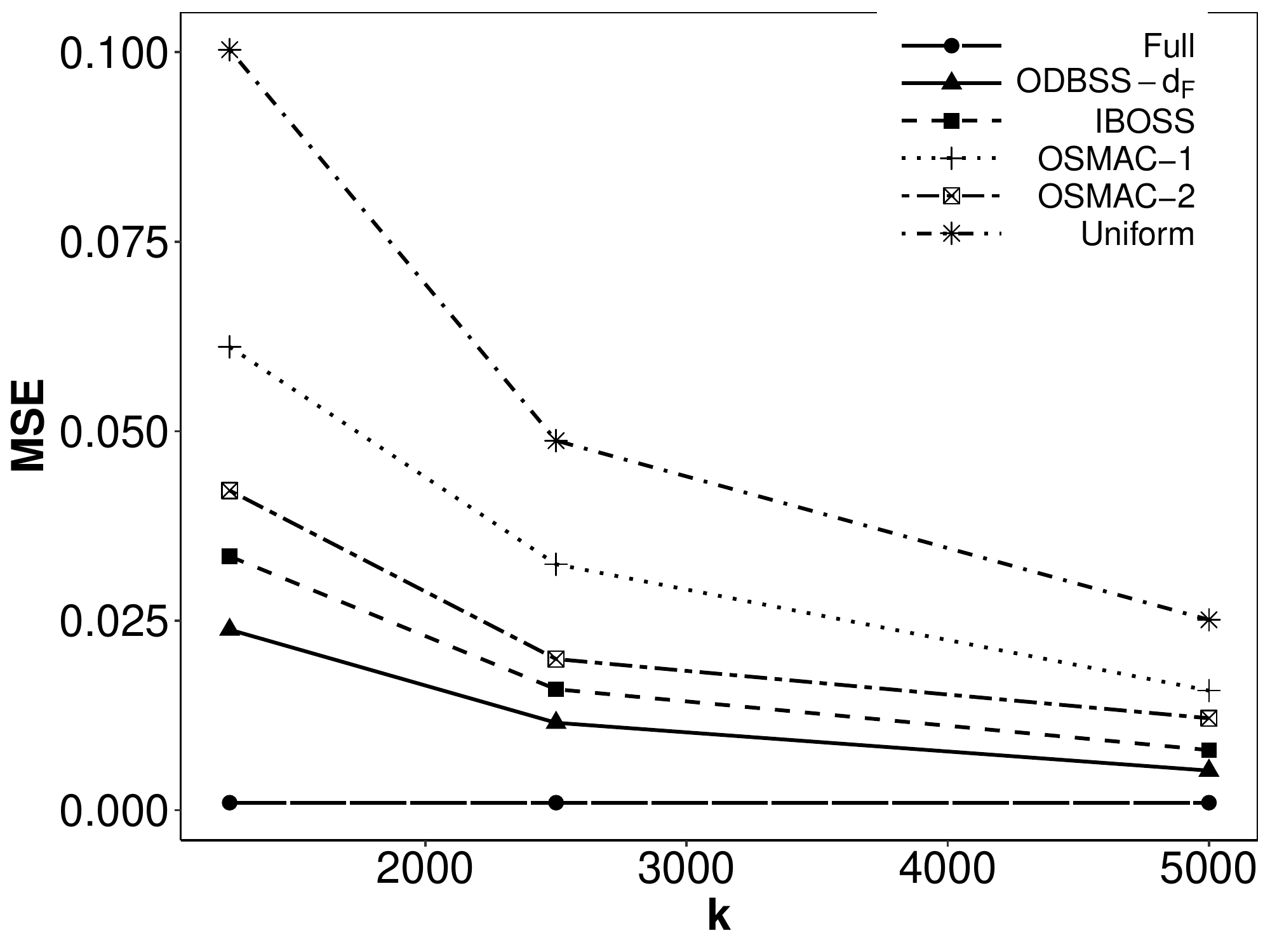}
		\caption{\scriptsize $\mathcal{T}_{p}(\pmb x; \pmb 0, \pmb \Sigma_2, 3)$}
		\label{fig3_5}
	\end{subfigure}
 \hfill
  \begin{subfigure}[b]{.3\textwidth}
		\centering
		\includegraphics[width=\linewidth]{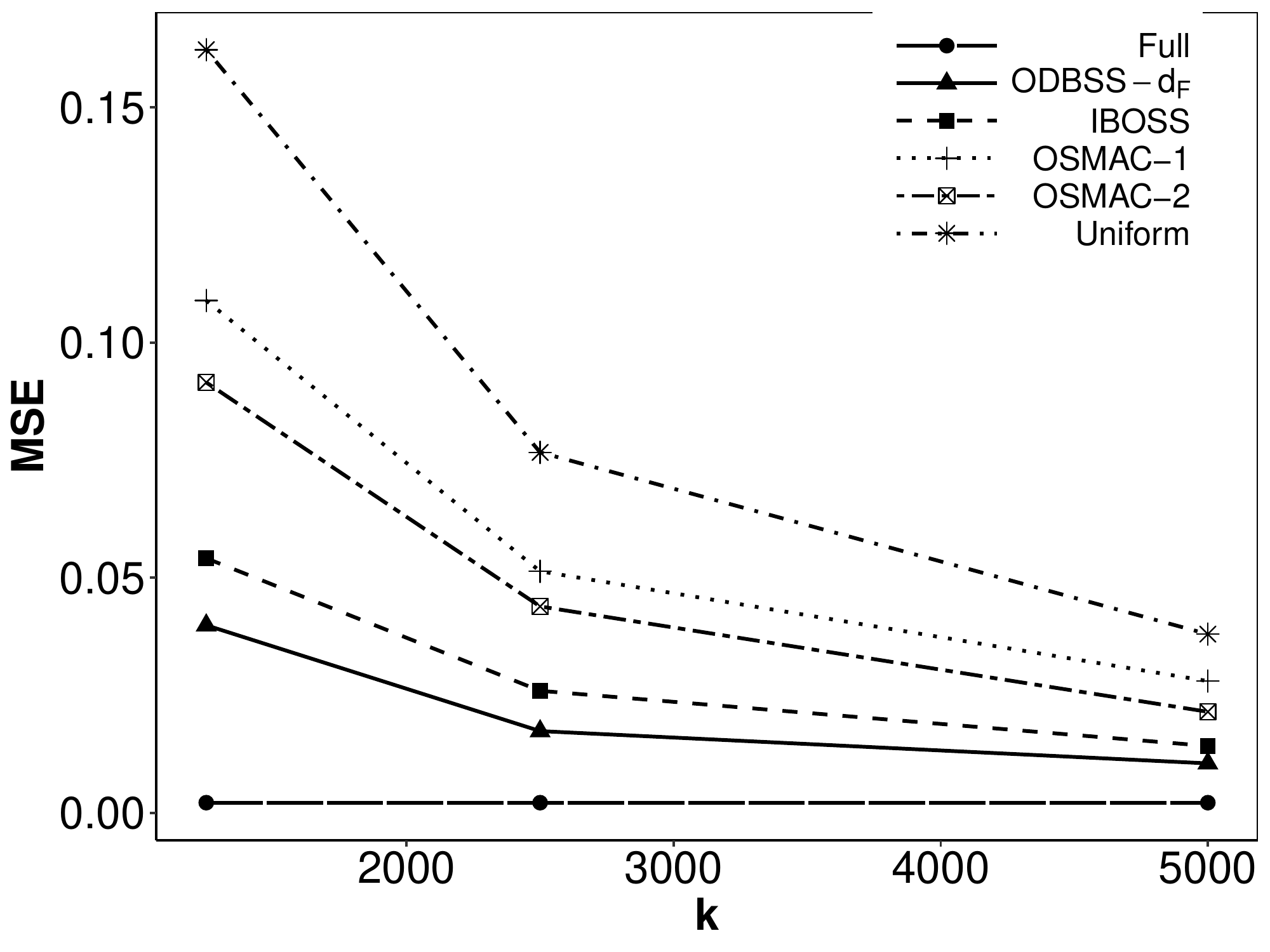}
		\caption{\scriptsize $ \mathcal{T}_{p}(\pmb x; \pmb 0, \pmb \Sigma_3, 3)$ }
		\label{fig3_6}
	\end{subfigure}
	\centering	\caption{ \it  The simulated  mean squared error of the parameter estimate in the  logistic regression model \eqref{41LogReg_NoIntercept} {with p = 7} based on
 subsamples obtained by
 ODBSS, IBOSS, OSMAC-1, OSMAC-2, uniform random sampling from $n = 10^5$ observations. 
 The covariates have centered normal and $t$-distribution with $3$ degrees of freedom with different covariances.  
 }
	\label{fig2a} 
\end{figure}

{First, we investigate the case where p = 7}. In Figure \ref{fig2a}, we display the simulated  MSE for the different subsampling procedures. We observe that in all cases under consideration, ODBSS has the best performance. It can be seen that the superiority of ODBSS is more pronounced when the size of the subsample is smaller ($k=1250$, which is 1.25\% and $k= 2500$, which is $2.5\%$ of the original sample size). Moreover, the advantages of  ODBSS are more pronounced if the covariates are ``concentrating''  on a  ``lower'' dimensional space defined by the matrices by ${\pmb \Sigma_2}$ or ${\pmb \Sigma_3}$. Interestingly, the mean squared error is smaller for $t$-distributed than for normally distributed predictors.

\HD{On the other hand, we do not observe large differences between IBOSS, {OSMAC-1, and OSMAC-2}, which partially contradicts the findings in \cite{LogisticCheng2020information}, 
who observed substantial advantages of IBOSS over OSMAC-1.
These different findings can be explained as follows. The subsample obtained by  {OSMAC-1 and OSMAC-2 consists of two parts, $k_0$ observations obtained by uniform subsampling and $k_1=k-k_0$ observations obtained by sampling with ``optimal'' probabilities  $\pi_i^{mVc}$ and $\pi_i^{mMSE}$ (calculated from the initial sample), respectively}.   }
\HD{ While \cite{LogisticCheng2020information} use OSMAC-1 with the subsample  $\mathcal{D}_{k_1}$ and weighted maximum likelihood with optimal weights $\pi_i^{mVc}$ to estimate the parameters, \cite{LogisticWang2018optimal} use the full sample $\mathcal{D}_{k}$ for this purpose. We argue that both approaches are sub-optimal. On the one hand, using only the sample  $\mathcal{D}_{k_1}$   
 does not take all available observations into account. On the other hand, if  the full sample 
 $\mathcal{D}_k$  is applied for the estimation, the weights $\pi_i^{mVc}$ ($\pi_i^{mMSE}$) in the likelihood function 
 do not reflect the nature of the random sampling mechanism. In fact, the subsample ${\cal D}_k = {\cal D}_{k_0} \cup {\cal D}_{k_1}  $ is  drawn from a mixture of a uniform and the $\pi^{mVc}$ (and $\pi_i^{mMSE}$) distribution.  Therefore we propose to estimate the parameters from the full subsample ${\cal D}_k$ by weighted logistic regression with the weights from the mixture distribution $\widetilde{\pi}_i^{mVc}= (k_0/k)~ (1/n)+ (k_1/k) ~\pi_i^{mVc}  $ (and $\widetilde{\pi}_i^{mMSE}= (k_0/k)~ (1/n)+ (k_1/k) ~\pi_i^{mMSE}$) for $i= 1, \ldots,n$. 
This procedure has been implemented in our comparison and we
 observe that it improves the procedures of \cite{LogisticWang2018optimal} substantially (these results are not displayed for the sake of brevity). In particular, the differences between the two subsampling algorithms are much smaller. Nevertheless, both procedures are outperformed by ODBSS.\\
}

\begin{figure}[H]
	\centering
	\begin{subfigure}[b]{.3\textwidth}
		\centering
		\includegraphics[width=\linewidth]{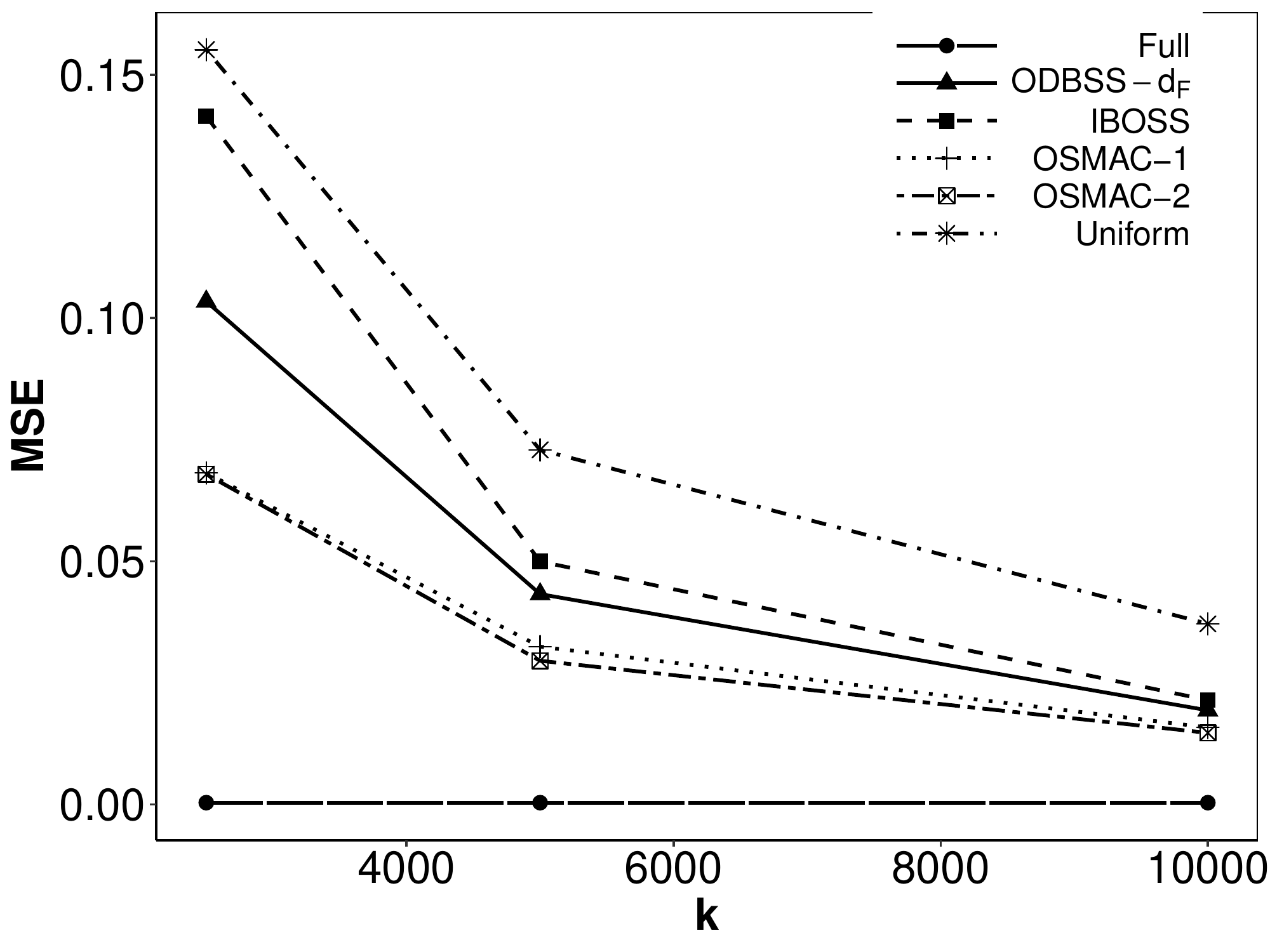}
		\caption{\scriptsize $\varphi_{p}(\pmb x; \pmb 0, \pmb \Sigma_1)$ }
		\label{fig3_1}
	\end{subfigure}%
 \hfill
   \begin{subfigure}[b]{.3\textwidth}
		\centering
		\includegraphics[width=\linewidth]{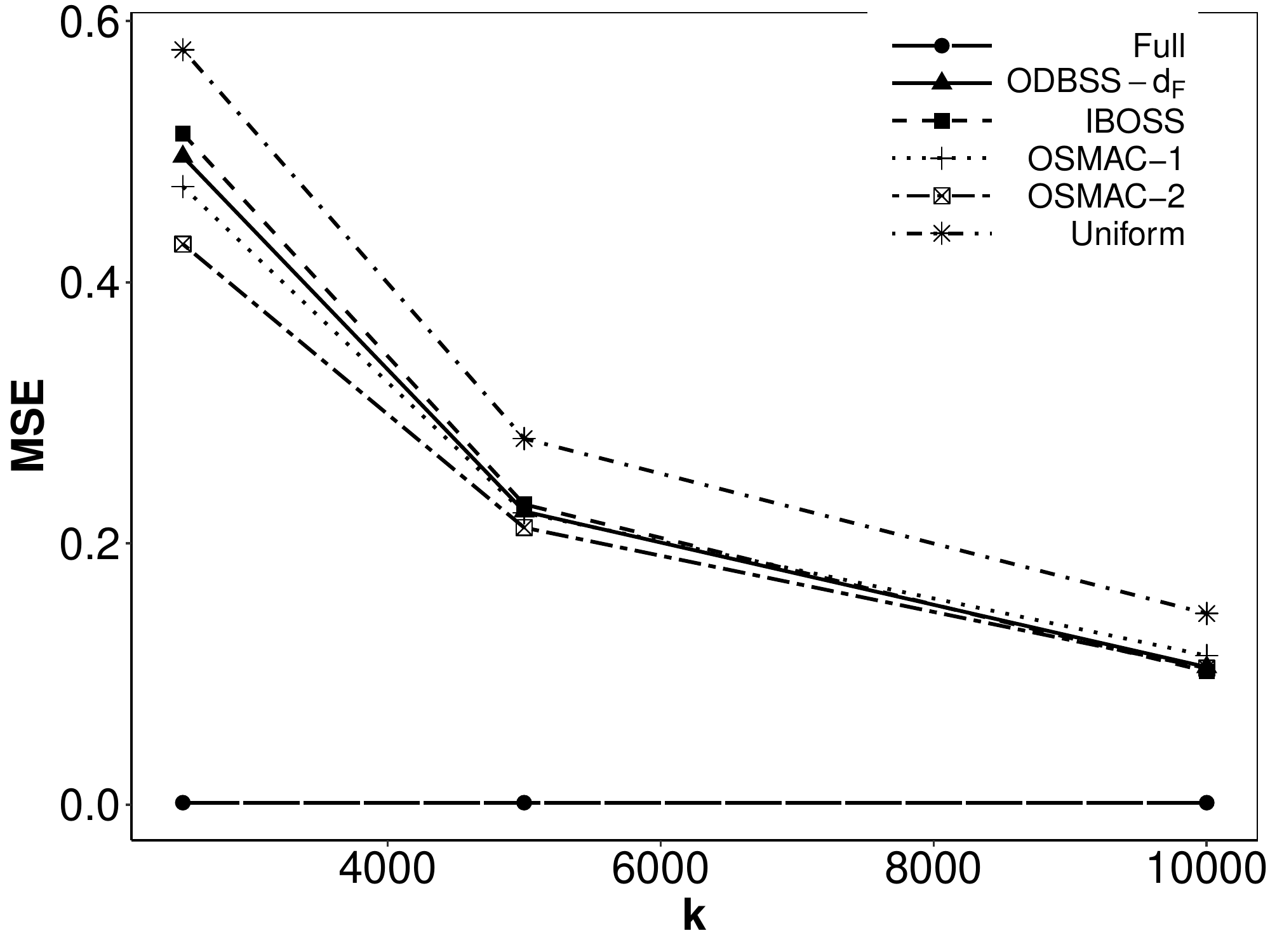}
		\caption{\scriptsize $ \varphi_{p}(\pmb x; \pmb 0, \pmb \Sigma_2)$ }
		\label{fig3_2}
	\end{subfigure}
 \hfill
  \begin{subfigure}[b]{.3\textwidth}
		\centering
		\includegraphics[width=\linewidth]{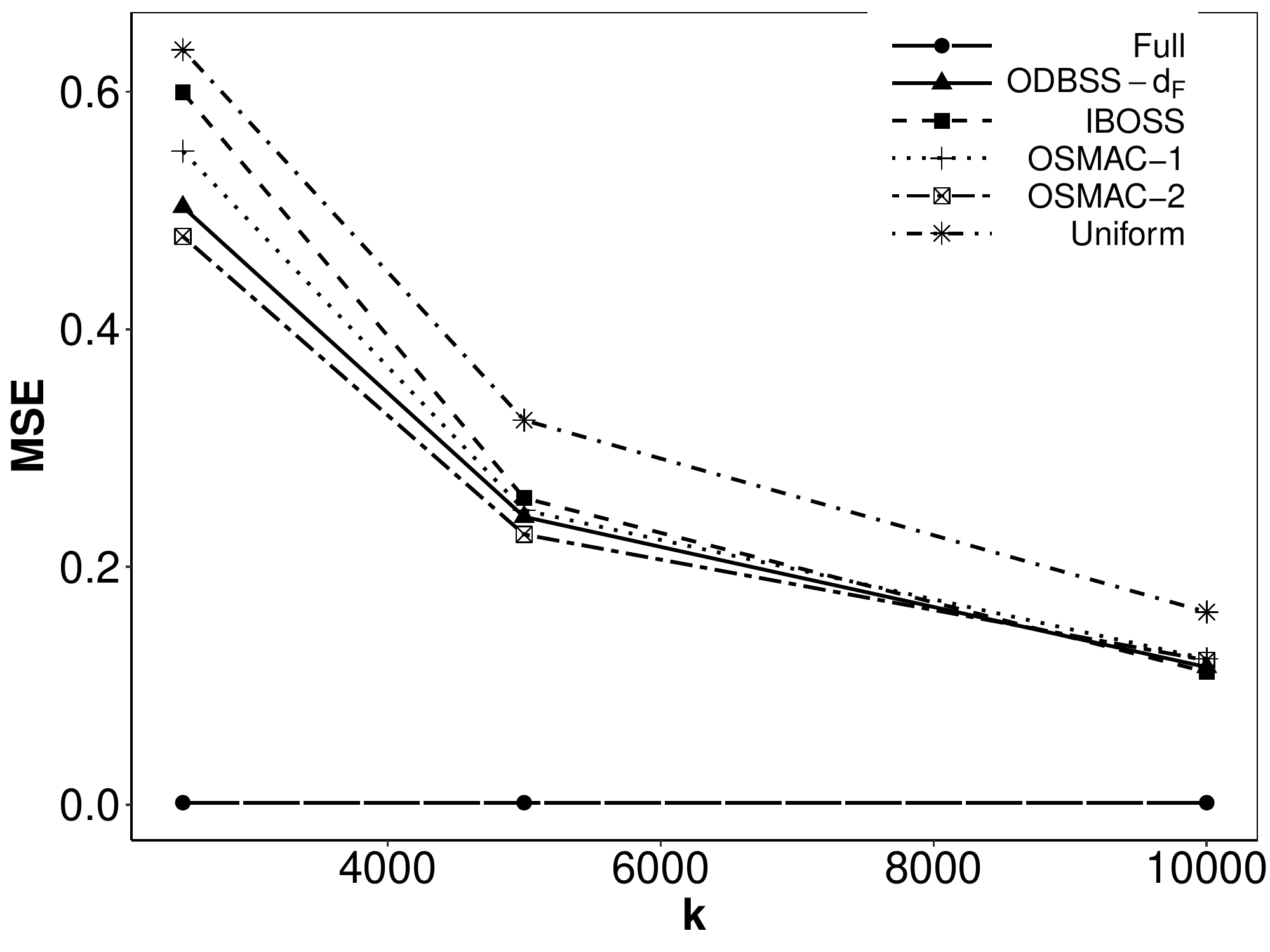}
		\caption{\scriptsize  $ \varphi_{p}(\pmb x; \pmb 0, \pmb \Sigma_3)$}
		\label{fig3_3}
	\end{subfigure}
 \vspace{1cm}
 
 \begin{subfigure}[b]{.3\textwidth}
		\centering
		\includegraphics[width=\linewidth]{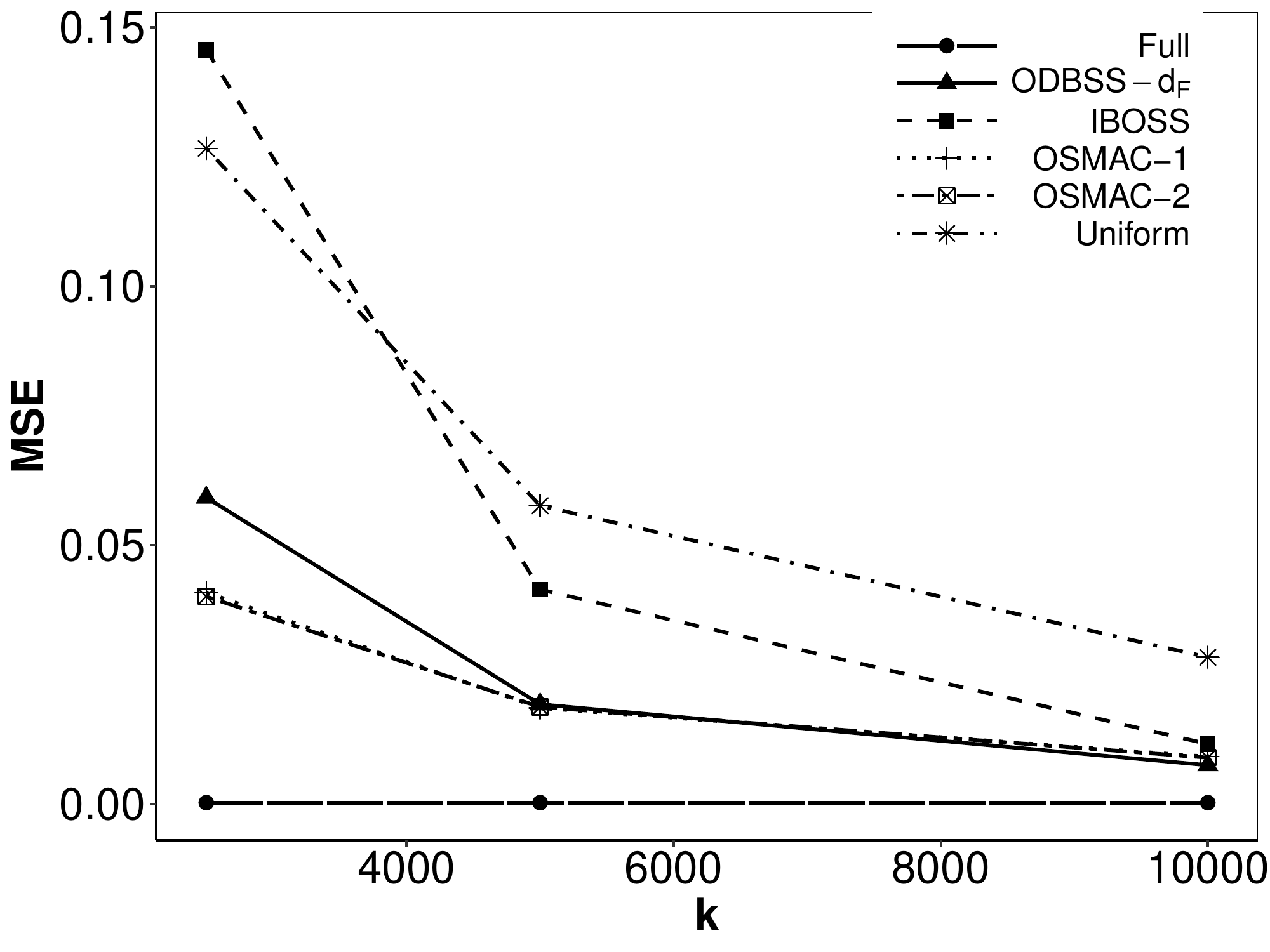}
		\caption{\scriptsize { $\mathcal{T}_{p}(\pmb x;\pmb 0, \pmb \Sigma_1,3)$} }
		\label{fig3_4}
	\end{subfigure}%
 \hfill
   \begin{subfigure}[b]{.3\textwidth}
		\centering
		\includegraphics[width=\linewidth]{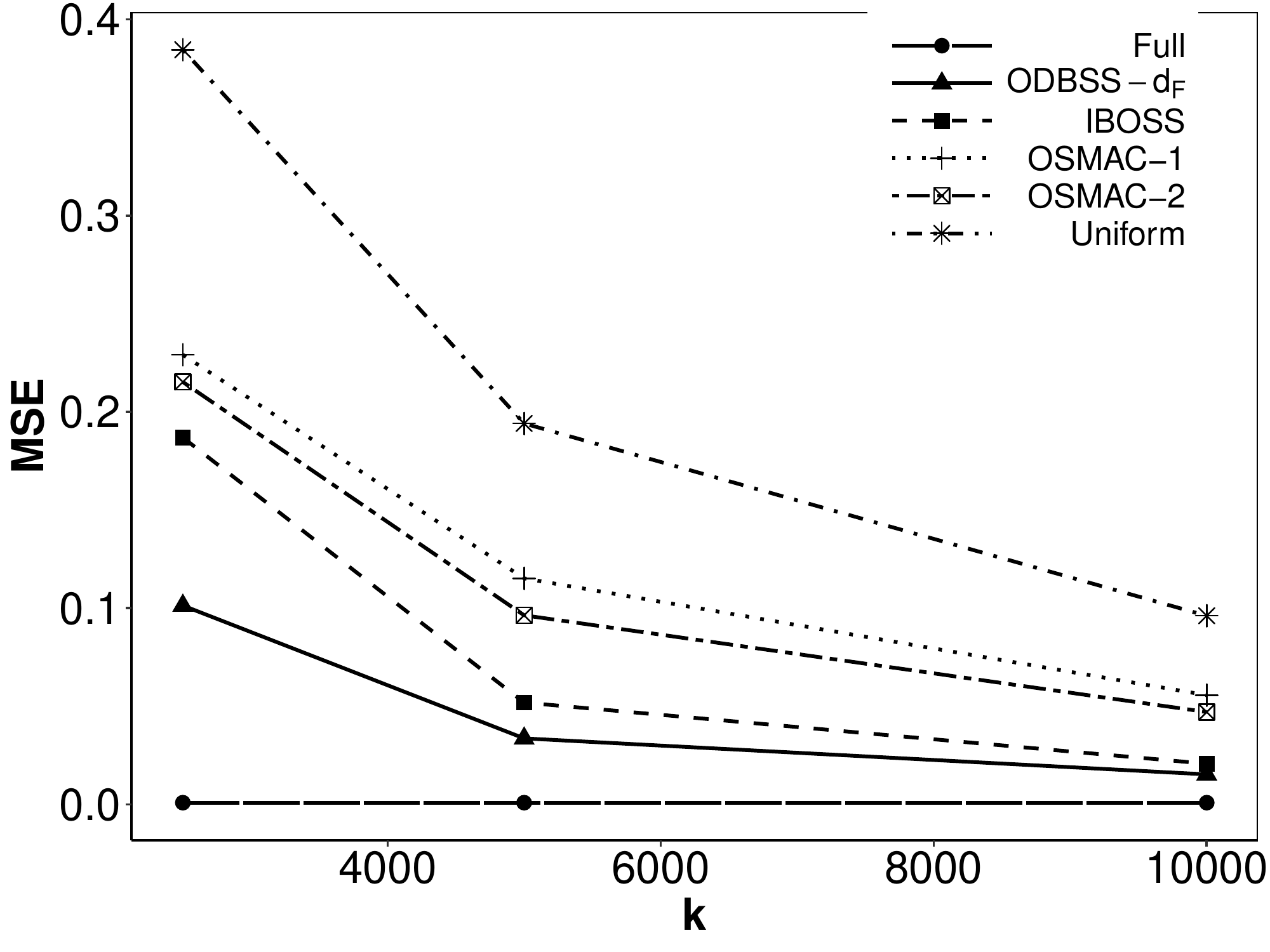}
		\caption{\scriptsize $ \mathcal{T}_{p}(\pmb x;\pmb 0, \pmb \Sigma_2,3)$}
		\label{fig3_5}
	\end{subfigure}
 \hfill
  \begin{subfigure}[b]{.3\textwidth}
		\centering
		\includegraphics[width=\linewidth]{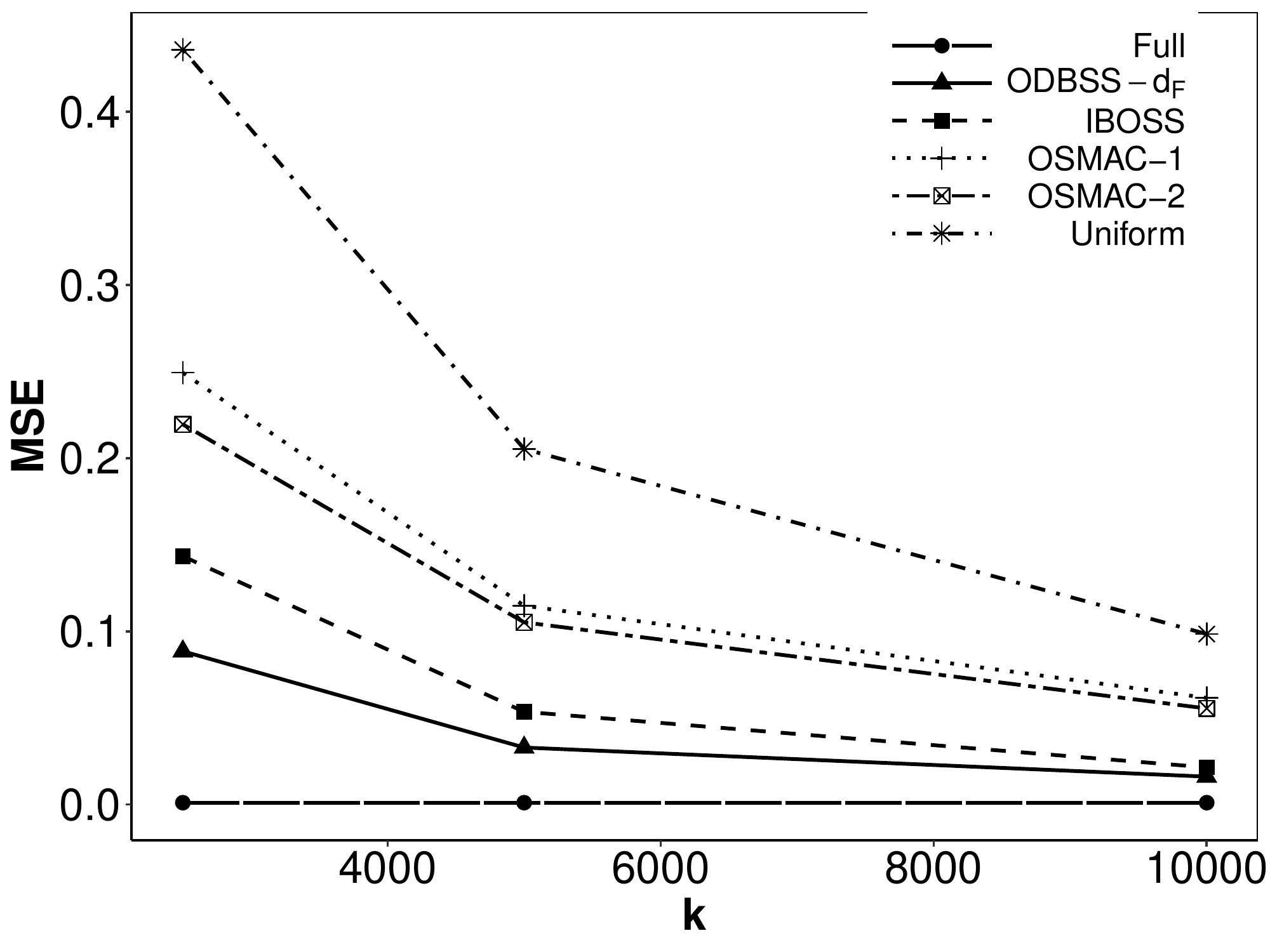}
		\caption{\scriptsize $ \mathcal{T}_{p}(\pmb x;\pmb 0, \pmb \Sigma_3,3)$ }
		\label{fig3_6}
	\end{subfigure}
	\centering	\caption{ \it  The simulated  mean squared error of the parameter estimate in the  logistic regression model \eqref{41LogReg_NoIntercept} {with p = 20} based on
 subsamples obtained by
 ODBSS, IBOSS, OSMAC-1, OSMAC-2, uniform random sampling from $n = 10^6$ observations. 
 The covariates have centered normal and $t$-distribution with $3$ degrees of freedom with different covariances.  
 }
	\label{fig2E} 
\end{figure}

\HD{ Next, we compare the different subsampling procedures in a high-dimensional logistic regression model. In Figure \ref{fig2E}, we display the simulated MSE for model \eqref{41LogReg_NoIntercept} with dimension $p=20$ and sample size $n=10^6$. For the size of the subsample, we use $k=$ 10000, 5000, and 2500. { For the normal distribution ODBSS and the OSMAC-1 (and OSAMC-2) procedures perform best (with no clear winner)} and the figures are very similar to the case $p=7$. For $t$-distributed covariates we also observe a better performance of ODBSS subsampling if $p=20$.}

\HD{\subsubsection{Complex scenarios}}

\HD{In this section, we evaluate the performance of ODBSS in various other scenarios: a more complex design space, unbalanced responses, and covariates from a skewed distribution.}

\HD{
{We begin investigating a situation  where the distribution of the design space has two modes (for the same logistic regression model \eqref{41LogReg_NoIntercept} with $\pmb \beta = (0.5,\ldots, 0.5)^\top $). To be precise we consider data  $\pmb x_1, \pmb x_2, \ldots, \pmb x_n$ are generated from the distribution 
 \begin{align}
 \label{det51}
 \frac{1}{2} \varphi_{p}(\pmb x;\pmb \mu_1, \pmb \Sigma_1) +  \frac{1}{2} \varphi_{p}(\pmb x; \pmb \mu_2, \pmb \Sigma_1),
  \end{align}
where $\pmb \mu_1 = - \pmb \mu_{2}  = (1, 1,\ldots, 1)^\top$ and $ p=7, 20$. A comparison of the different subsampling procedures for this data is shown in Figure \ref{Fig - Complex Space}}. We observe  that
estimation based on 
ODBSS,  OSMAC-1, and OSMAC-2 subsamples has a comparable performance and provide an improvement of IBOSS.  Note that  in 
Figure 	\ref{Fig - Complex Space} (b) we do not show  the MSEs corresponding to the uniform random and IBOSS subsamples as the MSE is larger than $0.6$ for these procedures. 
  \begin{figure}[H] 
 	\centering
	\begin{subfigure}[b]{.4\textwidth}
		\includegraphics[width=\linewidth]{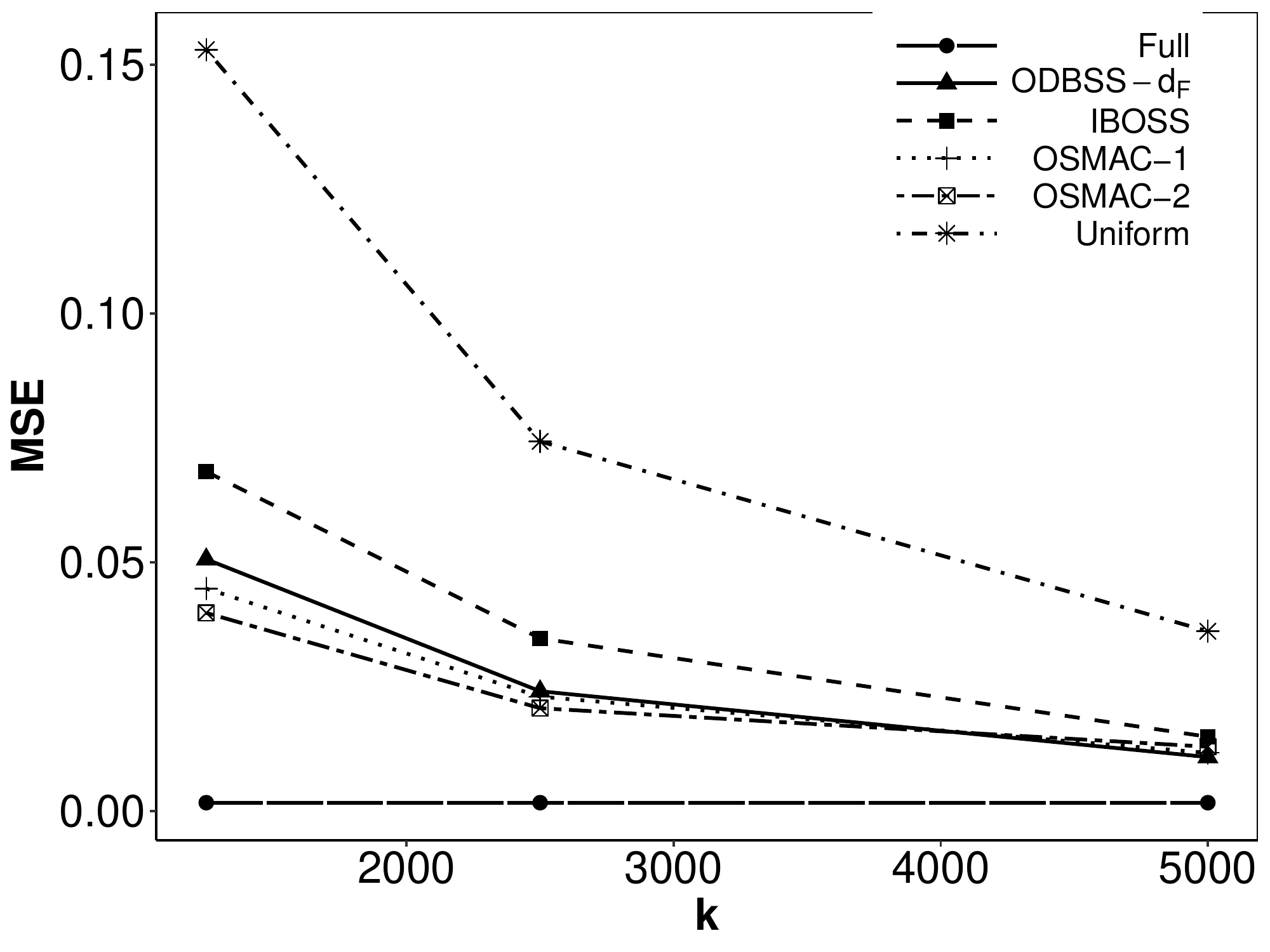}
		\caption{\scriptsize  $ \frac{1}{2} \varphi_{7}(\pmb x;\pmb \mu_1, \pmb \Sigma_1) +  \frac{1}{2} \varphi_{7}(\pmb x; \pmb \mu_2, \pmb \Sigma_1)$ and n= $10^5$}
	\end{subfigure}%
~~~~~~~~~~~~~~
	\begin{subfigure}[b]{.4\textwidth}
		\centering
		\includegraphics[width=\linewidth]{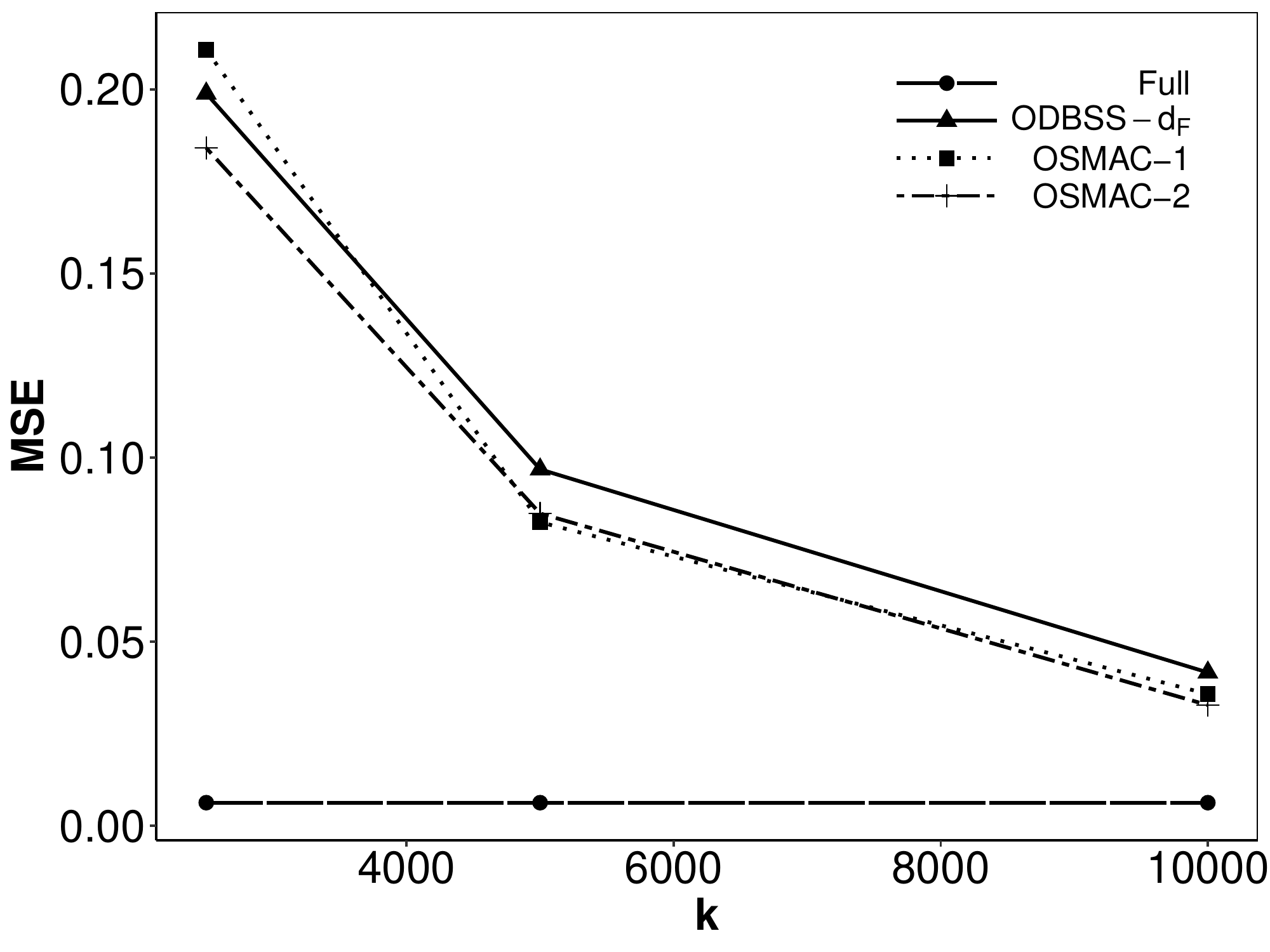}
		\caption{\scriptsize  $ \frac{1}{2} \varphi_{20}(\pmb x;\pmb \mu_1, \pmb \Sigma_1) +  \frac{1}{2} \varphi_{20}(\pmb x; \pmb \mu_2, \pmb \Sigma_1)$ and n= $10^6$}
	\end{subfigure}%
	 	\caption{ \it  \HD{ The simulated  mean squared error of the parameter estimate in the  logistic regression model \eqref{41LogReg_NoIntercept} {with p = 7 and 20} based on
 subsamples obtained by
 ODBSS,  OSMAC-1, OSMAC-2, and the full sample {with $n = 10^5$ and $n = 10^6$, respectively}. 
 The covariates are generated from the mixture of two normal distributions defined in \eqref{det51}.}}
	\label{Fig - Complex Space} 
\end{figure}
}

\begin{figure}[H]
  
	\centering
	\begin{subfigure}[b]{.3\textwidth}
		\centering
		\includegraphics[width=\linewidth]{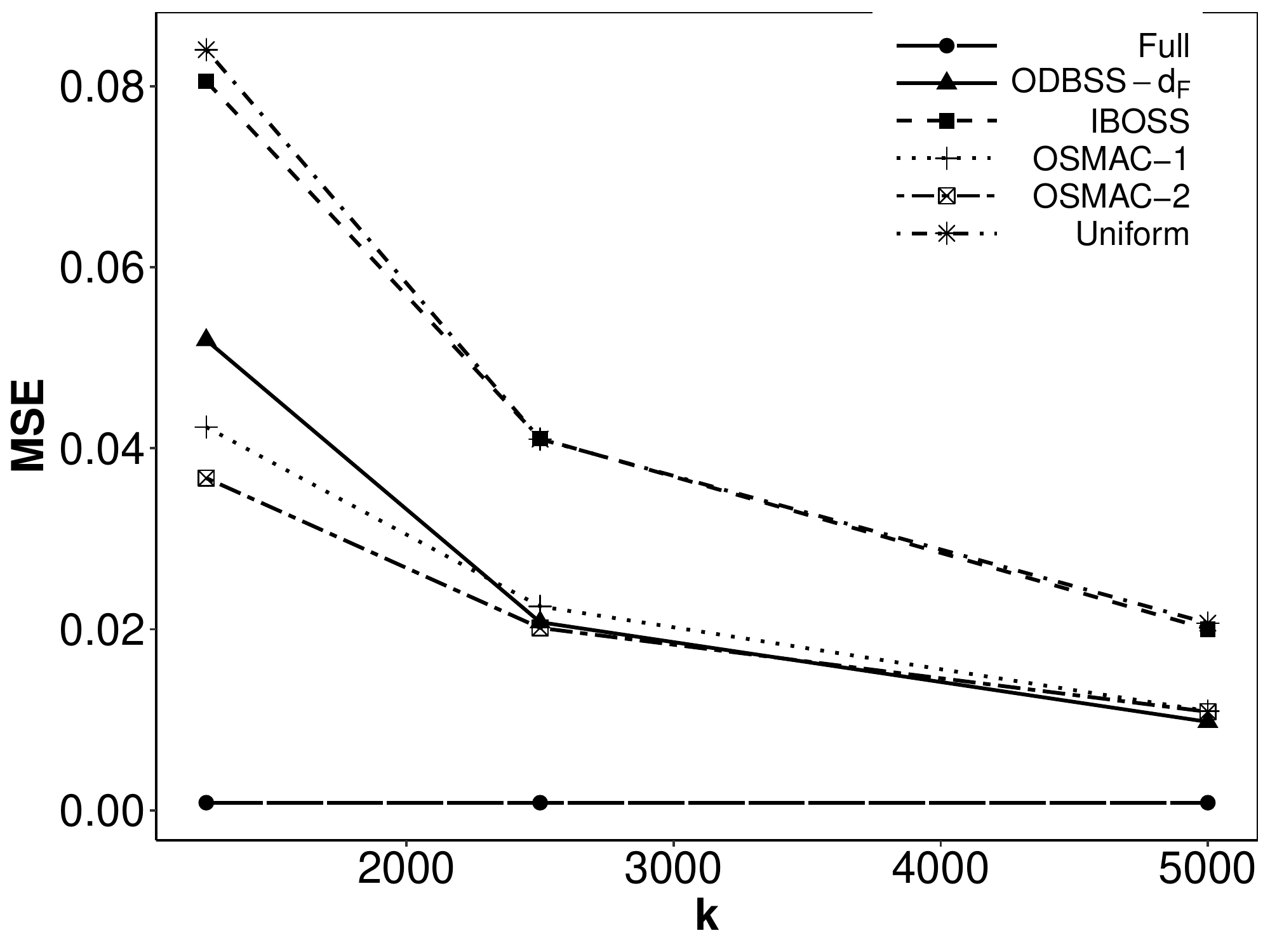}
		\caption{\scriptsize $
 \varphi_p( \pmb x; {0.5} \; \pmb 1, \pmb \Sigma_1)$ }
		\label{Fig - nzNormal - 2 a}
	\end{subfigure}%
\hfill
	\begin{subfigure}[b]{.3\textwidth}
		\centering
		\includegraphics[width=\linewidth]{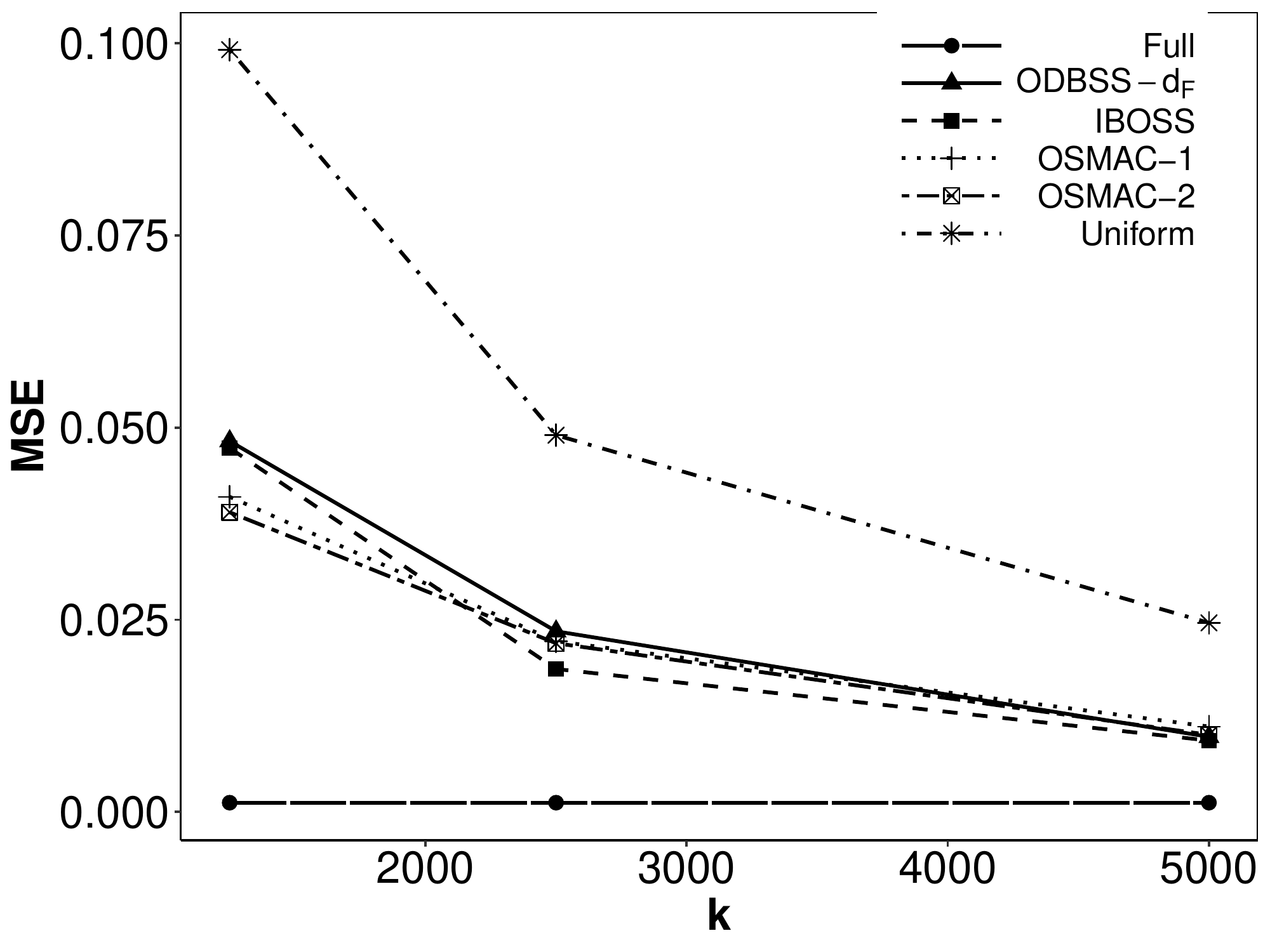}
		\caption{\scriptsize $ \varphi_p( \pmb x; {0.75} \; \pmb 1, \pmb \Sigma_1)  $ }
	\end{subfigure}%
 \hfill
	\begin{subfigure}[b]{.3\textwidth}
		\centering
		\includegraphics[width=\linewidth]{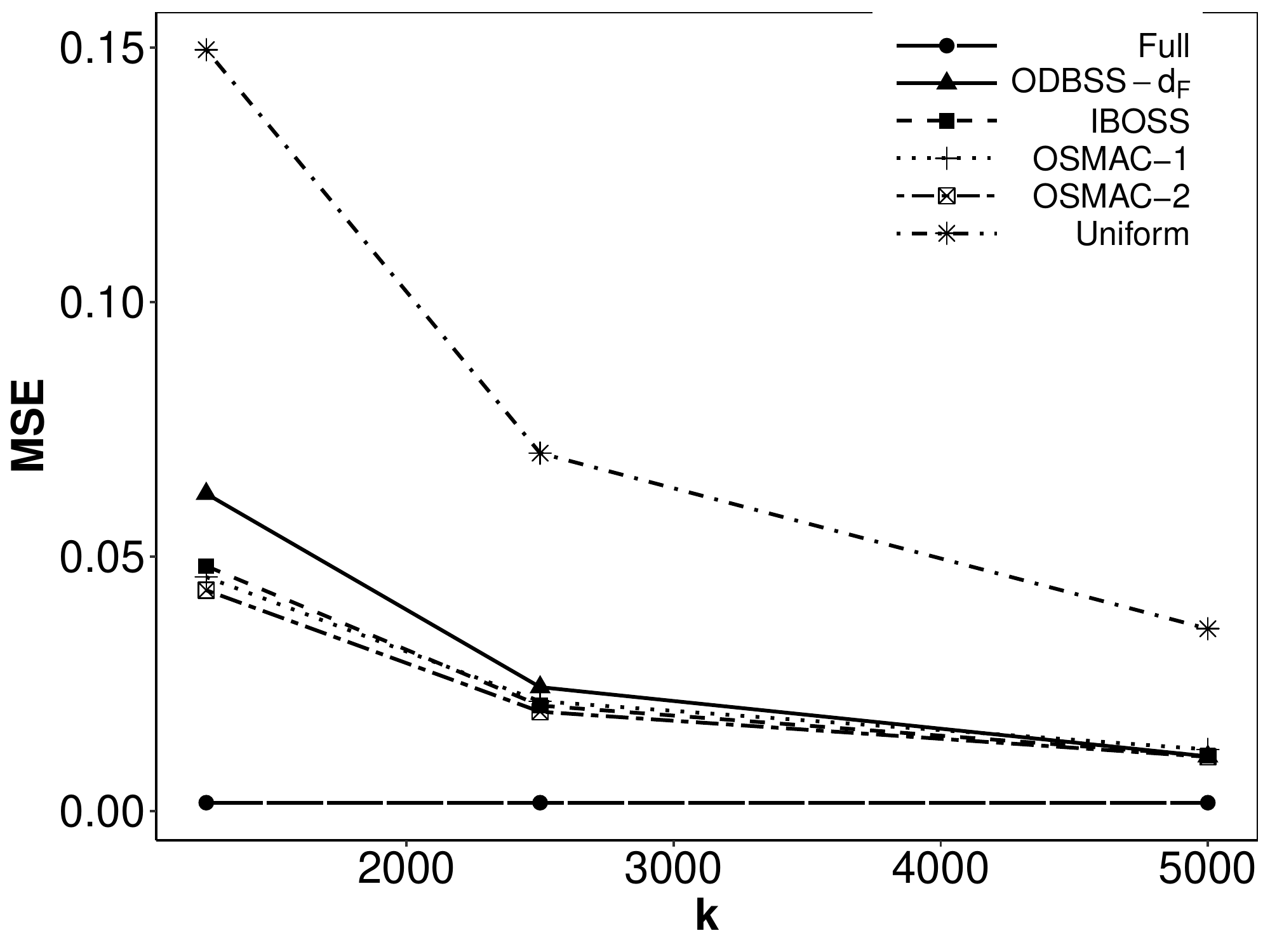}
		\caption{\scriptsize  $\varphi_p(  \pmb x;  \textbf{1}, \pmb \Sigma_1)$ }
	\end{subfigure}%
 	\caption{ \textit{\HD{The simulated  mean squared error of the parameter estimate in the  logistic regression model \eqref{41LogReg_NoIntercept} {with p = 7} based on
 subsamples obtained by
 ODBSS, IBOSS, OSMAC-1, OSMAC-2, uniform random sampling from {$n = 10^5$} observations for the unbalanced response cases. 
 }} 
  }
	\label{Fig - nzNormal} 
\end{figure}

\begin{figure}[H]
  
	\centering
	\begin{subfigure}[b]{.3\textwidth}
		\centering
		\includegraphics[width=\linewidth]{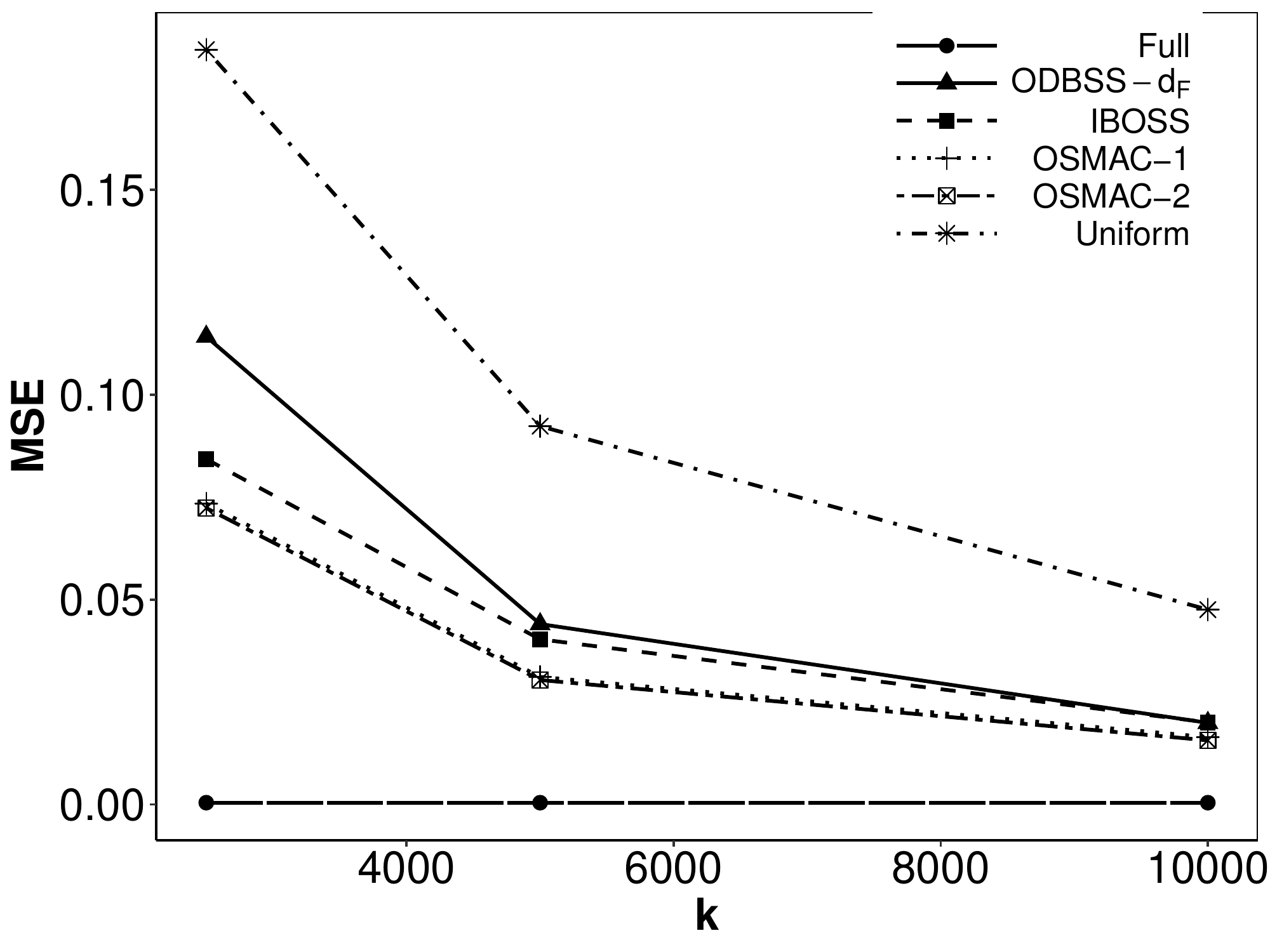}
		\caption{\scriptsize $ \varphi_{p}( \pmb x;  0.3 \; \textbf{1} , \pmb \Sigma_1) $ }
	\end{subfigure}%
\hfill
	\begin{subfigure}[b]{.3\textwidth}
		\centering
		\includegraphics[width=\linewidth]{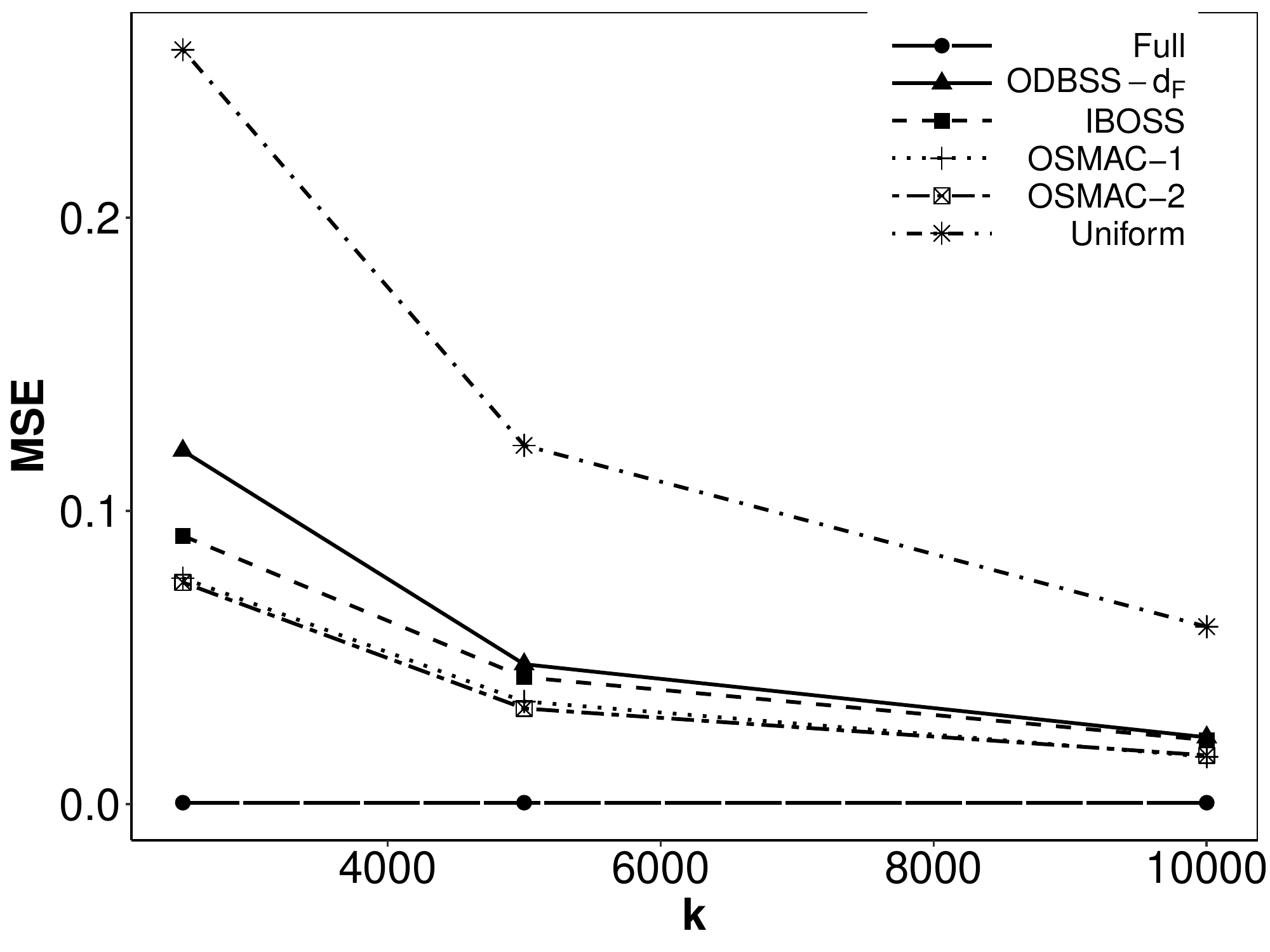}
		\caption{\scriptsize  $ \varphi_{p}( \pmb x;  0.4 \; \textbf{1} , \pmb \Sigma_1) $ }
	\end{subfigure}%
 \hfill
	\begin{subfigure}[b]{.3\textwidth}
		\centering
		\includegraphics[width=\linewidth]{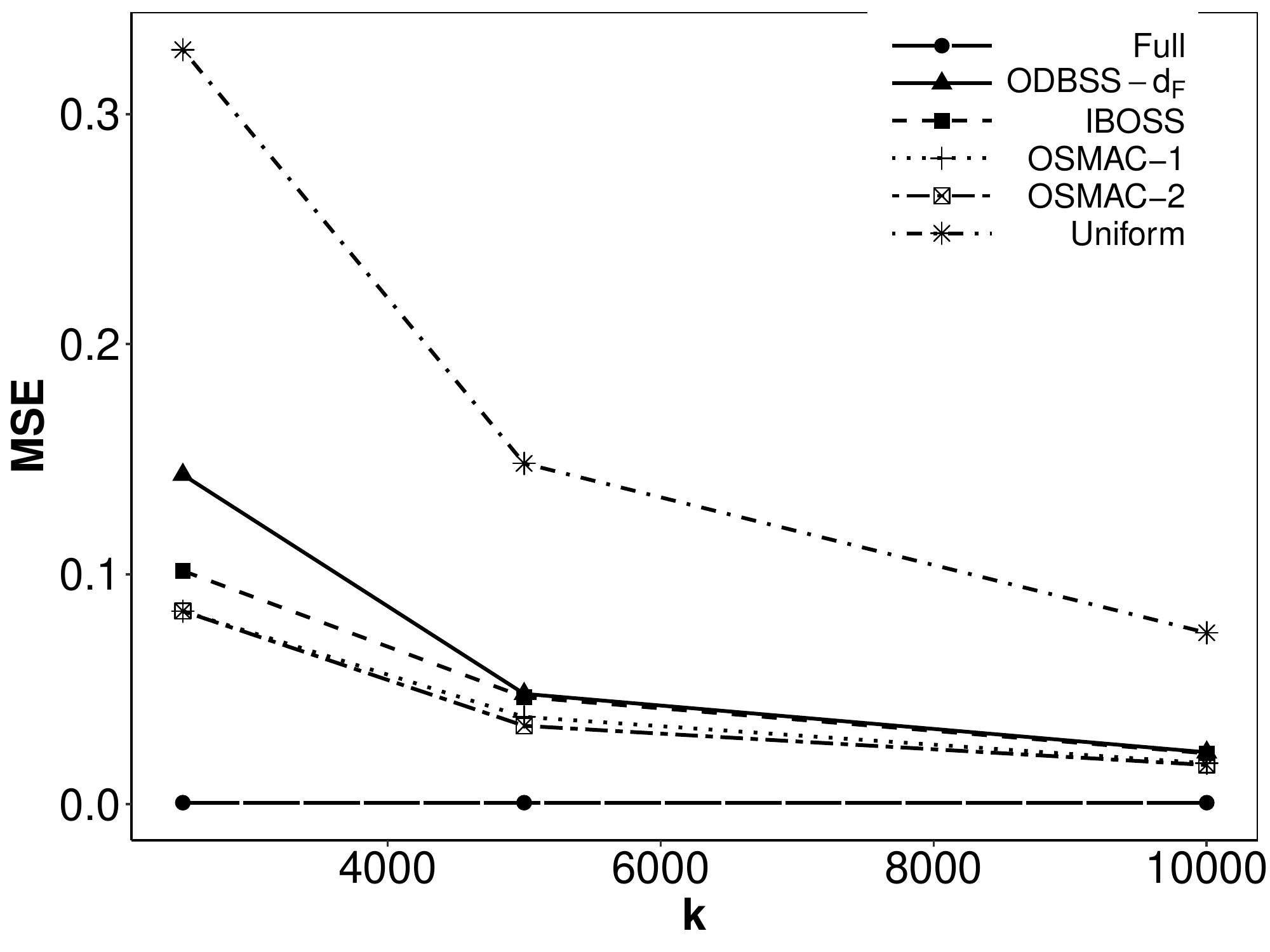}
		\caption{\scriptsize  $ \varphi_{p}( \pmb x;  0.5 \; \textbf{1} , \pmb \Sigma_1) $ }
	\end{subfigure}%

 	\caption{ \textit{\HD{The simulated  mean squared error of the parameter estimate in the  logistic regression model \eqref{41LogReg_NoIntercept} {with p = 20} based on
 subsamples obtained by
 ODBSS, IBOSS, OSMAC-1, OSMAC-2, uniform random sampling, and the full sample from {$n = 10^6$} observations for the unbalanced response cases. 
 }}   }
	
	\label{Fig - nzNormal - 2} 
\end{figure}

\HD{Next, we investigate some cases where the responses in the logistic regression model \eqref{41LogReg_NoIntercept} are unbalanced. For this purpose considering $p$-dimensional normal distributions with $p=7$ and $p=20$ and covariance $\pmb \Sigma_1$ for the predictor, where we use  different centers to ensure an increasing percentage of positive responses in the data. More precisely, for $p=7$, we consider three distributions, $ 
 \varphi_7( \pmb x; {0.5} \; \pmb 1, \pmb \Sigma_1)$, $ 
 \varphi_7(\pmb x; {0.75} \; \pmb 1, \pmb \Sigma_1)$, and $ 
 \varphi_7(  \pmb x; \; \pmb 1, \pmb \Sigma_1)$, where $\pmb 1 = (1,\ldots,1)^\top$. Here the percentage of positive responses is 75\%, 85\%, and 90\%, respectively. For $p=20$, we consider $
 \varphi_{20}( \pmb x;{0.3} \; \pmb 1, \pmb \Sigma_1)$, $\varphi_{20}( \pmb x;{0.4} \; \pmb 1, \pmb \Sigma_1)$, and $
 \varphi_{20}( \pmb x;  0.5 \; \pmb 1, \pmb \Sigma_1)$ with 75\%, 85\%, and 90\% positive responses, respectively.\\
 We display the simulated mean squared error for the different subsampling procedures in Figure \ref{Fig - nzNormal} and \ref{Fig - nzNormal - 2} and observe that ODBSS is comparable to all other algorithms and performs well when subsample sizes are \HD{higher (that is, $k=2500, 5000$ when $p=7$ and  $k=5000, 10000$ when $p=20$)}. In case, the data has unbalanced responses, we recommend taking subsamples that are not very small for ODBSS to work well. Both, OSMAC-1 and OSMAC-2 give the best estimation results and this is understandable as OSMAC-1 and OSMAC-2 are designed to choose points that have a higher probability of mis-classification. }

\HD{ Finally, we investigate the performance of various subsampling algorithms when the covariates are generated from a skewed distribution.
We consider two classes of skewed distributions studied in \cite{azzalini2013skew}, a centered $p$-dimensional skew-normal distribution with covariance $\boldsymbol{\Sigma}$  and slant parameter $\boldsymbol{\alpha} \in \mathbb{R}^p$, defined by the density  
\begin{equation}
\varphi_p(\pmb x; \pmb 0, \boldsymbol{\Sigma}, \boldsymbol{\alpha}) =   2  \; \varphi_p(\boldsymbol{x}; \pmb 0, \boldsymbol{\Sigma}) \; \varphi_1( \boldsymbol{\alpha}^\top \boldsymbol{x};0, 1) 
\end{equation}
and a centered $p$-dimensional skew-t distribution with covariance $\boldsymbol{\Sigma}$, slant parameter $\boldsymbol{\alpha} \in \mathbb{R}^p$, and degrees of freedom  $\kappa$, defined by the density 
\begin{equation}
\mathcal{T}_{p}(\pmb x; \pmb 0, \pmb \Sigma, \kappa, \boldsymbol{\alpha} ) =   2  \;  \mathcal{T}_{p}(\pmb x; \pmb 0, \pmb \Sigma, \kappa) \; \;
\mathcal{T}_{ 1} \left( \boldsymbol{\alpha}^\top \boldsymbol{x} \; \sqrt{  \frac{\kappa+ p}{\kappa+ \pmb x^\top \boldsymbol{\Sigma}^{-1} \pmb x }   } ; 0, 1, \kappa+p \right  ).
\end{equation}
For our simulation studies, we use $p=7$, $\pmb \Sigma = \pmb \Sigma_1, \pmb \Sigma_2, \pmb \Sigma_3$ as we did in all previous studies, and set the slant parameter $\pmb \alpha = (20,1,1,1,1,1,1)$.
We use the package $sn$ in $R-software$ \citep{snPackage} to generate multivariate skew-normal and skew-t distributed predictors.
}

\begin{figure}[H]
	\centering
	\begin{subfigure}[b]{.3\textwidth}
		\centering
		\includegraphics[width=\linewidth]{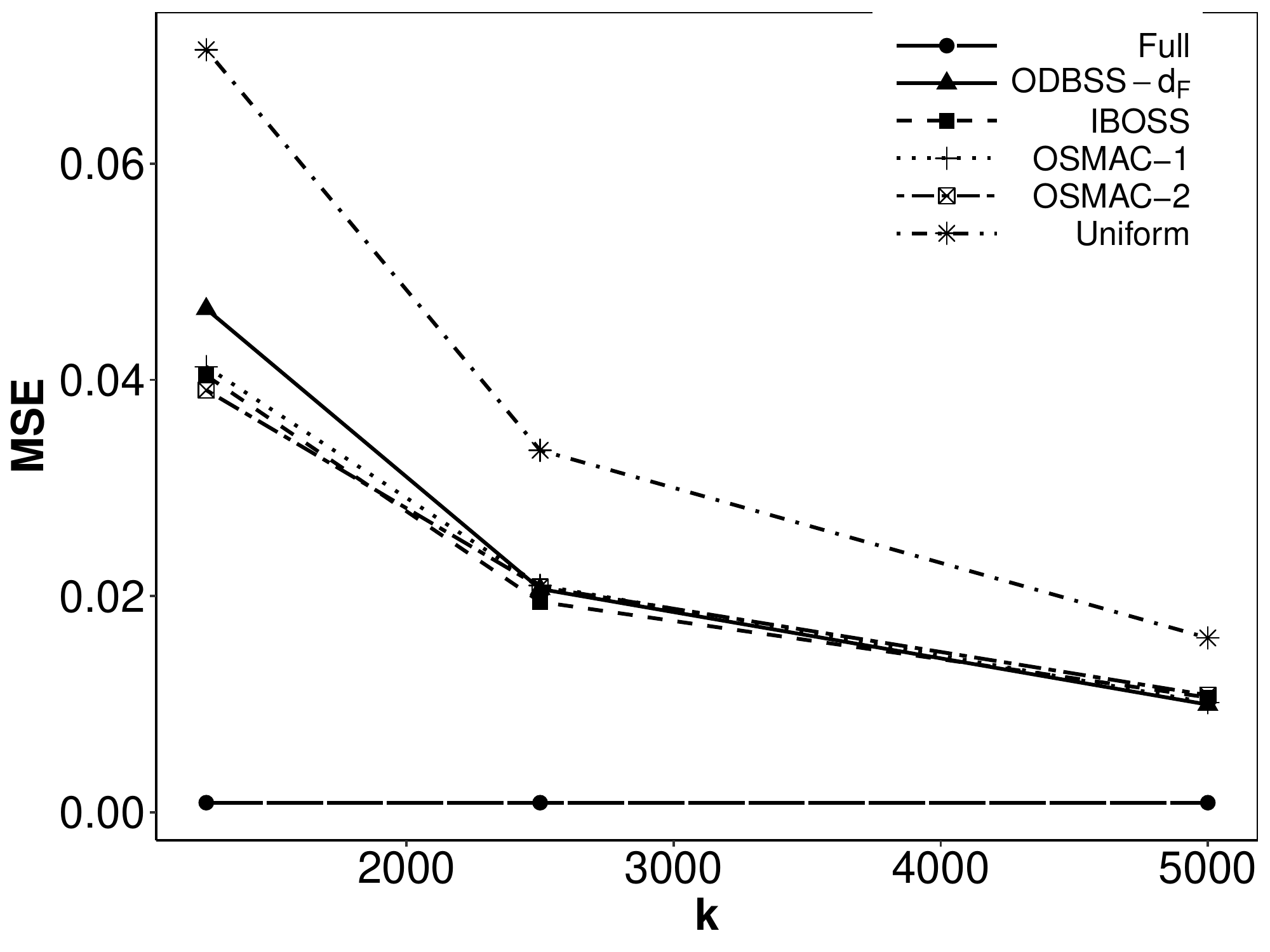}
		\caption{\scriptsize $ \varphi_p(  {\pmb x; \pmb 0, \Sigma_1}, \pmb \alpha )$ }
		\end{subfigure}%
 \hfill
   \begin{subfigure}[b]{.3\textwidth}
		\centering
		\includegraphics[width=\linewidth]{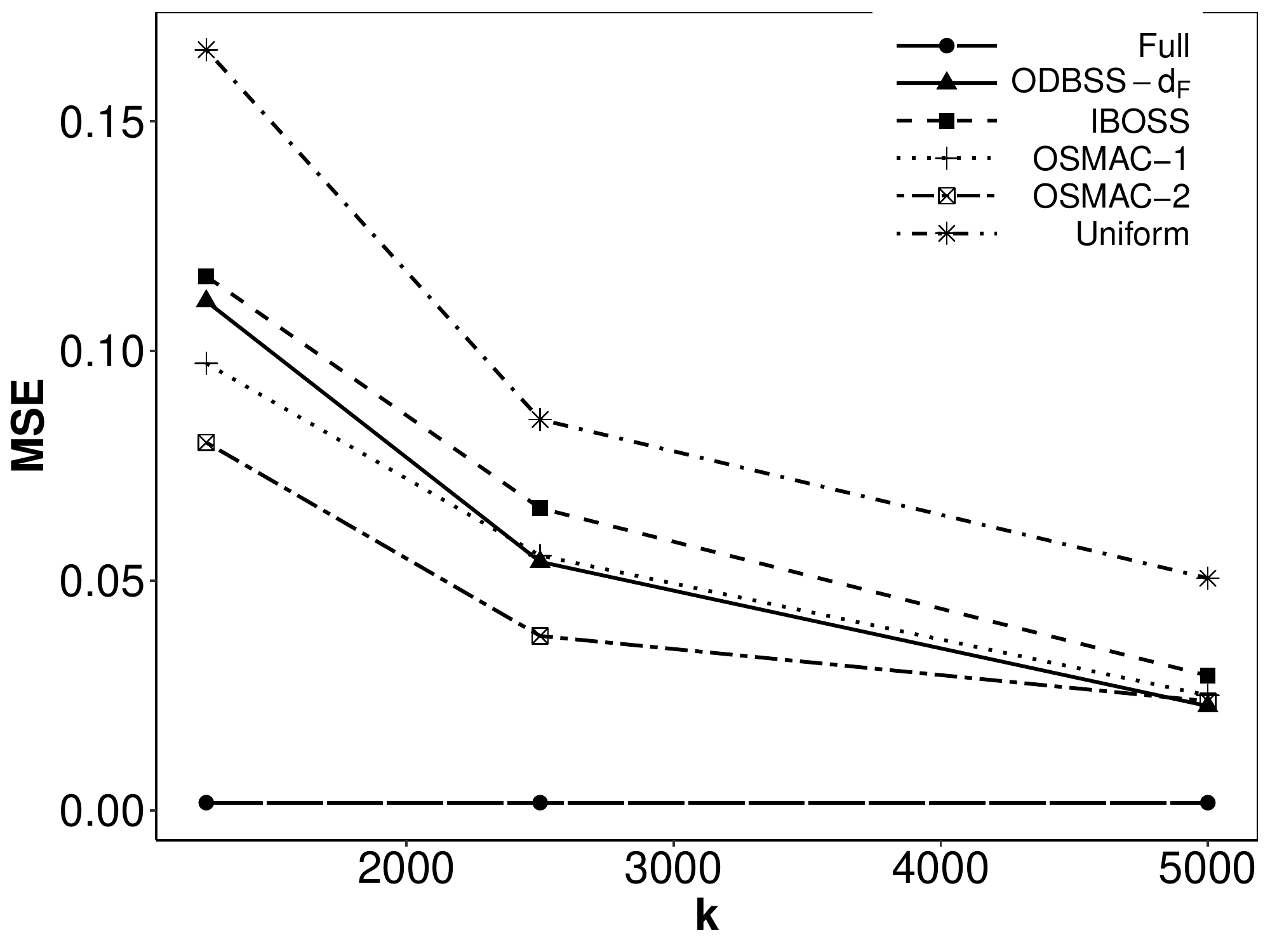}
		\caption{\scriptsize $ \varphi_p(  {\pmb x;\pmb 0, \Sigma_2}, \pmb \alpha )$}
		\end{subfigure}
 \hfill
  \begin{subfigure}[b]{.3\textwidth}
		\centering
		\includegraphics[width=\linewidth]{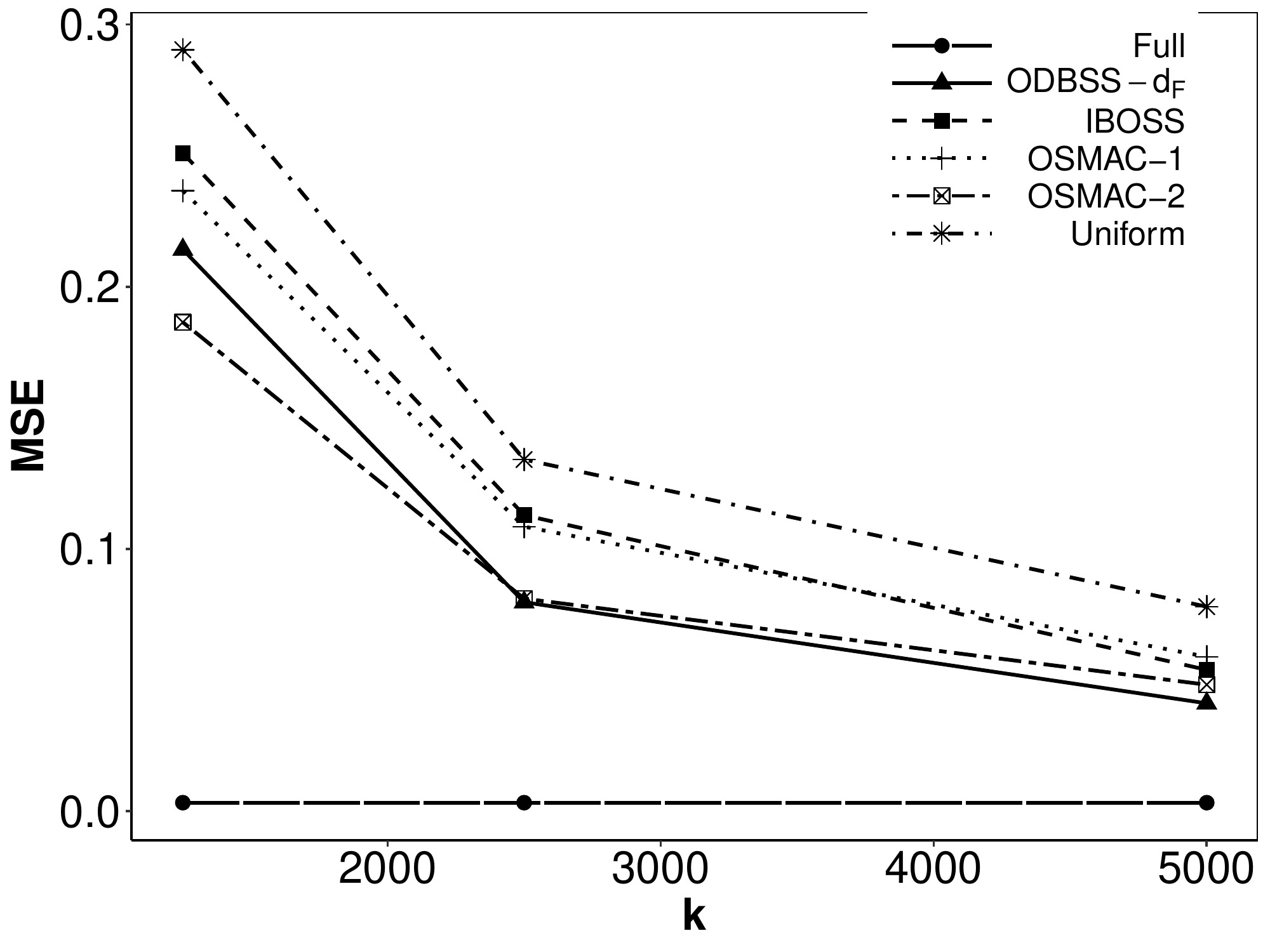}
		\caption{\scriptsize  $ \varphi_p( {\pmb x;\pmb 0, \Sigma_3}, \pmb \alpha )$}
		\end{subfigure}
 \vspace{1cm}
 
 \begin{subfigure}[b]{.3\textwidth}
		\centering
		\includegraphics[width=\linewidth]{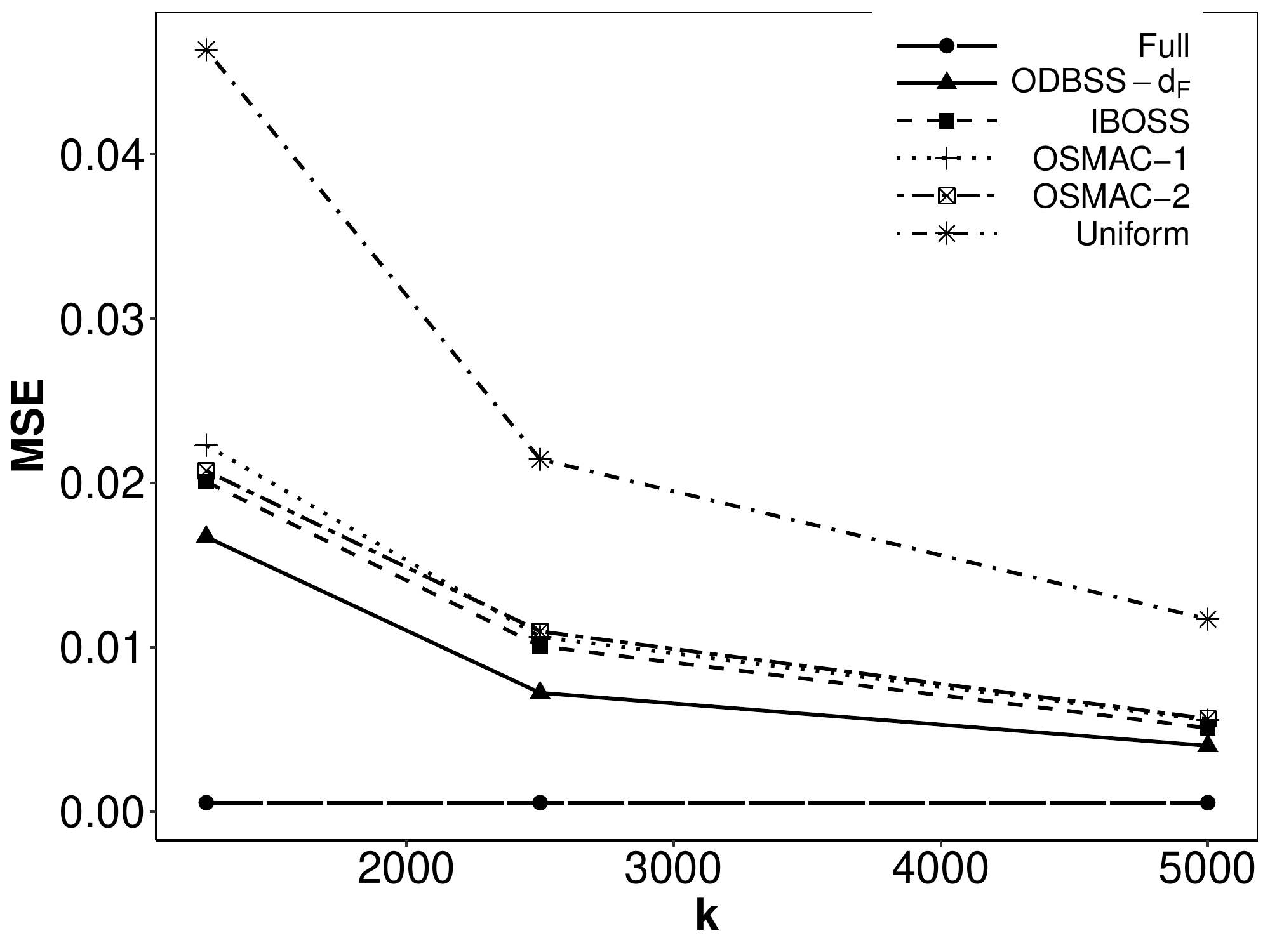}
		\caption{\scriptsize $ \mathcal{T}_p( \pmb x; \pmb 0, {\pmb \Sigma_1}, 3,\pmb \alpha )$ }
	\end{subfigure}%
 \hfill
   \begin{subfigure}[b]{.3\textwidth}
		\centering
		\includegraphics[width=\linewidth]{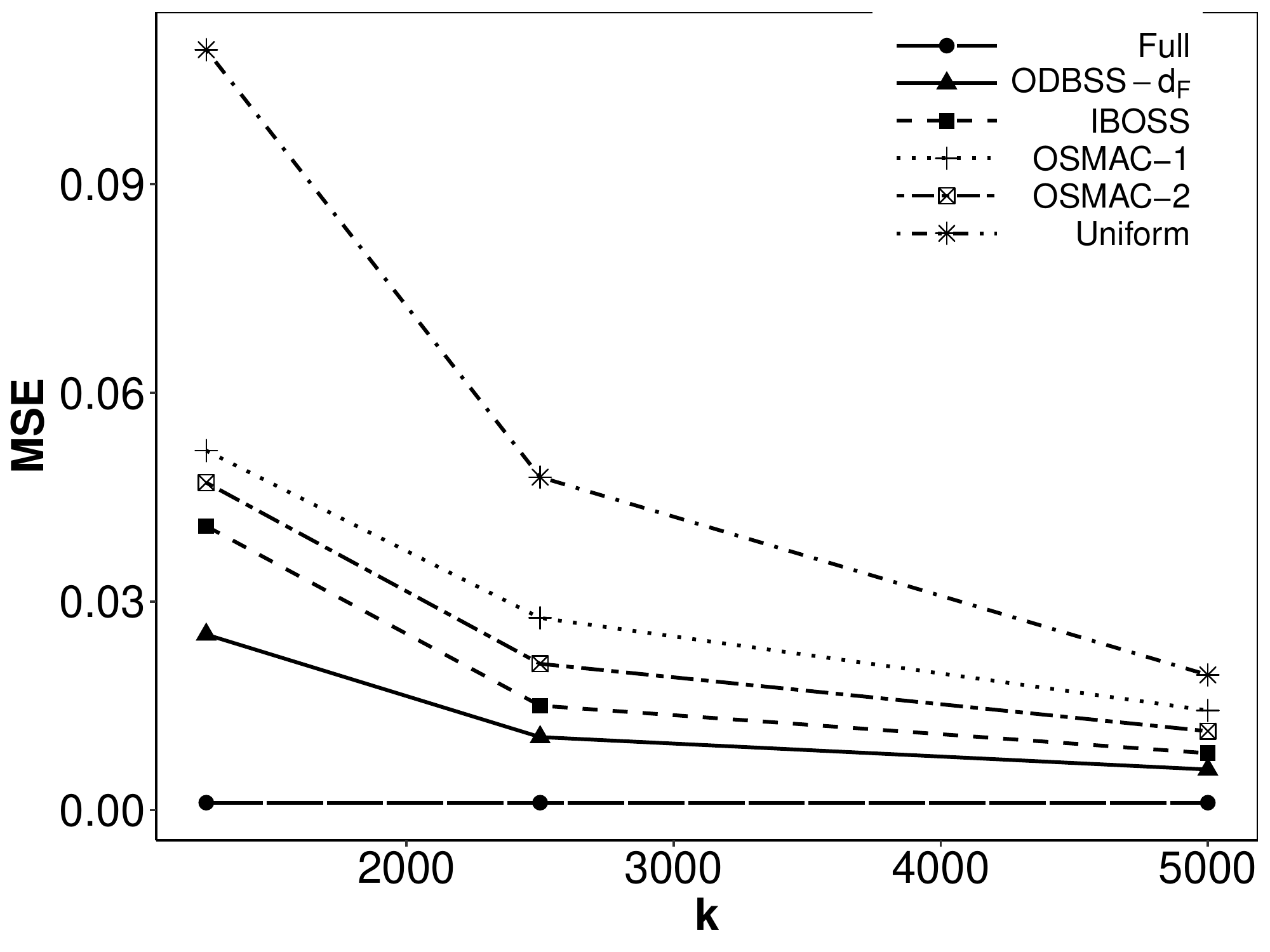}
		\caption{\scriptsize $ \mathcal{T}_p( \pmb x;\pmb 0 , {\pmb \Sigma_2}, 3,\pmb \alpha )$}
	\end{subfigure}
 \hfill
  \begin{subfigure}[b]{.3\textwidth}
		\centering
		\includegraphics[width=\linewidth]{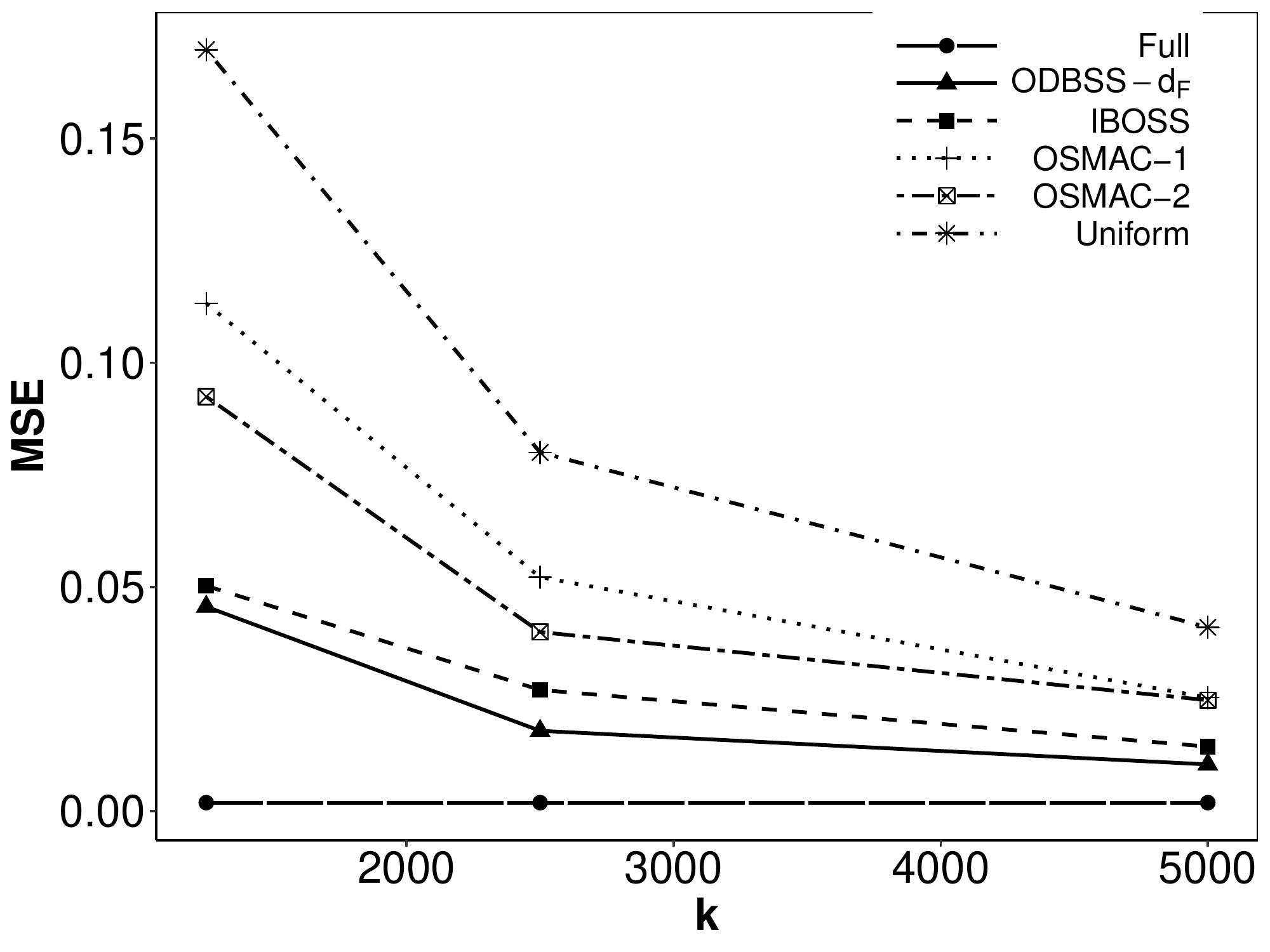}
		\caption{\scriptsize $ \mathcal{T}_p( \pmb x;\pmb 0 , {\pmb \Sigma_3} ,3, \pmb \alpha)$ }
	\end{subfigure}
	\centering	\caption{ \it  The simulated  mean squared error of the parameter estimate in the  logistic regression model  \eqref{41LogReg_NoIntercept} {with p = 7} based on
 subsamples obtained by
 ODBSS, IBOSS, OSMAC-1, OSMAC-2, uniform random sampling, from {$n = 10^5$} observations for skew-normal and skew-t distributions with slant parameter $\pmb \alpha = (20,1,1,1,1,1,1)$.
 }
	\label{Figure Supplementary 4} 
\end{figure}

\HD{The corresponding results are given in Figure \ref{Figure Supplementary 4} and can be compared with Figure \ref{fig2a} where we display the corresponding results non-skewed distributions. We observe that ODBSS (and also the other procedures) are rather robust with respect to the skewness, and that for skew-t distriution ODBSS shows superioriority. }

\HD{The results of the simulation studies in this section and Section \ref{sec412} indicate that ODBSS is very competitive and performs better than existing algorithms in many cases (especially, if the covariates are generated from a heavy tailed distribution).  Moreover, ODBSS is robust to various covariate distributions and imbalance in the responses.}

\subsubsection{ Using only the most important points of the optimal design} 
\label{sec413}
In Section \ref{sec32}, we proposed Algorithm \ref{algred}, which removes support points of the optimal design with small weights. As this procedure would yield substantial computational advantages of ODBSS it is of interest to investigate its impact on the quality of the estimates based on the resulting subsample. In the left part of Figure \ref{fig4} we display the simulated mean squared error of the estimator from a subsample obtained by ODBSS, where we apply Algorithm \ref{Algorithm3}  with efficiency thresholds $\zeta = 100\%,$ $97.5\%,$  $95\%,$ $90\%,$ \HD{$85\%$ } 
and  $80\%$ to reduce the number of support points of the optimal design in Step 2 of Algorithm \ref{Algorithm1}. We consider centered
 normal distributed predictors with covariance matrices $\pmb \Sigma_1$, $\pmb \Sigma_2$, and $\pmb \Sigma_3$. We observe that the MSE increases only very slowly.
For example, compared to the normal distribution with covariance matrix $\pmb \Sigma_2$ the MSE is $0.0279$ if we apply ODBSS using all support points ($100\%$ efficiency) and $0.0257$ if ODBSS uses only the support points of the design with $80\%$ efficiency. 
In the right panel, we display the corresponding average numbers of support points, which decrease from about 25 to about 10 for all three normal distributions. We observe that the number of support points decreases sharply, as the $A$-efficiency decreases. The designs with  $97.5\%$  or $95\%$ $A$-efficiency seem to be a very good choice. In this case, there is not much of an increase in MSE but the computational complexity is reduced to almost half as the number of support points of the optimal design is almost halved at this cutoff. 
{      \HD{ We recommend using a design with A-efficiency in the range of 90\%-95\% as this leads to significant decrease in computational complexity}.  \\
       \HD{Finally, we briefly comment  on the  - on a first glance -  counter-intuitive decrease in MSE for the matrix $\Sigma_3$ in the left panel of Figure \ref{Fig - nzNormal - 2}. An $A$-optimal design minimizes the variances of the  estimates for the coefficients of the parameter vector. While for linear models the estimates are unbiased and 
      these variances coincide with the mean squared errors, the bias of the estimates in  non-linear models vanishes only  asymptotically. Therefore we conjecture  that the decrease in MSE is caused by a bias, which makes $A$-optimality and minimization of the MSE not equivalent.   }

\begin{figure}[H]
	\centering
	\begin{subfigure}[b]{.5\textwidth}
		\centering
		\includegraphics[width=\linewidth]{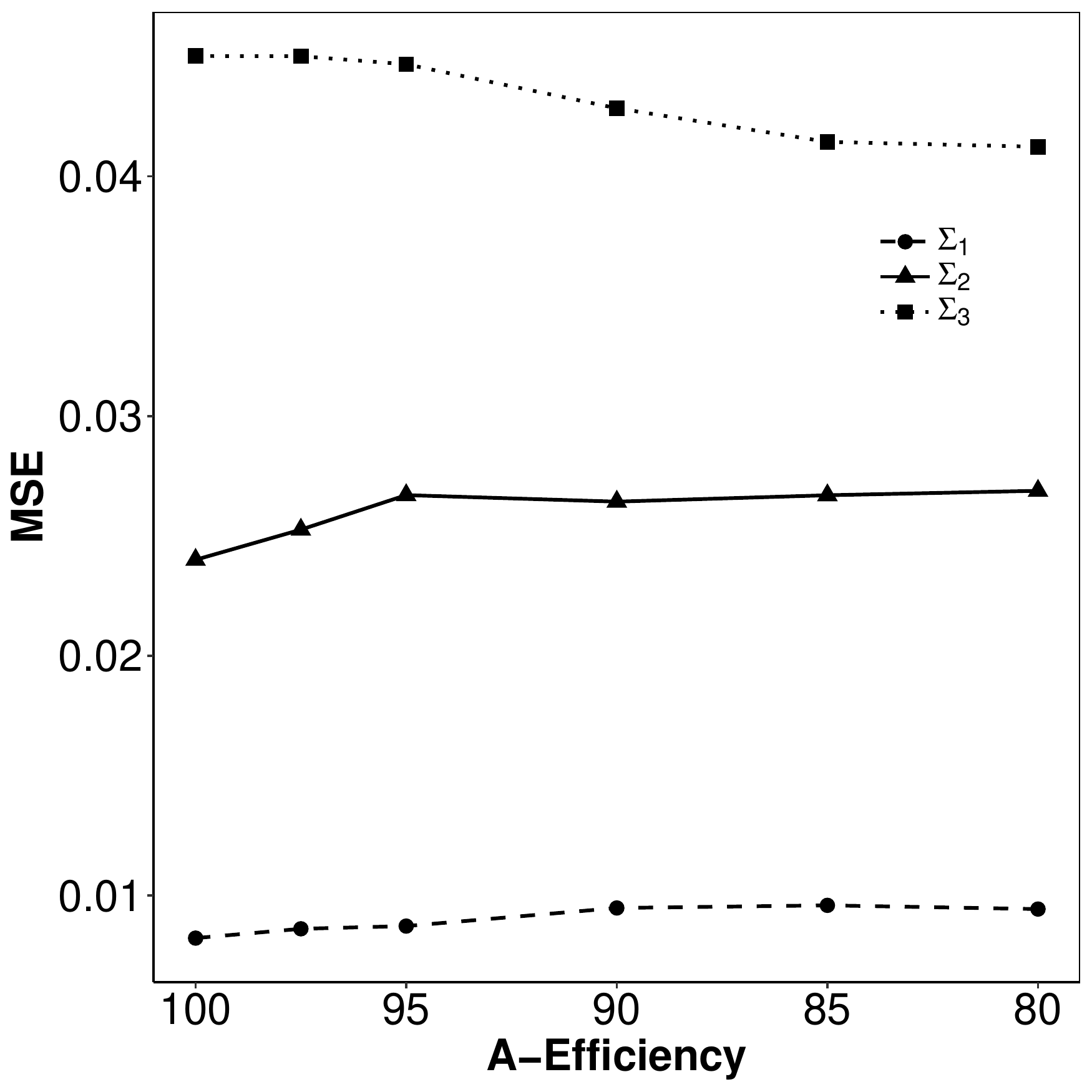}
	\end{subfigure}%
 \hfill
   \begin{subfigure}[b]{.5\textwidth}
		\centering
		\includegraphics[width=\linewidth]{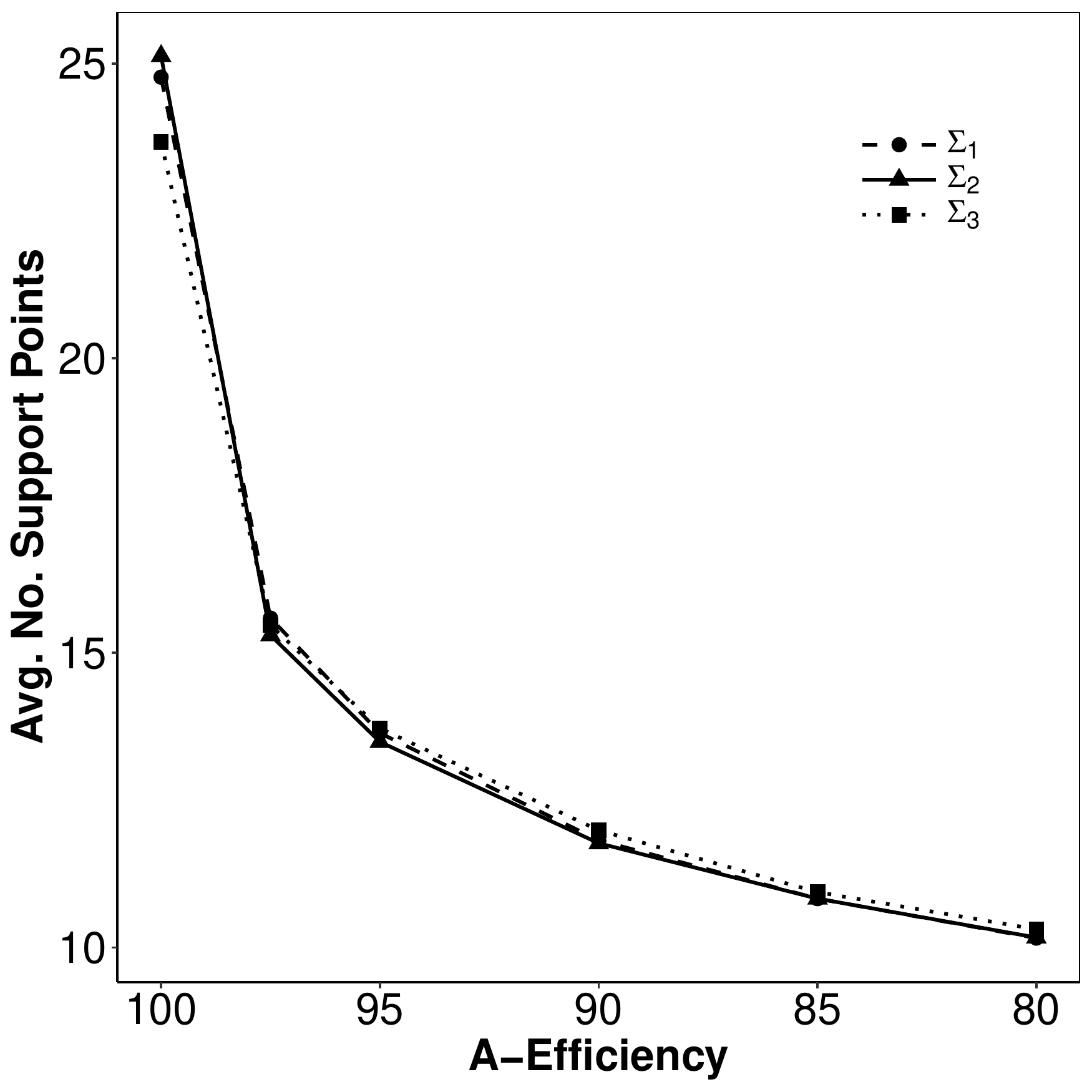}
	\end{subfigure}
 \hfill
	\centering	\caption{\it Consider using ODBSS for the logistic regression model \eqref{41LogReg_NoIntercept} {with $p = 7$ } and  $n = 10^5$ to determine a subsample of size $k = 5000$. The covariates are generated from $\varphi_p(\pmb x; \pmb 0, \pmb \Sigma_i)$  $i=1,2,3$. 
  Left:
The simulated mean squared of estimates obtained by  ODBSS with a reduced number of support points (see Algorithm \ref{algred}). 
 Right: The average number of support points for the optimal design determined in ODBSS.
 \label{fig4}}
\end{figure}

\subsubsection{Alternative design space estimation}
\label{sec414} 
The estimation of the design space proposed in Section \ref{sec31} requires additional computational costs in ODBSS. 
{An obvious alternative here is, to use the full sample $\mathcal{D}$
rather than using a clustering-based area approximation procedure as an approximation for design space, which 
is required 
 for the calculation of the optimal design in step (2) of Algorithm \ref{Algorithm1}.}
This will eliminate the area approximation step in step (1) of Algorithm \ref{Algorithm1} and hence, the time corresponding to this operation. This modification of ODBSS will be denoted by ODBSS-2 in the following discussion. A similar approach was also proposed in \cite{deltom2022}. We observe in our numerical studies that the performance of ODBSS and ODBSS-2 is very similar in terms of accuracy for parameter estimation. There are also several cases where ODBSS performs slightly better. Exemplary, we display in Figure \ref{fig3} the mean squared error of parameter estimates in the logistic regression model \eqref{41LogReg_NoIntercept} \HD{with p = 7 }using a subsample obtained from ODBSS and ODBSS-2. 

On the other hand, we compare in Table \ref{Tab_1} the run times of the two versions of ODBSS for finding a subsample of size $k = 5\% \cdot n $. 
{ We observe that for $n=10^5$
the average run-time of ODBSS-2 is smaller compared to ODBSS.   However,  
the average run-time of

ODBSS-2 sharply increases as $n$ increases, while the changes for ODBSS are rather small. These
observations can be explained by taking a closer look 
at the design space estimation in step (1) and at the optimal design determination in step (2) of Algorithm \ref{Algorithm1}.
As only the initial sample 
subsample $\mathcal{D}_{k_0}$ is used 
in ODBSS for the  design space estimation by density-based clustering, the  contribution from this step does not increase drastically even if the sample size is increased
(the size of the estimated design space ${{\pmb\chi}}_{k_0}$
does not depend on the sample size rather it depends on the grid size and is at most $L^p$, see Section \ref{sec34}).
On the other hand, ODBSS-2 uses the full sample ${\cal D} $ as an estimate of design space and the number of points in this set is $n$. The differences can be quite substantial. For the example considered above,  the number of points  of the estimated design space using density-based clustering increases 
 from from $21875$ to $32187$ ($47\%$), if the full sample size increases from $10^5$ to $4*10^5$
 ($400 \%$).  
As the time complexity for the optimal design determination in step (2) of Algorithm \ref{Algorithm1} depends on the number of points of the 
estimated design space
(in fact, this is a  cubic dependency), the computation time of ODBSS-2
increases sharply with $n$, while it only increases slightly for ODBSS.
}

\begin{figure}[H]
	\centering
	\begin{subfigure}[b]{.3\textwidth}
		\centering
		\includegraphics[width=\linewidth]{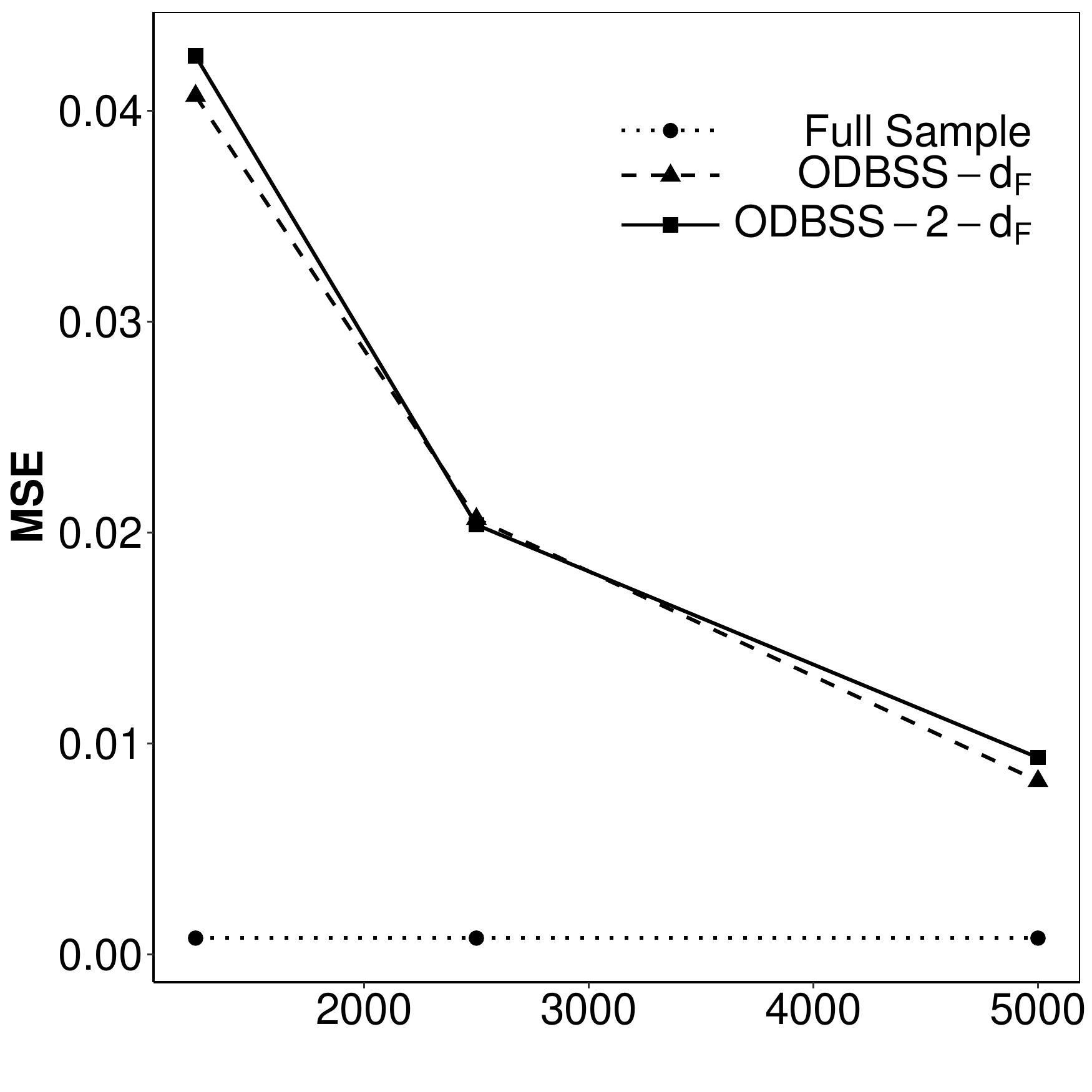}
		\caption{\scriptsize $ \varphi_p( {\pmb x;\pmb 0, \Sigma_1})$  }
		\label{fig5_1}
	\end{subfigure}%
 \hfill
   \begin{subfigure}[b]{.3\textwidth}
		\centering
		\includegraphics[width=\linewidth]{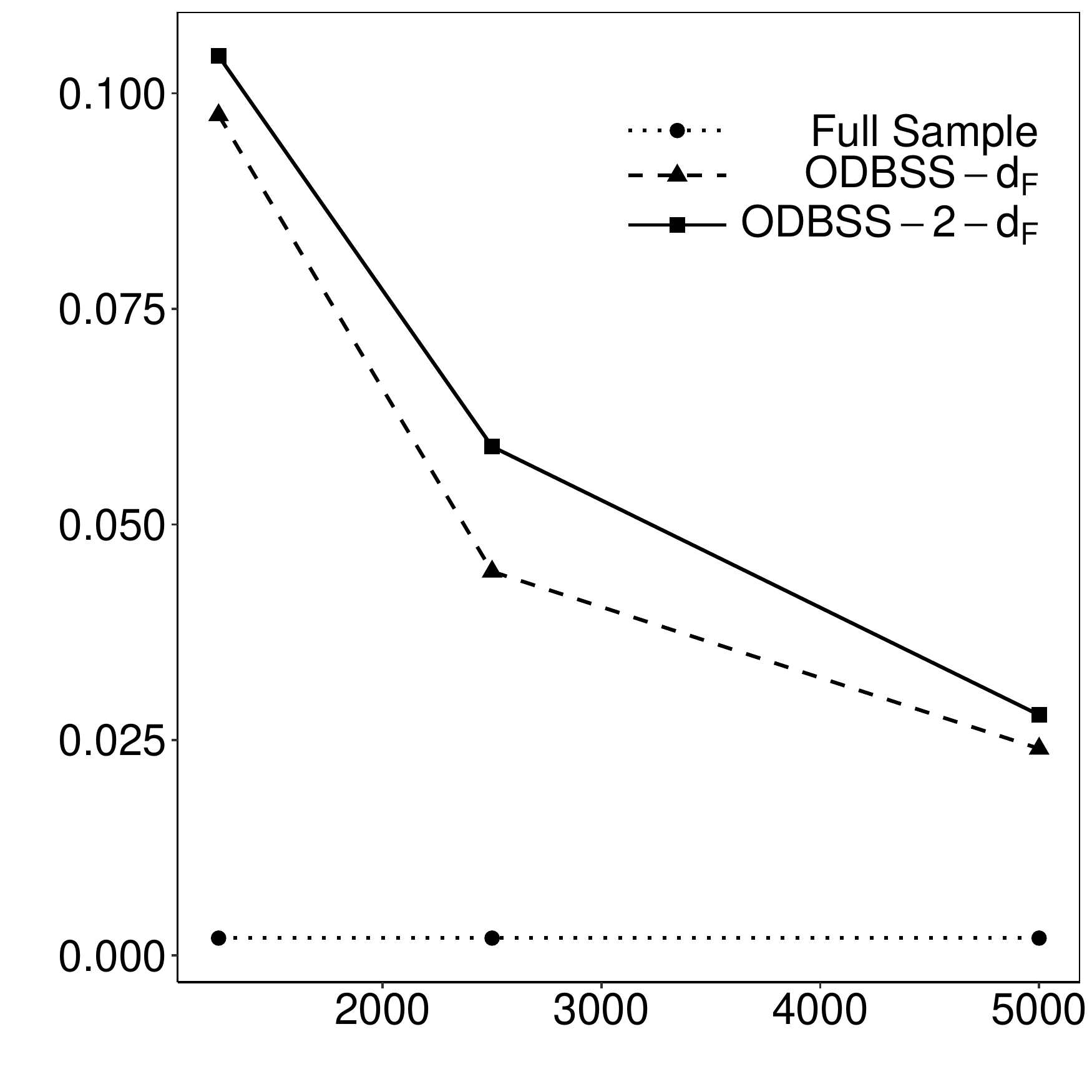}
		\caption{\scriptsize $ \varphi_p( {\pmb x;\pmb 0, \Sigma_2})$  }
		\label{fig5_2}
	\end{subfigure}
 \hfill
  \begin{subfigure}[b]{.3\textwidth}
		\centering
		\includegraphics[width=\linewidth]{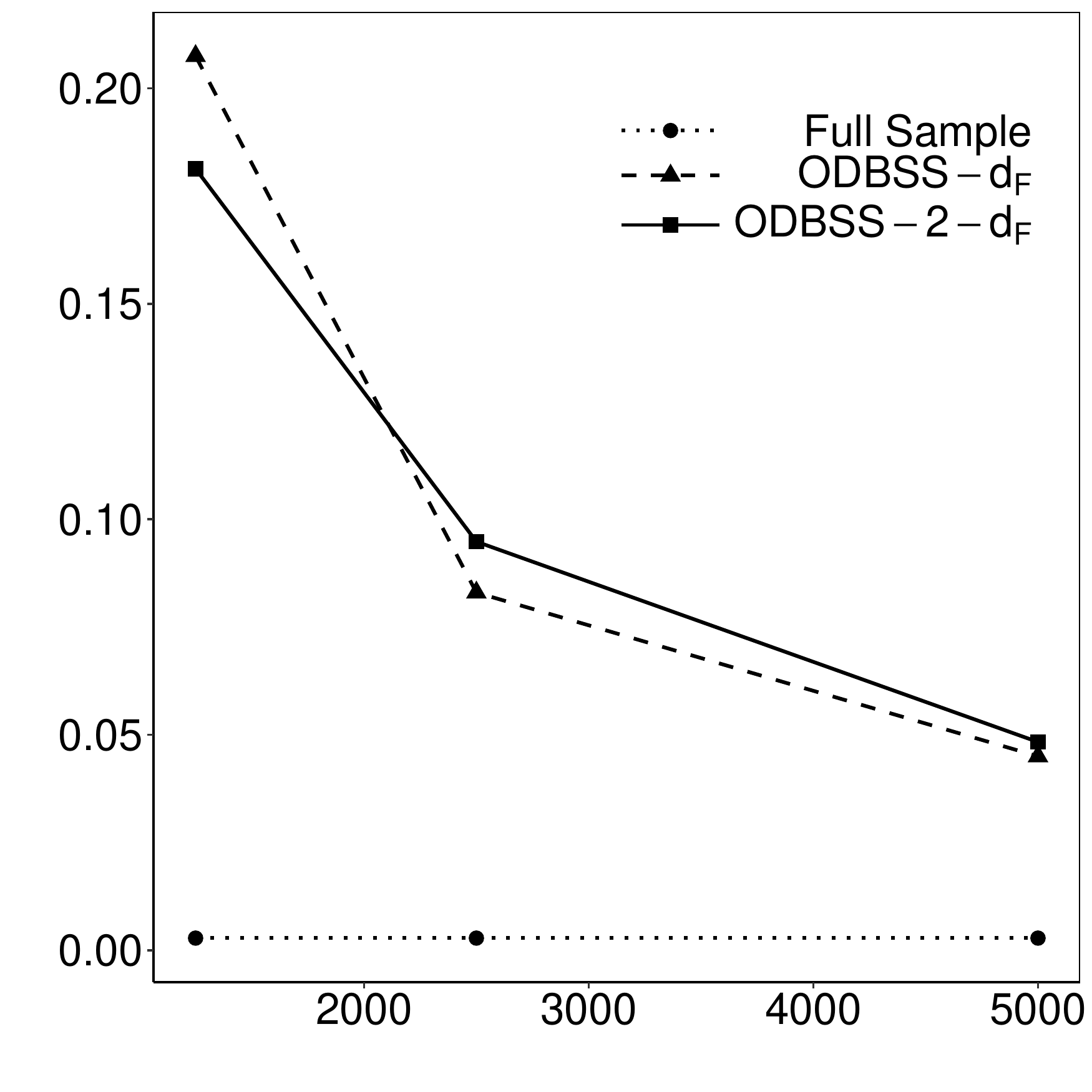}
		\caption{\scriptsize  $ \varphi_p( {\pmb x;\pmb 0, \Sigma_3})$ }
		\label{fig5_3}
	\end{subfigure}
\centering	\caption{\it 
The simulated mean squared error of estimates in the  logistic regression model \eqref{41LogReg_NoIntercept} \HD{with p = 7 }   based on   a
 subsample by 
 ODBSS and ODBSS-2 from {$n = 10^5$} observations. 
 A centered normal distribution with different covariance matrices.
 }
	\label{fig3} 
\end{figure}

\begin{table}[h]
	\begin{center}
			{
				\begin{tabular}{  |c|c |c c c c    |}
					\hline
      & & \multicolumn{4}{c|}{$n$} \\
      \hline
   &   &$10^5$& $2*10^5$&$3*10^5$&$4*10^5$ \\
      \hline
      $ \varphi_p( {\pmb x;\pmb 0, \Sigma_1})$    &ODBSS&  6.62 & 	8.30 	& 7.78 & 	8.65 \\		
   & ODBSS-2& 5.69 &	8.76 &	9.43 & 	13.07 \\
   \hline
  $ \varphi_p( {\pmb x;\pmb 0, \Sigma_2})$  & ODBSS & 8.21 	& 7.31 &	8.47 &	8.04  \\		
                                            &ODBSS-2 & 5.33 	& 8.00 	& 10.85 	& 11.52 \\
  \hline
 $ \varphi_p( {\pmb x;\pmb 0, \Sigma_3})$  & ODBSS & 5.91 &	6.26 &	6.75 &	9.01 \\		
  & ODBSS-2 & 3.87 &	7.03 &	9.30 &	11.22 \\		
  \hline
				\end{tabular}				
			}
		\end{center}
		\caption{\it Comparison of the average run-times (seconds) of two versions of Algorithm  \ref{Algorithm1} corresponding to ODBSS and ODBSS-2. }
		\label{Tab_1}
	\end{table}

\subsection{Heteroskedastic regression - rank $2$ case}
\label{sec43}

So far the literature subsampling strategies consider models where the rank of the Fisher information matrix is $1$. In this section, we demonstrate that Algorithm \ref{Algorithm1}  can also deal with more general cases without any modification. Consider the model in Example \ref{ex3}, 
where the variance is a deterministic function of the expectation, that is
\begin{align}
    g(\VectorX, \VectorBeta) &=    \pmb z^\top \pmb \beta ~, \label{hd42}\\
    \sigma^2({\pmb x, \VectorBeta}) &=   \exp{(\pmb x^\top  (\beta_1 , \dots ,\beta_p)^\top)}. \label{hd43} 
\end{align}
In this model, the maximum likelihood estimator is defined by  
\begin{align}
 \hat{\pmb \beta} =   \arg \max_{\pmb \beta}  
 \Big\{ -\dfrac{n}{2} \log(2 \pi) -\dfrac{1}{2} \sum_{i=1}^{n} \log \sigma^2(\pmb x_i, \pmb \beta)- \dfrac{1}{2} \sum_{i=1}^{n}  \dfrac{(y_i - \pmb z_i^\top \pmb \beta)^2}{\sigma^2(\pmb x_i, \pmb \beta)}  \Big\}. 
\end{align}
The Fisher-information matrix at point $\pmb x$ is given by
$$
\mathcal{I}(\pmb \beta, \pmb x) = \dfrac{1}{\sigma^2({\pmb x, \VectorBeta})}   \begin{pmatrix}
    1 \\ \pmb x 
\end{pmatrix}   ~ (1 ~~~\pmb x^\top) + \dfrac{1}{2}
\begin{pmatrix}
    0 \\ \pmb x  
\end{pmatrix} ~ (0 ~~~ \pmb x^\top) 
$$
and has  rank equal $2$ if $ \pmb  x \not =0$. We investigate the performance of ODBSS (Algorithm \ref{Algorithm1}) for finding the most informative subsamples. \HD{ For this purpose we have implemented ODBSS in MATLAB as currently the $R$-package $OptimalDesign$ does not compute optimal designs for this type of model}. In the simulation experiment, we consider the case $p=7$ and the parameter $\VectorBeta = (0.25,\ldots, 0.25)^\top $ and three setups: covariates are generated from a normal distribution with the three different covariance matrices from Section \ref{sec411}. The sample size is fixed at $n=10^5$ and the subsample sizes are varied from $k= 5000, 2500, 1250$. As expected, the ODBSS is much better than the uniform random subsampling.
\begin{figure}[H]
	\centering
	\begin{subfigure}[b]{.3\textwidth}
		\centering
		\includegraphics[width=\linewidth]{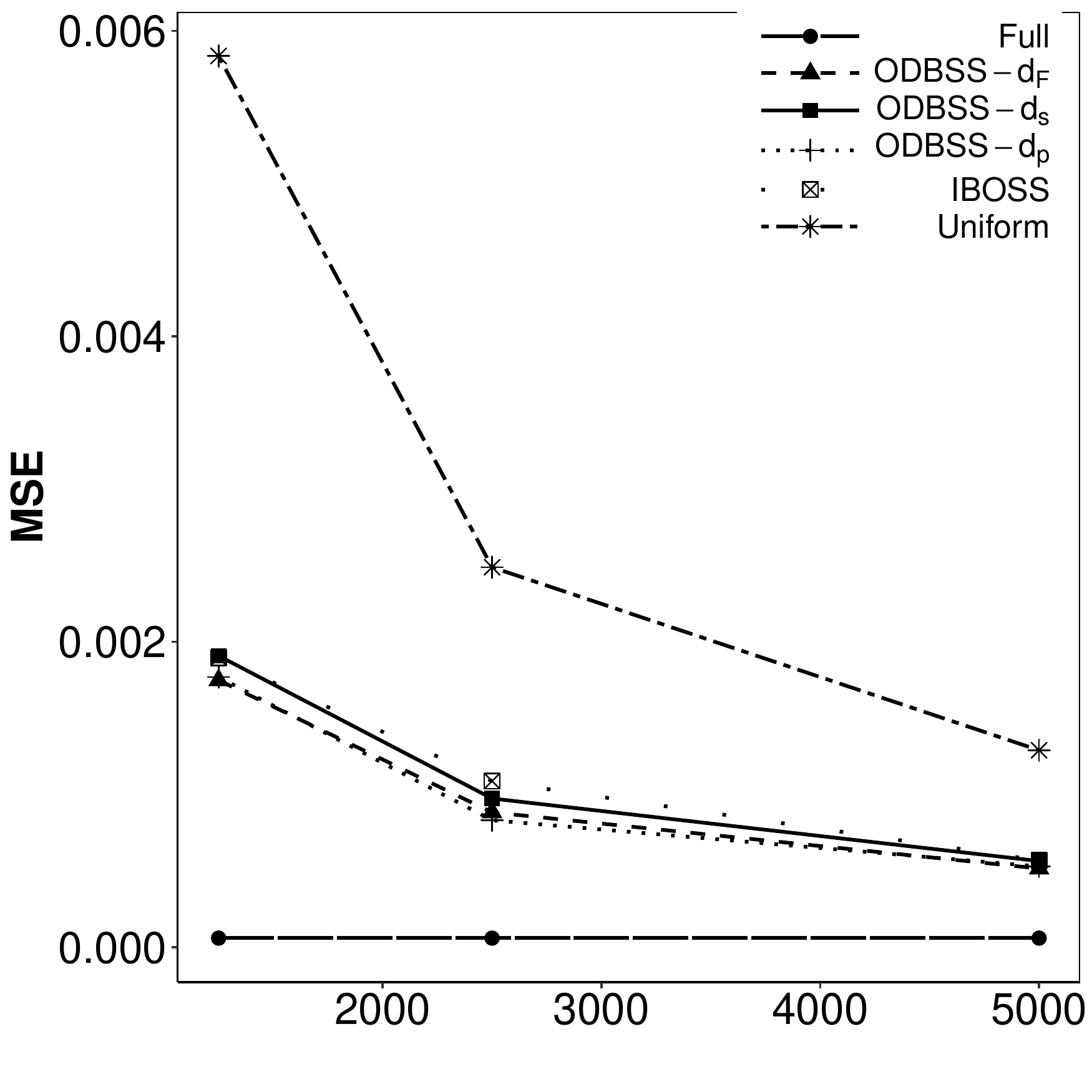}
		\caption{\scriptsize $ \varphi_p( {\pmb x;\pmb 0, \Sigma_1})$}
		\label{fig6_1}
	\end{subfigure}%
 \hfill
   \begin{subfigure}[b]{.3\textwidth}
		\centering
		\includegraphics[width=\linewidth]{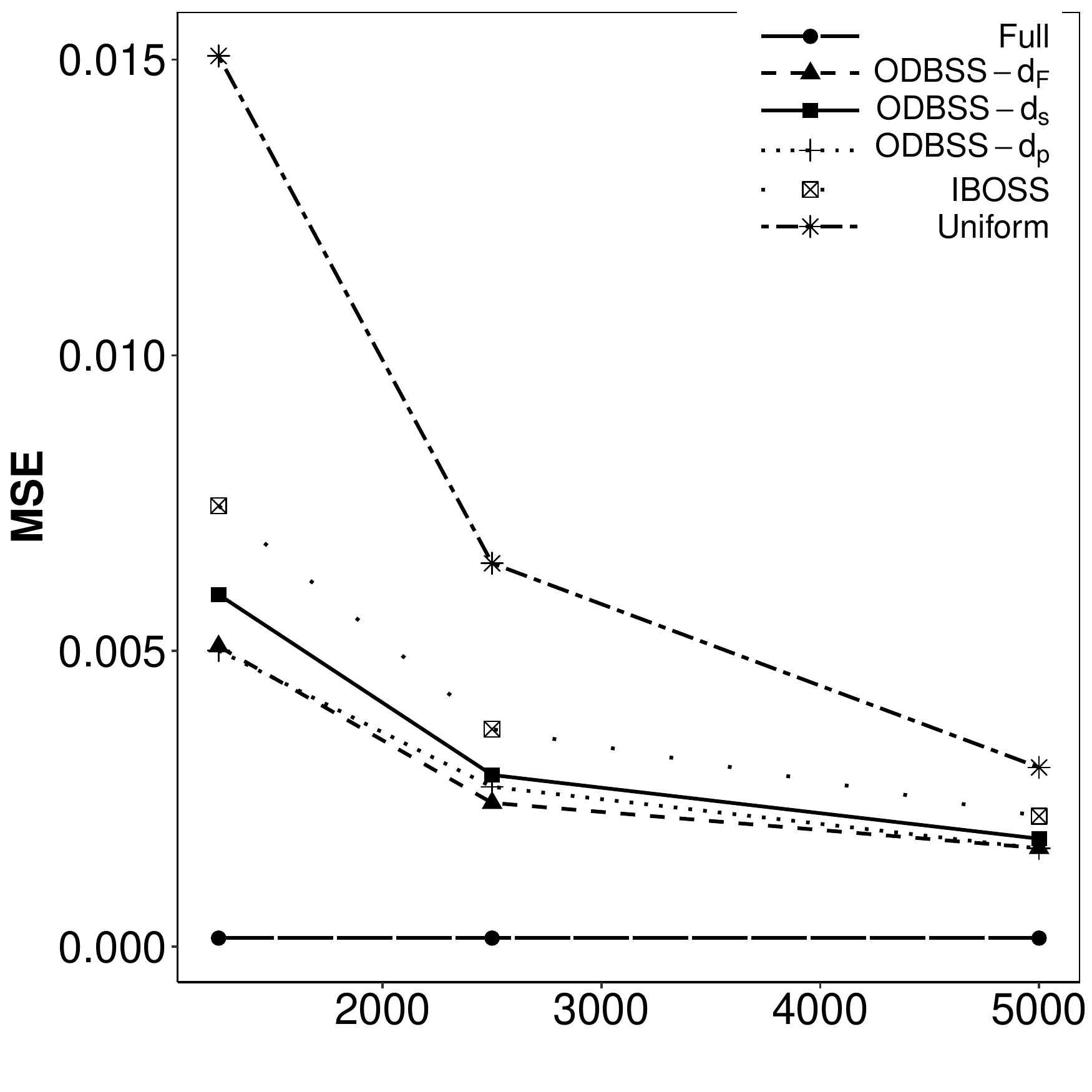}
		\caption{\scriptsize $ \varphi_p( {\pmb x;\pmb 0, \Sigma_2})$ }
		\label{fig6_2}
	\end{subfigure}
 \hfill 
  \begin{subfigure}[b]{.3\textwidth}
		\centering
		\includegraphics[width=\linewidth]{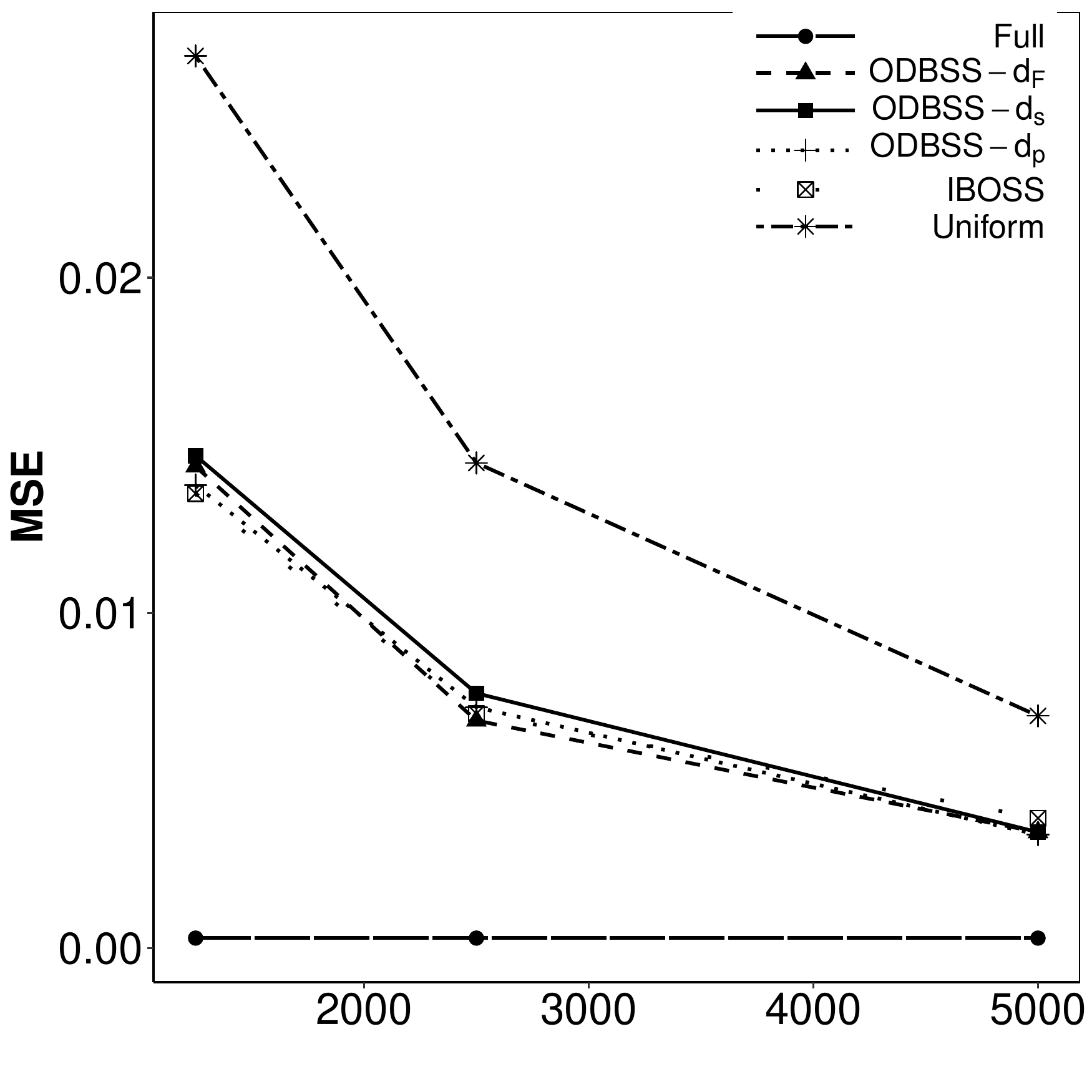}
		\caption{\scriptsize  $ \varphi_p( {\pmb x;\pmb 0, \Sigma_3})$}
		\label{fig6_3}
	\end{subfigure}
	\centering	\caption{ {\it  Simulated mean squared error of the parameter estimate in the heteroskedastic regression model given by equation \eqref{hd42} and \eqref{hd43} using ODBSS subsampling with the metric $d_F(\cdot)$, $d_s(\cdot)$, and $d_p(\cdot)$ defined in \eqref{d1}, \eqref{d2} and \eqref{d3}, respectively. A comparison of ODBSS (all three distances) with an IBOSS and uniform random  subsample is done. The covariates are normally distributed with different covariance matrices (as in section \ref{sec41})}.
 }
	\label{fig6} 
\end{figure}

From Figure \ref{fig6},  we observe that all three matrix distances yield a very comparable performance of ODBSS. Using ODBSS with the Procrustes and Frobenius metric gives slightly better results than with square root distance. For the sake of comparison we have also included the results of IBOSS  as proposed by  \cite{wang2019information}.    
We observe that in most cases ODBSS has a better performance than IBOSS   as well. Its superiority is more pronounced on the  subspaces defined by covariance matrices  $\Sigma_1$ and $\Sigma_2$.

\section{Concluding remarks and future research }

In this paper, we develop a new deterministic subsampling strategy (ODBSS), which is applicable for finding maximum likelihood estimates for a large class of models. Subsampling is carried out in three steps: i) design space estimation from a small initial sample, ii)  the optimal design determination for the estimated design space, and iii) subsample allocation.
Simulation experiments show an improved performance of the new method over competing algorithms. \HD{We also observe that ODBSS is very robust to change in covariate distributions and is very consistent with it performance in each case. Its performance is much better compared to existing algorithms in the case of heavy-tailed distributions and when the design space is complex. } The 
 structure of the algorithm allows modification to an online subsampling setting easily. 
 For this setup steps i) and ii), which determine the optimal design could be used adaptively (in batches) to determine subsamples that should be retained (depending upon proximity to the optimal design). These extensions will be investigated in future research. {A further 
interesting question is statistical guarantees for the estimates based on the sample of the proposed algorithm.   
}

\bigskip

\noindent 
\textbf{{Software} used for computation}\\
We have implemented the ODBSS algorithm for the simulation studies in  $R$-software({\it R-4.2.0}) and {\it MATLAB-R2022} on a MacBook Air.

\bigskip

\begin{appendix}
    \appendix
 \setcounter{section}{0}
 \setcounter{figure}{0}
 \setcounter{table}{0}
 \setcounter{illustration}{0}

  \renewcommand{\theillustration}{S\arabic{illustration}}
 \renewcommand{\thesection}{S\arabic{section}}
 \renewcommand{\thefigure}{S\arabic{figure}}
 \renewcommand{\thetable}{S\arabic{table}}

\section*{Supplementary} \label{Main Appendix}

\section{Effect of initial subsample size on parameter estimation} \label{Supplementary Section 1}

The ODBSS algorithm is a two-stage algorithm. In the first stage, a random subsample ($\mathcal{D}_{k_0}$) of size $k_0$ is drawn, where $k_0$ is much smaller than the size $k$ of the total subsample. This initial subsample is used for obtaining the initial parameter estimate $\hat{\pmb \beta}_{\mathcal{D}_{k_0}}$ and to estimate design space $\pmb \chi_{k_0}$. In this section, we investigate the impact of the choice of $k_0$ on ODBSS. In Figure \ref{FigSupplementary_1}, we then display the MSE of ODBSS for various values of $k_0$ in the logistic regression model in equation (4.2). We observe the following:
\begin{itemize}
\item  There is a gradual increase in MSE as the $k_0$ increases. This is as expected; as higher values of $k_0$ mean the majority of the subsample is constituted of randomly chosen points.
\item For $0.1k \leq k_0 \leq 0.5k$, the difference in MSE are relatively small. This also indicates that a rough estimate of $\mathcal{D}_{k_0}$ and $\hat{\beta}_{\mathcal{D}_{k_0}}$ is good enough for ODBSS to work well. 
\item For the $t$-distribution, even larger values of $k_0$ do not yield to a significant increase in the $MSE$. Even for $k_0= 0.9 k$ the MSE of ODBSS is much smaller than that of uniform subsampling. This shows that taking even a small number of informative points leads to good estimation results. 
\end{itemize}
To summarize, values between $0.2k$ and $0.5k$ are a good choice for the size $k_0$ of the initial subsample. We also recommend keeping more informative points than random subsamples in the final subsample and therefore propose to use $k_0= 0.2k$.

\begin{figure}[H]
	\centering
	\begin{subfigure}[b]{.3\textwidth}
		\centering
		\includegraphics[width=\linewidth]{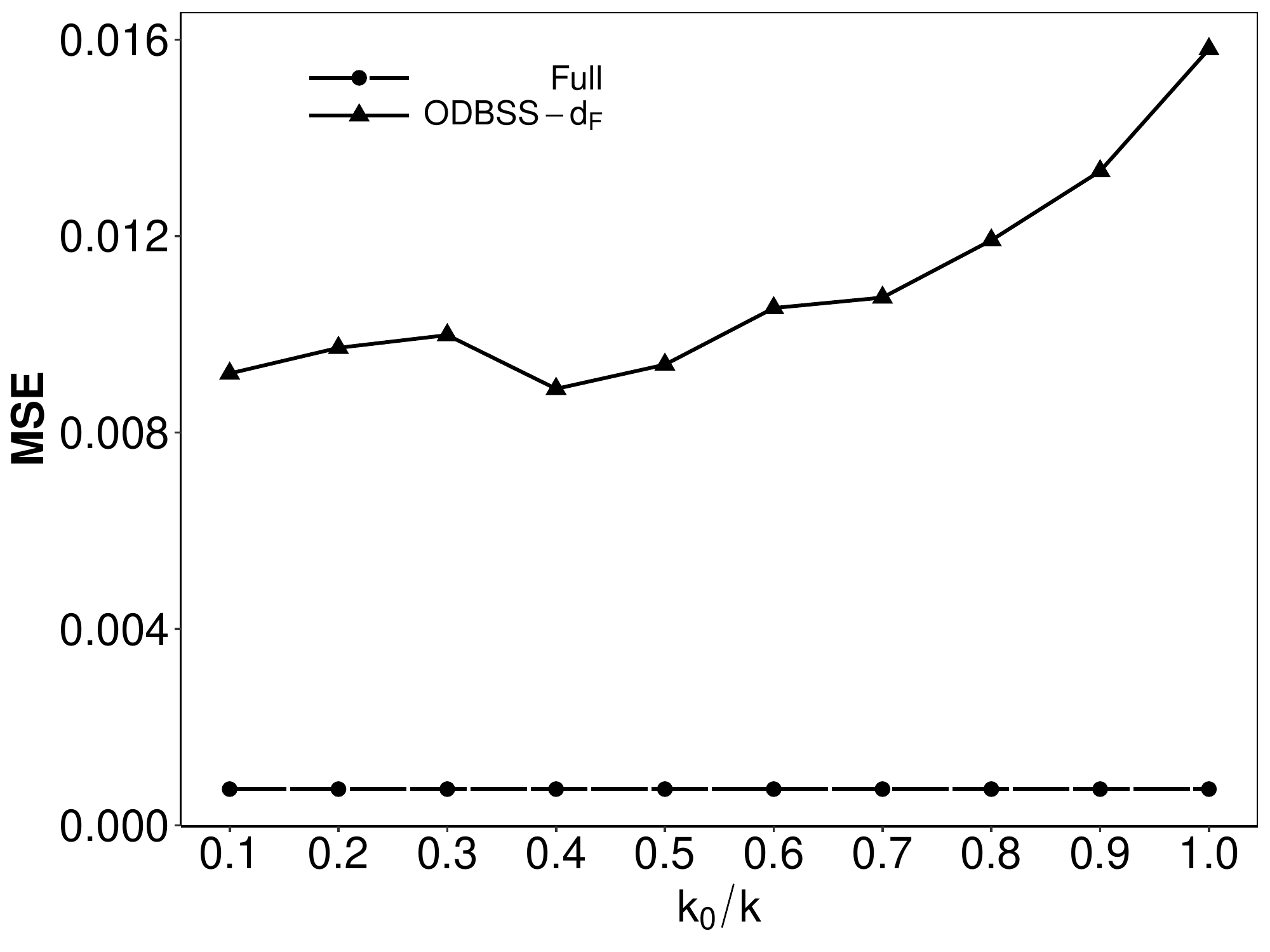}
		\caption{\scriptsize $ \varphi_{p}( \pmb x;  \pmb 0 , \pmb \Sigma_1) $ }
		\end{subfigure}%
 \hfill
   \begin{subfigure}[b]{.3\textwidth}
		\centering
		\includegraphics[width=\linewidth]{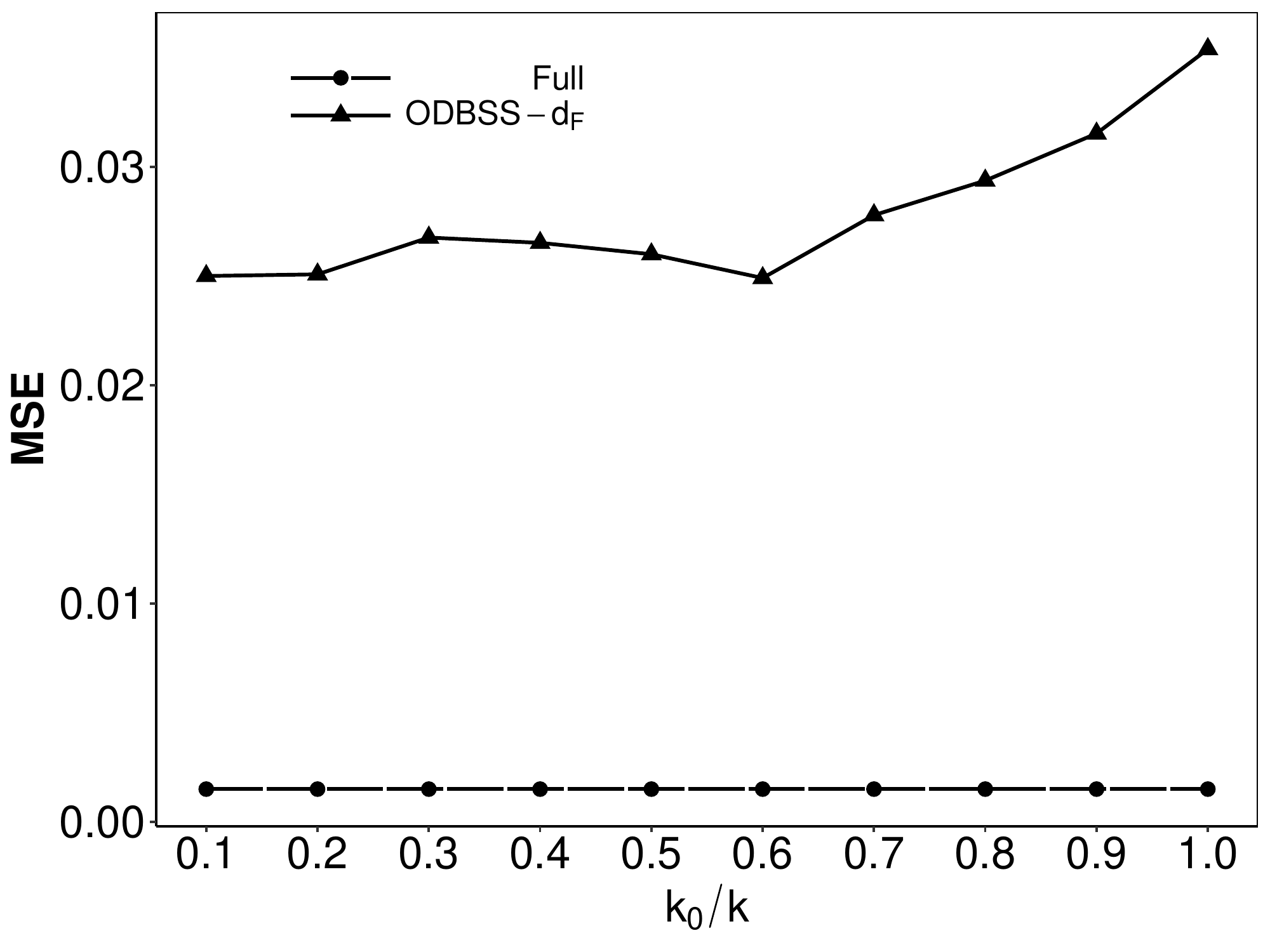}
		\caption{\scriptsize $ \varphi_{p}( \pmb x;  \pmb 0 , \pmb \Sigma_2) $ }
		\end{subfigure}
 \hfill
  \begin{subfigure}[b]{.3\textwidth}
		\centering
		\includegraphics[width=\linewidth]{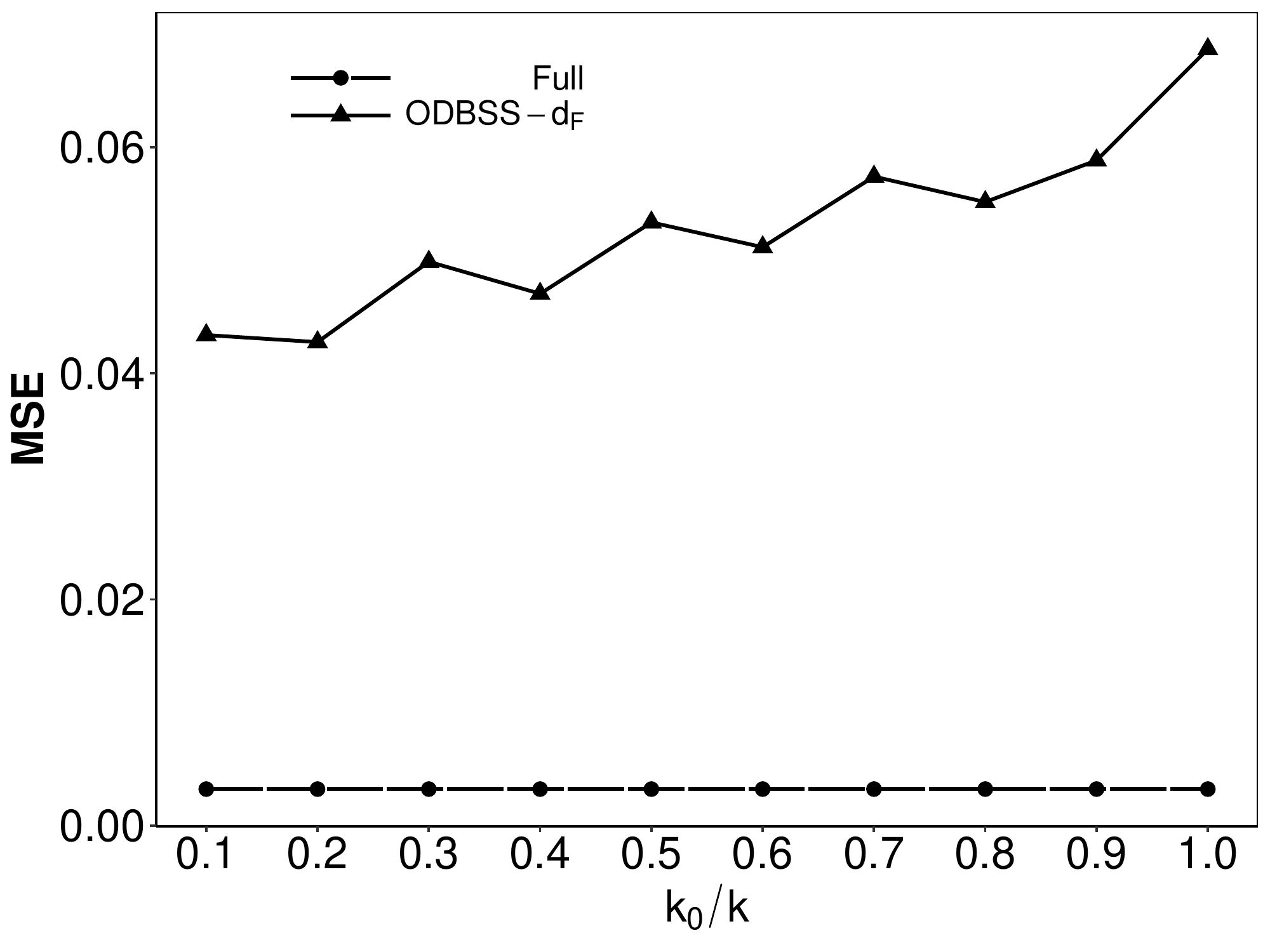}
		\caption{\scriptsize  $ \varphi_{p}( \pmb x;  \pmb 0 , \pmb \Sigma_3) $}
		\end{subfigure}
 \vspace{1cm}
 
 \begin{subfigure}[b]{.3\textwidth}
		\centering
		\includegraphics[width=\linewidth]{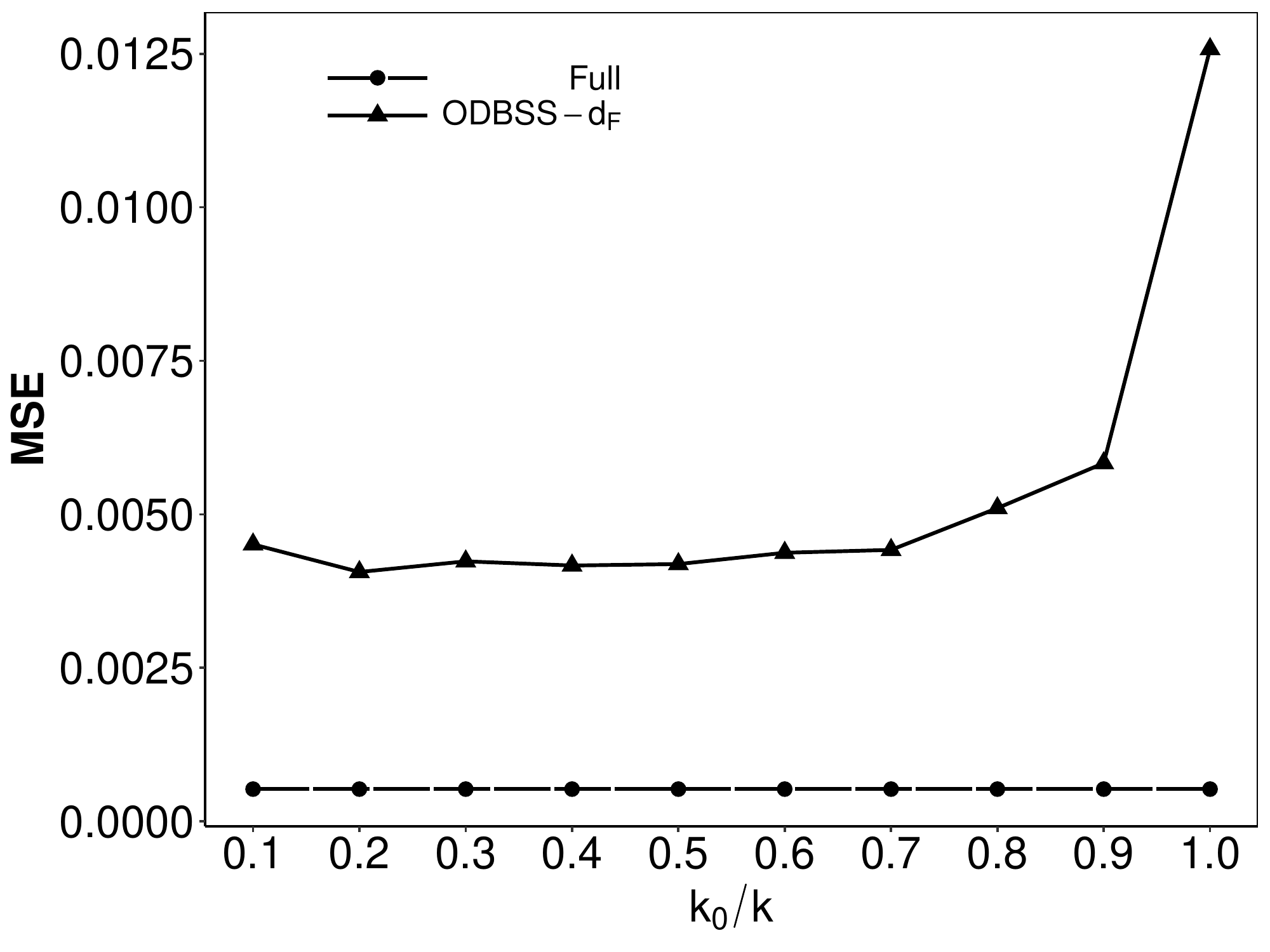}
		\caption{\scriptsize $\mathcal{T}_{p}(\pmb x;\pmb 0, \pmb \Sigma_1,3)$ }
		\end{subfigure}%
 \hfill
   \begin{subfigure}[b]{.3\textwidth}
		\centering
		\includegraphics[width=\linewidth]{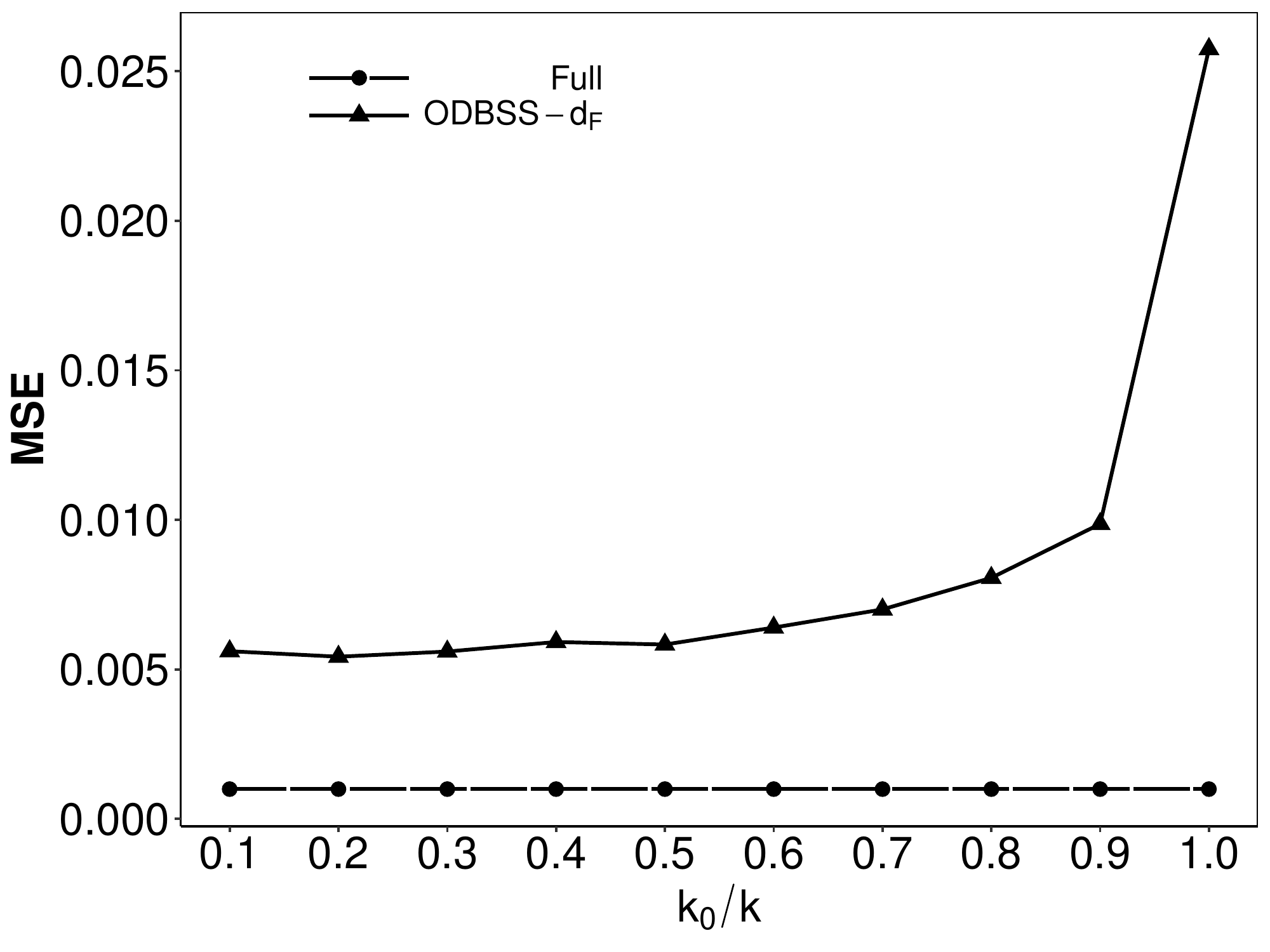}
		\caption{\scriptsize $\mathcal{T}_{p}(\pmb x;\pmb 0, \pmb \Sigma_2,3)$}
		\end{subfigure}
 \hfill
  \begin{subfigure}[b]{.3\textwidth}
		\centering
		\includegraphics[width=\linewidth]{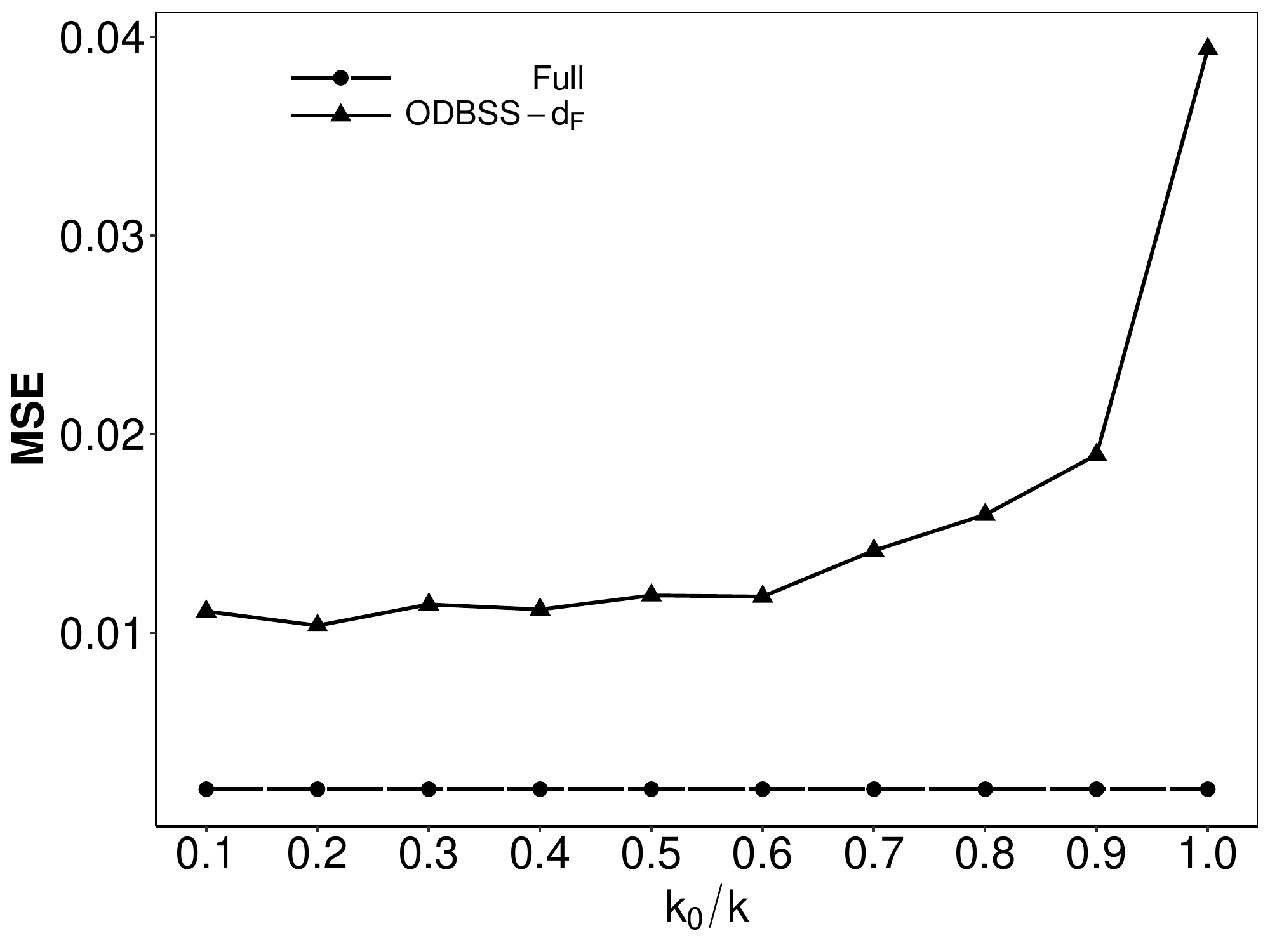}
		\caption{\scriptsize$\mathcal{T}_{p}(\pmb x;\pmb 0, \pmb \Sigma_3,3)$}
		\end{subfigure}
	\centering	\caption{ \it \footnotesize MSE versus $k_0/k$ in the logistic regression model in equation (4.2) of the main paper when p = 7,  $n = 10^5$, for the subsample of size $k = 5000$ obtained from ODBSS for when covariates have centered normal and $t$-distribution with $3$ degrees of freedom with different covariances.  }
	\label{FigSupplementary_1} 
\end{figure}

\section{Parameter Tuning for DBSCAN}

An important part of ODBSS is to estimate the design space $\pmb \chi_{k_0}$ applying DBSCAN to the initial sample $\mathcal{D}_{k_0}$. The algorithm requires two tuning parameters $\epsilon$ and $m_p$. Section \ref{Supplementary Section 2.1} and \ref{Supplementary Section 2.2} investigate the sensitivity of ODBSS with respect to their choice and develop a recommendation. 

\subsection{Tuning $\epsilon$} \label{Supplementary Section 2.1}

{We first investigate the sensitivity of ODBSS(with Frobenius norm) with respect to the choice of the tuning parameter $\epsilon$. For this purpose consider the logistic regression discussed in Section 4 of the main paper with 
\begin{equation}
    \epsilon_{opt} = \min \Bigg\{0.1 (p-1) \big(\max_{i,j} \pmb {(X_{\mathcal{D}_{k_0}})}_{ij}- \min_{i,j} \pmb {(X_{\mathcal{D}_{k_0}})}_{ij}\big), \max_{\pmb x \in  \mathcal{D}_{k_0}} dist_4(\pmb x) \Bigg\} \label{rule1}
\end{equation} in DBSCAN, where $\pmb X_{\mathcal{D}_{k_0}}$  is 
the design matrix corresponding to the initial subsample.
In Figure \ref{Figure Supplementary 2-1}, for $p=7$ 
we compare the performance of the ODBSS (in particular the DBSCAN component) by varying values of $\epsilon$ such that $ \frac{\epsilon}{\epsilon_{opt}} = 0.75, 0.9, 1, 1.25, 1.5,2 $. We observe that ODBSS is not very sensitive to the variations in parameter in the range $\epsilon_{opt}$ that is, $(0.75 \;  \epsilon_{opt},  1.5 \; \epsilon_{opt})$, especially when subsample sizes are not too small ($k=2500$ and $5000$). \\
However, we found in our simulation studies that for high dimensional settings (p=20), the choice $\epsilon < \epsilon_{opt}$ leads to various problems. In this case, DBSCAN partitions the initial subsample into many clusters because the set $\mathcal{D}_{k_0}$ is a very sparse set in $\mathbb{R}^p$ if $p$ is large. The creation of these smaller (false) clusters leads to problems in the discretization of the design space as the Metropolis-Hastings (MH) algorithm becomes extremely slow. 
On similar lines, if we set $\epsilon > 2\; \epsilon_{opt}$, for low and high dimensional problems, the DBSCAN cannot distinguish between two separate clusters (in many cases). \\
Summarizing, based on numerical studies we recommend setting $\epsilon = \epsilon_{opt}$ for the ODBSS algorithm.
   }

\begin{figure}[H]
	\centering
	\begin{subfigure}[b]{.3\textwidth}
		\centering
		\includegraphics[width=\linewidth]{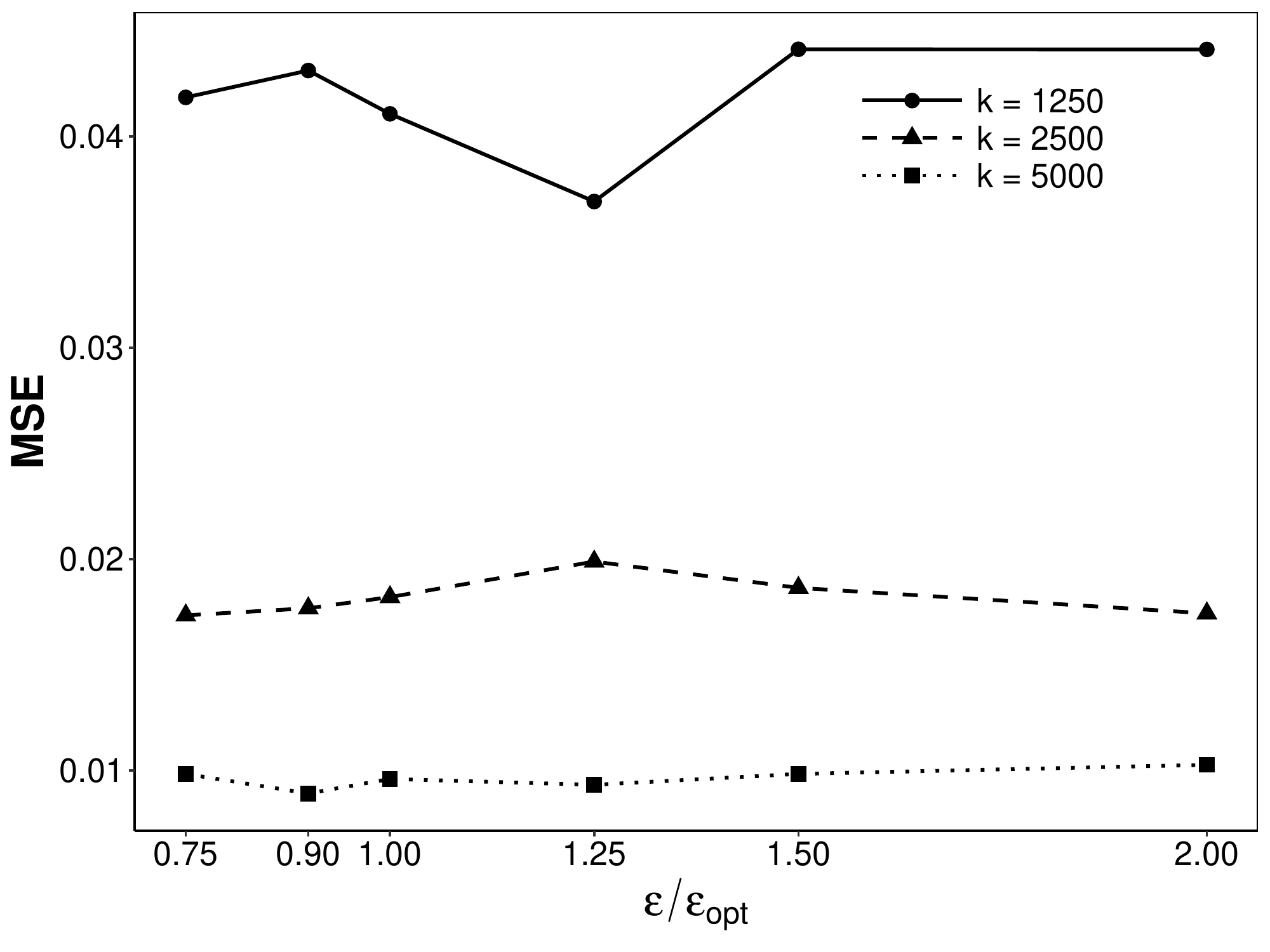}
		\caption{\scriptsize $ \varphi_{p}( \pmb x;  \pmb 0 , \pmb \Sigma_1) $ }
		\end{subfigure}%
 \hfill
   \begin{subfigure}[b]{.3\textwidth}
		\centering
		\includegraphics[width=\linewidth]{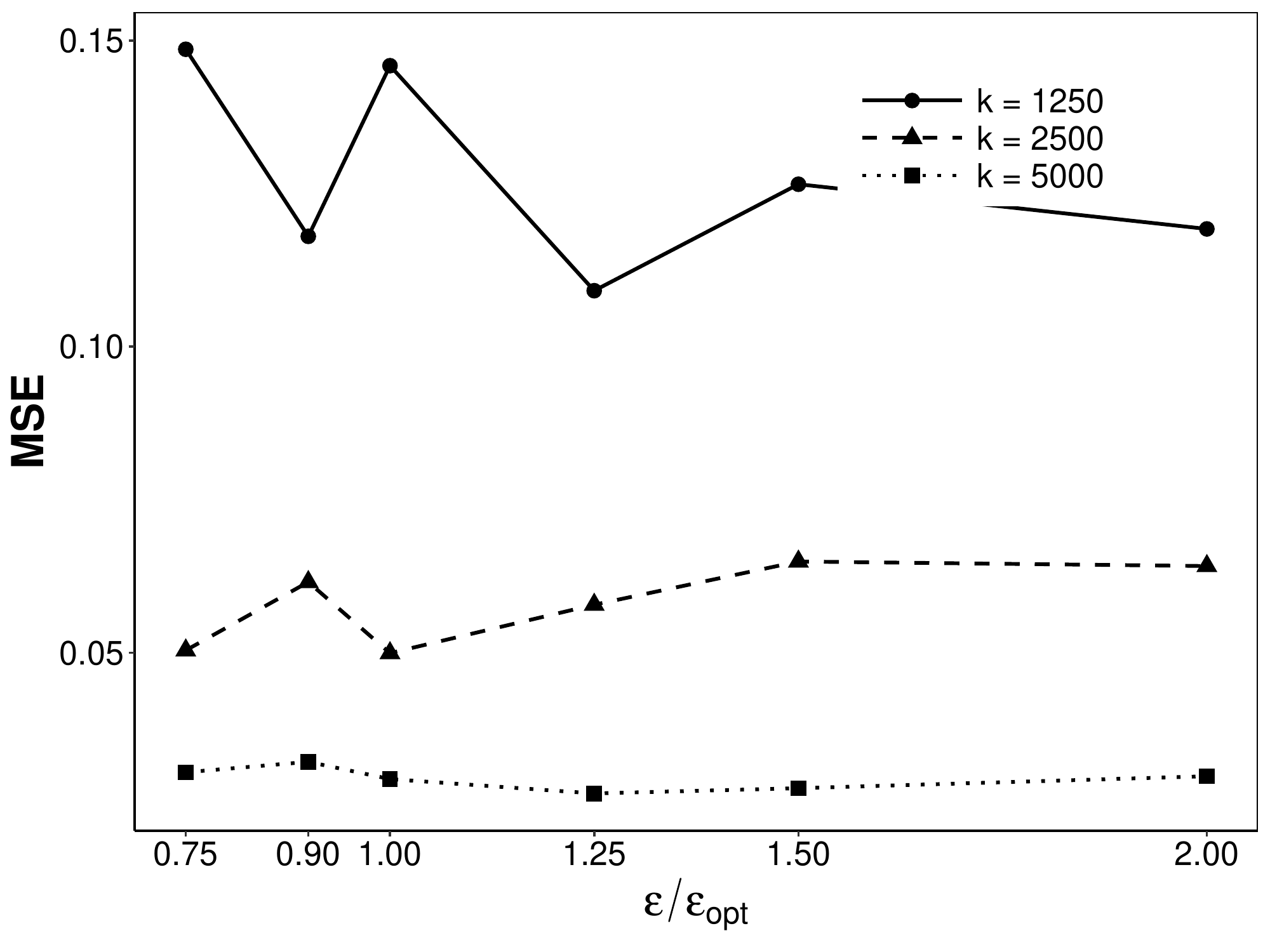}
		\caption{\scriptsize $ \varphi_{p}( \pmb x;  \pmb 0 , \pmb \Sigma_2) $}
	\end{subfigure}
 \hfill
  \begin{subfigure}[b]{.3\textwidth}
		\centering
		\includegraphics[width=\linewidth]{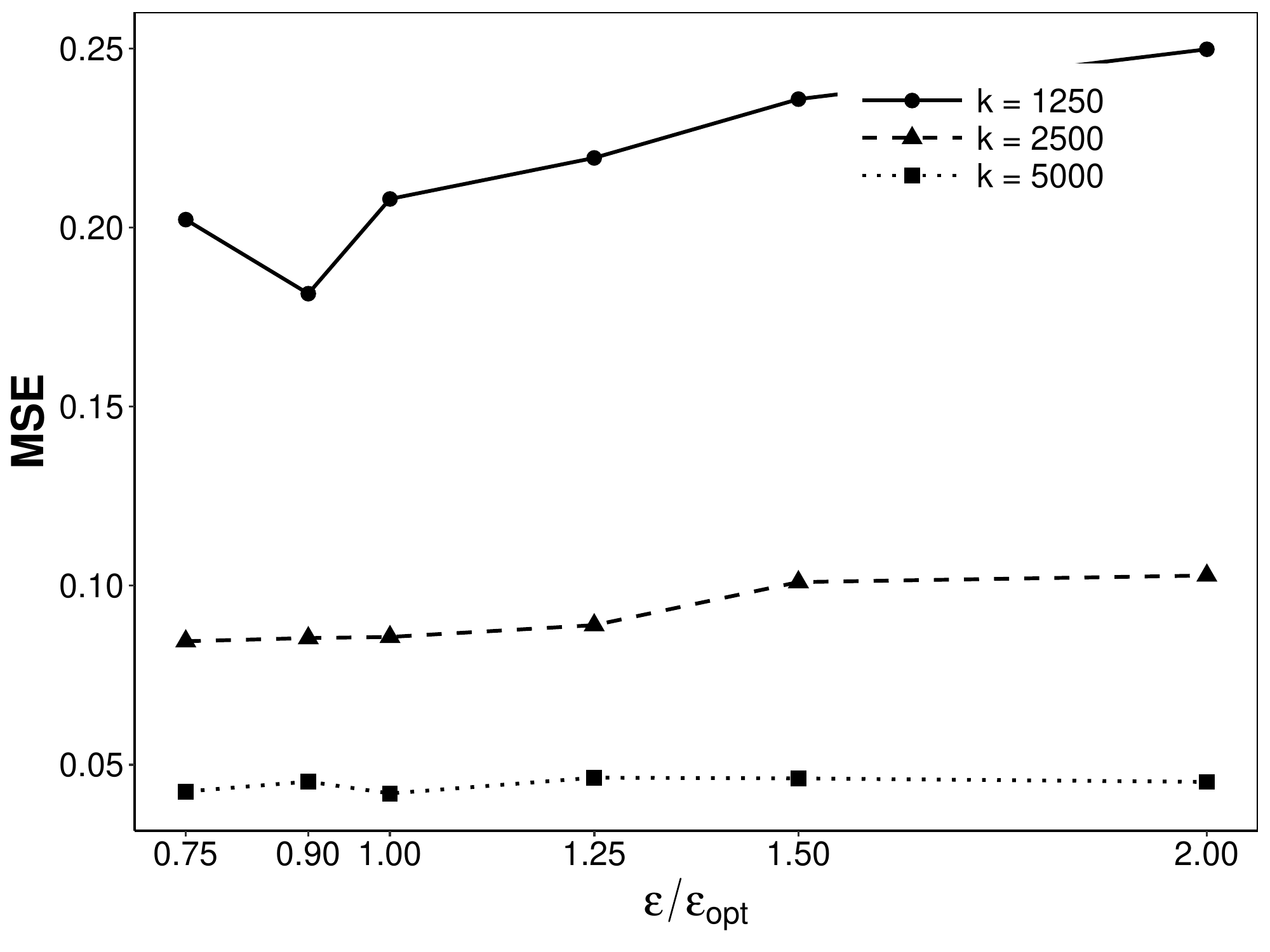}
		\caption{\scriptsize  $ \varphi_{p}( \pmb x;  \pmb 0 , \pmb \Sigma_3) $}
	\end{subfigure}
 \vspace{1cm}
 
 \begin{subfigure}[b]{.3\textwidth}
		\centering
		\includegraphics[width=\linewidth]{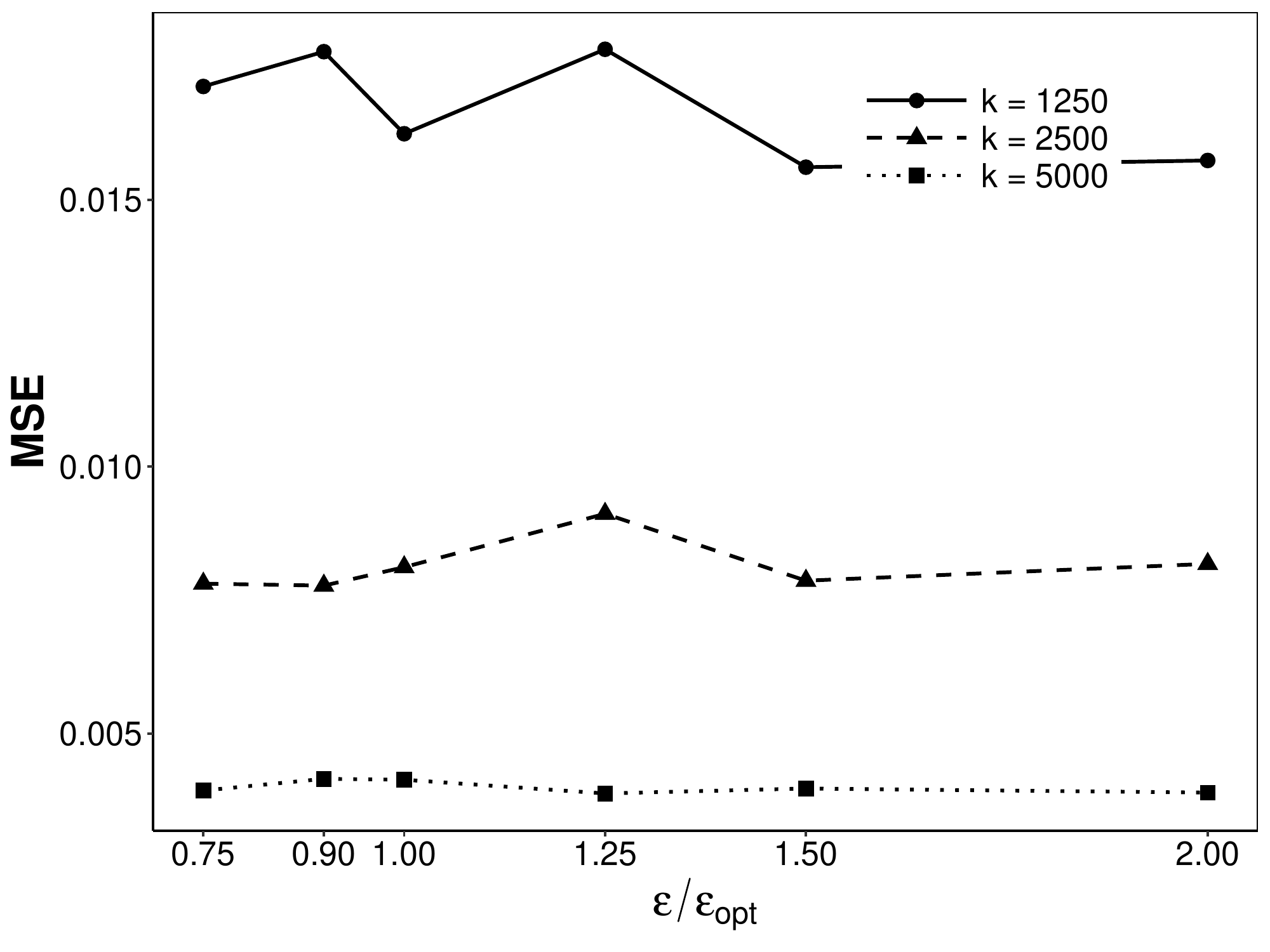}
		\caption{\scriptsize $\mathcal{T}_{p}(\pmb x;\pmb 0, \pmb \Sigma_1,3)$ }
		\end{subfigure}%
 \hfill
   \begin{subfigure}[b]{.3\textwidth}
		\centering
		\includegraphics[width=\linewidth]{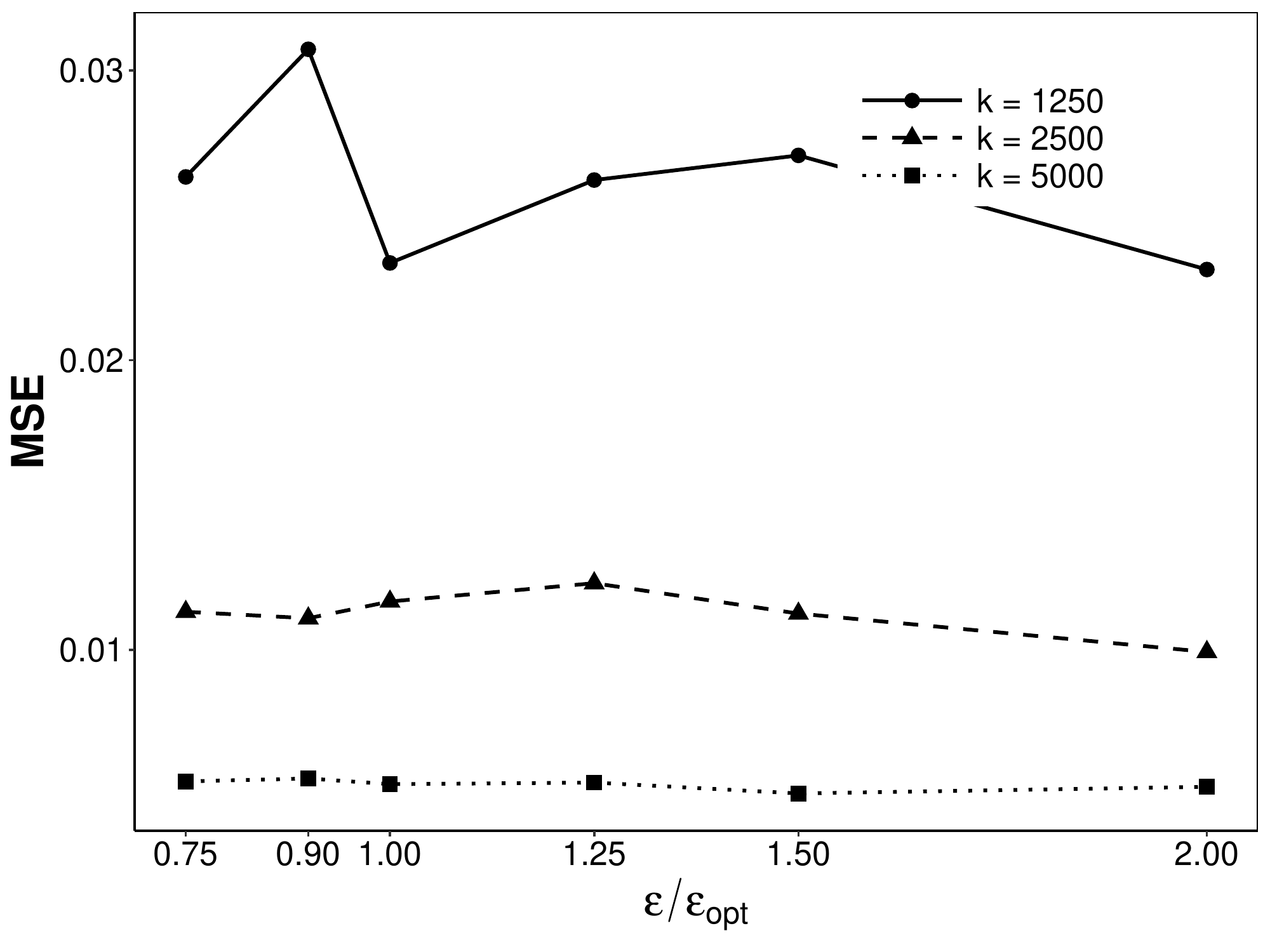}
		\caption{\scriptsize $\mathcal{T}_{p}(\pmb x;\pmb 0, \pmb \Sigma_2,3)$}
		\end{subfigure}
 \hfill
  \begin{subfigure}[b]{.3\textwidth}
		\centering
		\includegraphics[width=\linewidth]{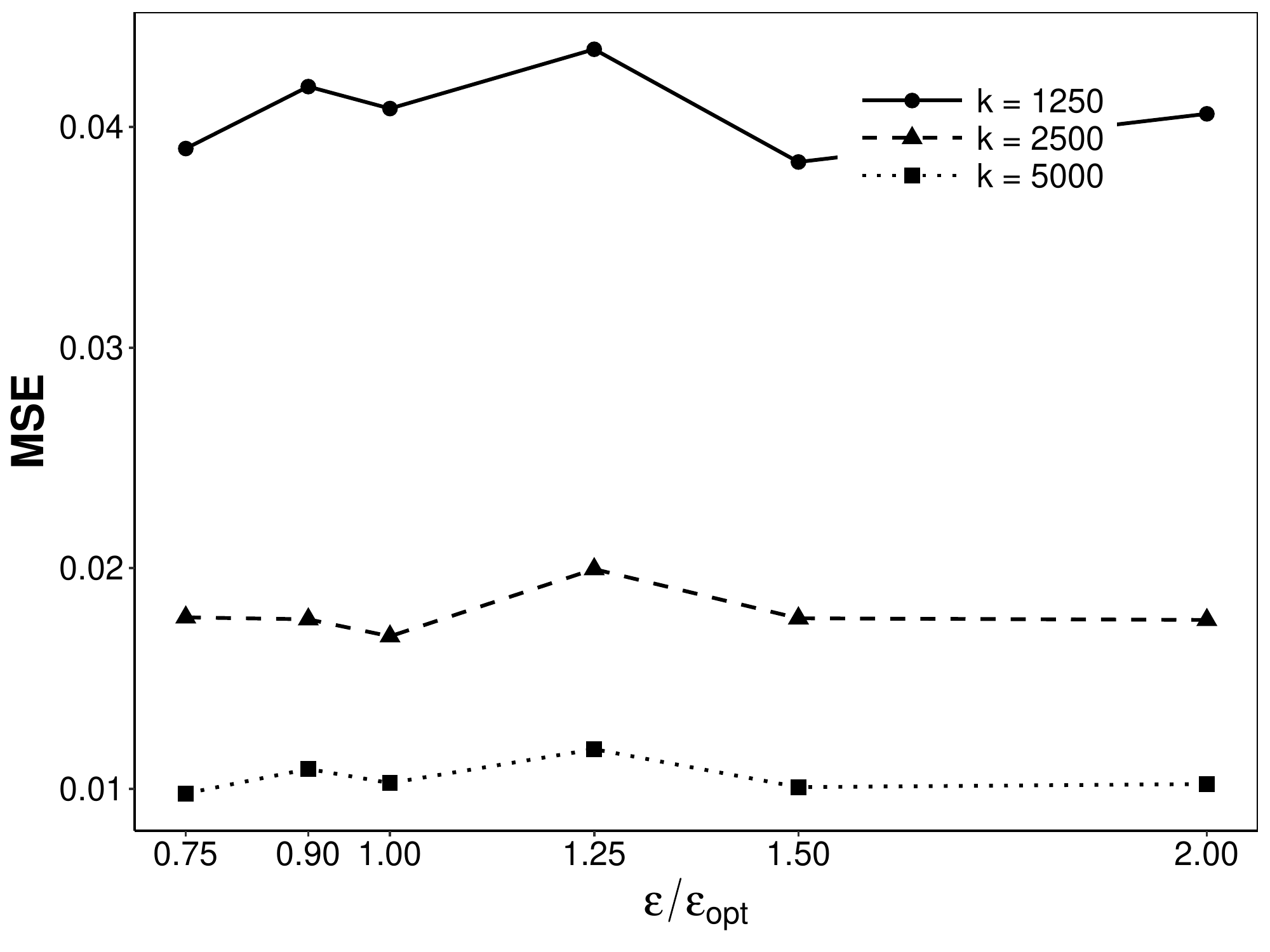}
		\caption{\scriptsize $\mathcal{T}_{p}(\pmb x;\pmb 0, \pmb \Sigma_3,3)$}
		\end{subfigure}
	\centering	\caption{ \it   MSE versus $\dfrac{\epsilon}{\epsilon_{opt}}$ for the logistic regression model in equation (4.2) {with p = 7 }, $n = 10^5$ for subsample sizes $k=5000, 2500, 1250$ obtained by ODBSS (with Frobenius norm) for centered normal and $t$-distribution with $3$ degrees of freedom with different covariance matrices.    
 }
	\label{Figure Supplementary 2-1} 
\end{figure}

\subsection{Tuning $m_p$} \label{Supplementary Section 2.2}
As recommended in Section \ref{Supplementary Section 2.1} we fix $\epsilon$ by the rule \ref{rule1}. We use six different values of the tuning parameter $m_p=  1, 2, 4,5,6,10$ for running the ODBSS algorithm. Figure \ref{Figure Supplementary 3-1} 
show the sensitivity of the ODBSS (with Frobenius norm) with respect to the choice of the tuning parameter $m_p$. 
\begin{figure}[H]
	\centering
	\begin{subfigure}[b]{.3\textwidth}
		\centering
		\includegraphics[width=\linewidth]{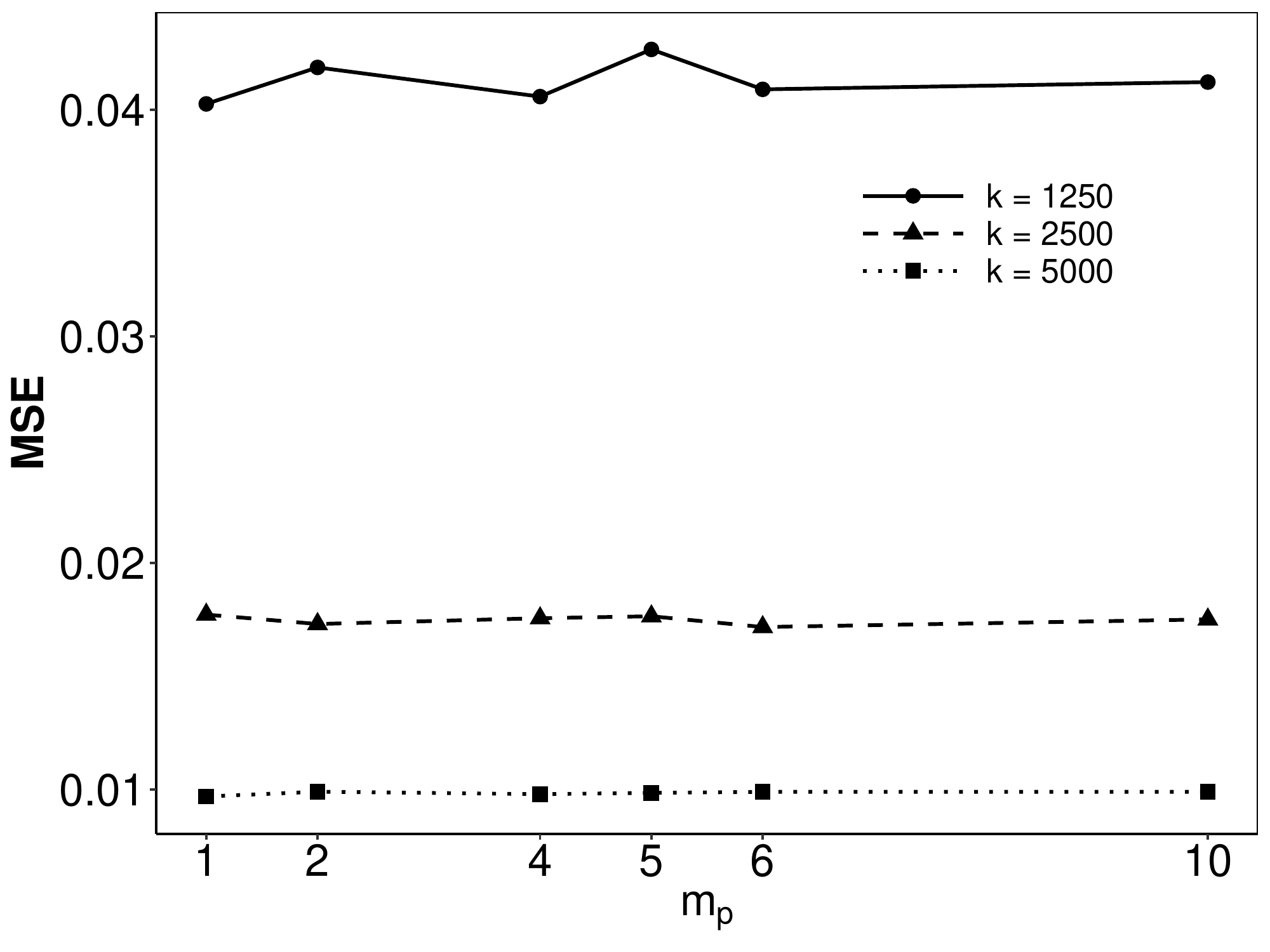}
		\caption{\scriptsize  $ \varphi_{p}( \pmb x;  \pmb 0 , \pmb \Sigma_1) $ }
		\label{fig3_1}
	\end{subfigure}%
 \hfill
   \begin{subfigure}[b]{.3\textwidth}
		\centering
		\includegraphics[width=\linewidth]{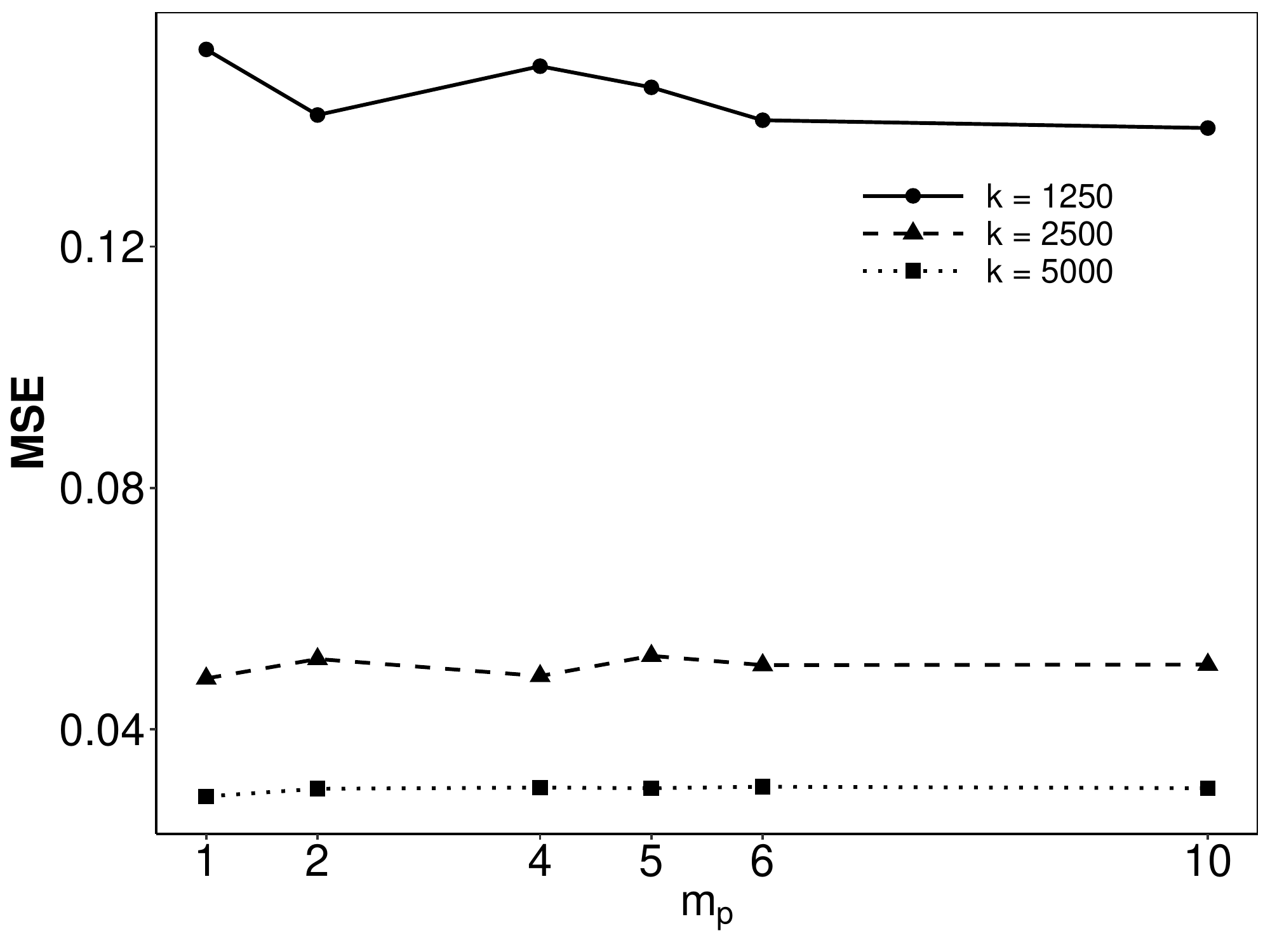}
		\caption{\scriptsize  $ \varphi_{p}( \pmb x;  \pmb 0 , \pmb \Sigma_2) $ }
		\label{fig3_2}
	\end{subfigure}
 \hfill
  \begin{subfigure}[b]{.3\textwidth}
		\centering
		\includegraphics[width=\linewidth]{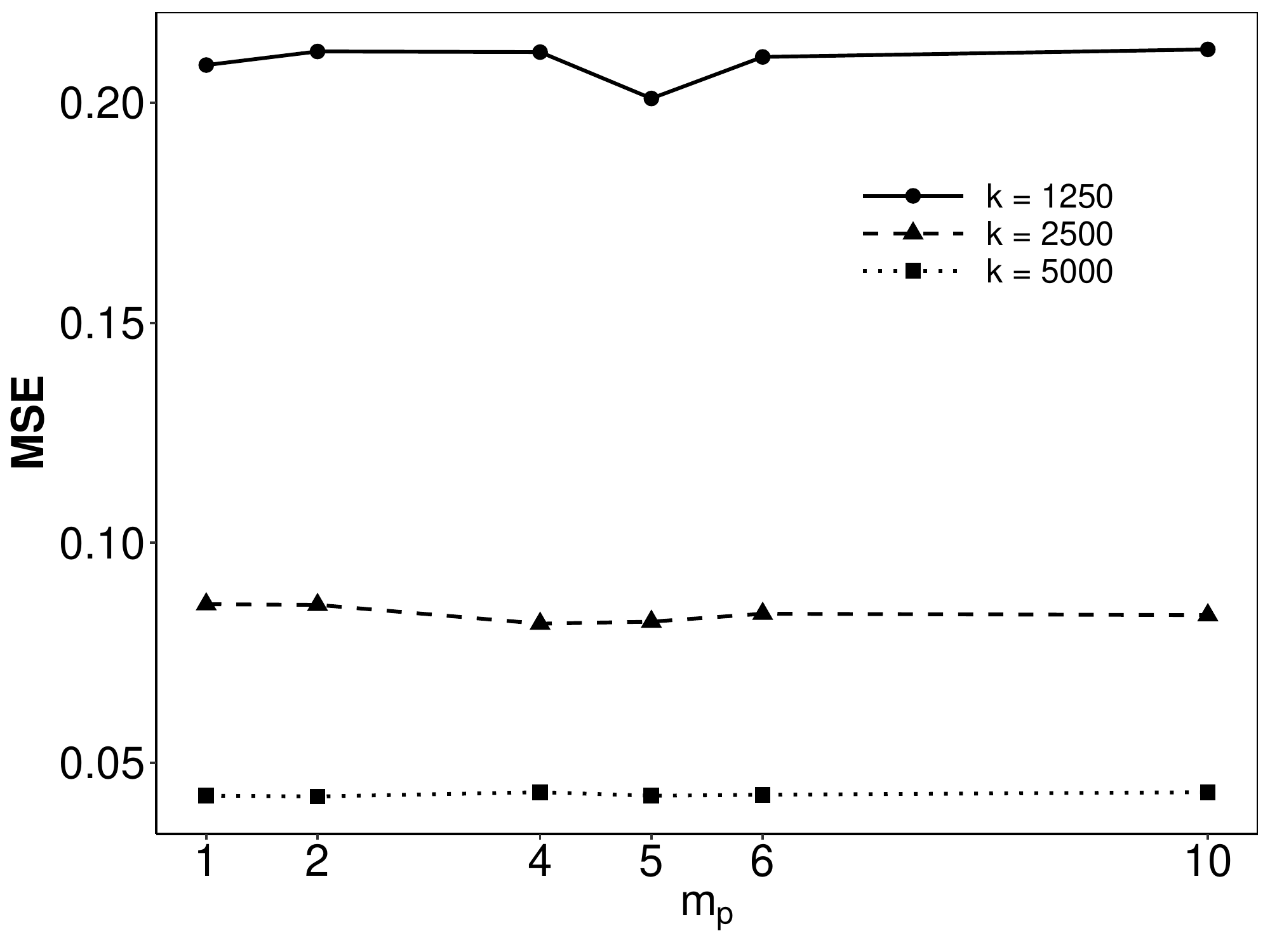}
		\caption{\scriptsize   $ \varphi_{p}( \pmb x;  \pmb 0 , \pmb \Sigma_3) $}
		\label{fig3_3}
	\end{subfigure}
 \vspace{1cm}
 
 \begin{subfigure}[b]{.3\textwidth}
		\centering
		\includegraphics[width=\linewidth]{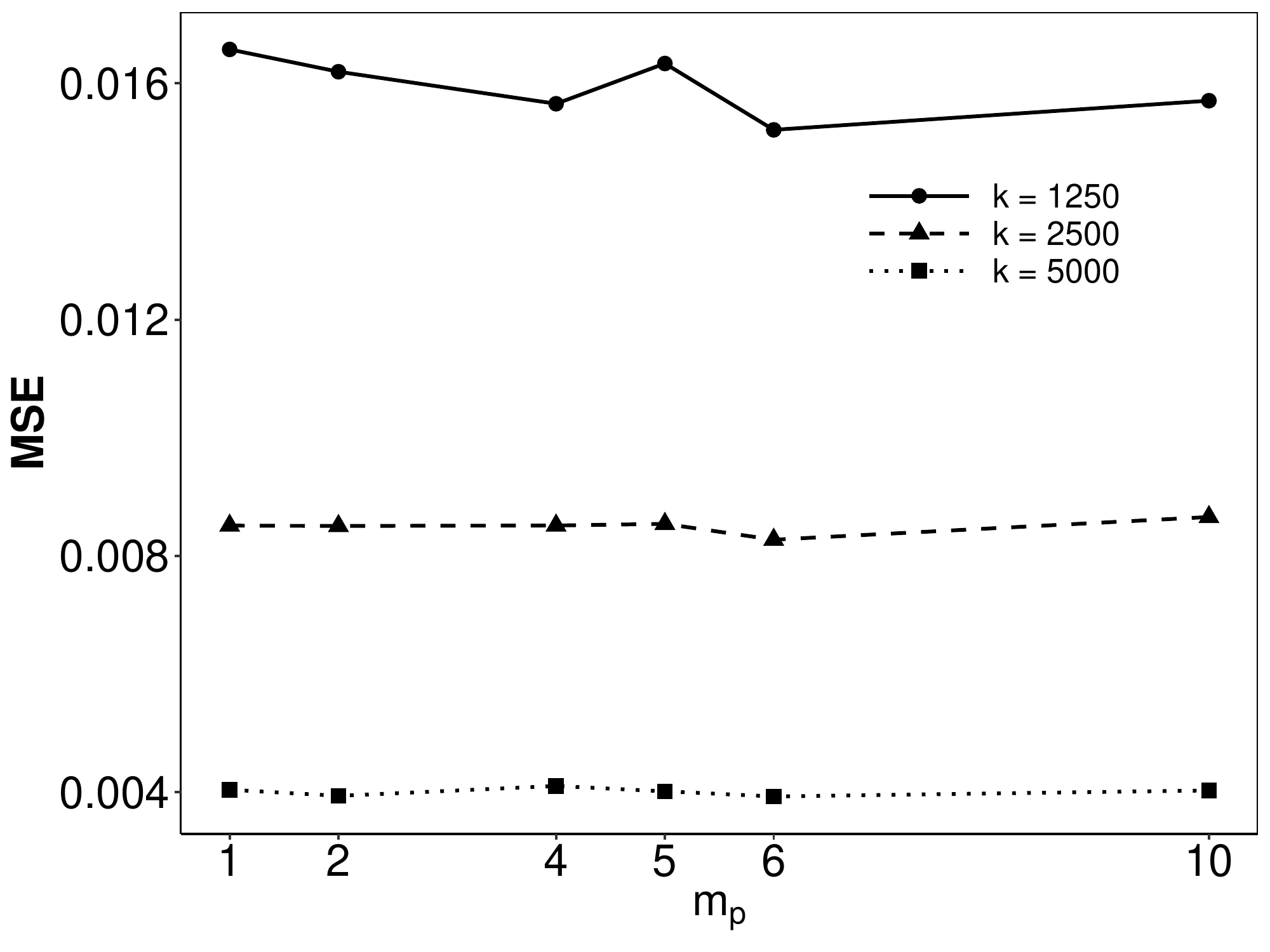}
		\caption{\scriptsize $\mathcal{T}_{p}(\pmb x;\pmb 0, \pmb \Sigma_1,3)$ }
		\label{fig3_4}
	\end{subfigure}%
 \hfill
   \begin{subfigure}[b]{.3\textwidth}
		\centering
		\includegraphics[width=\linewidth]{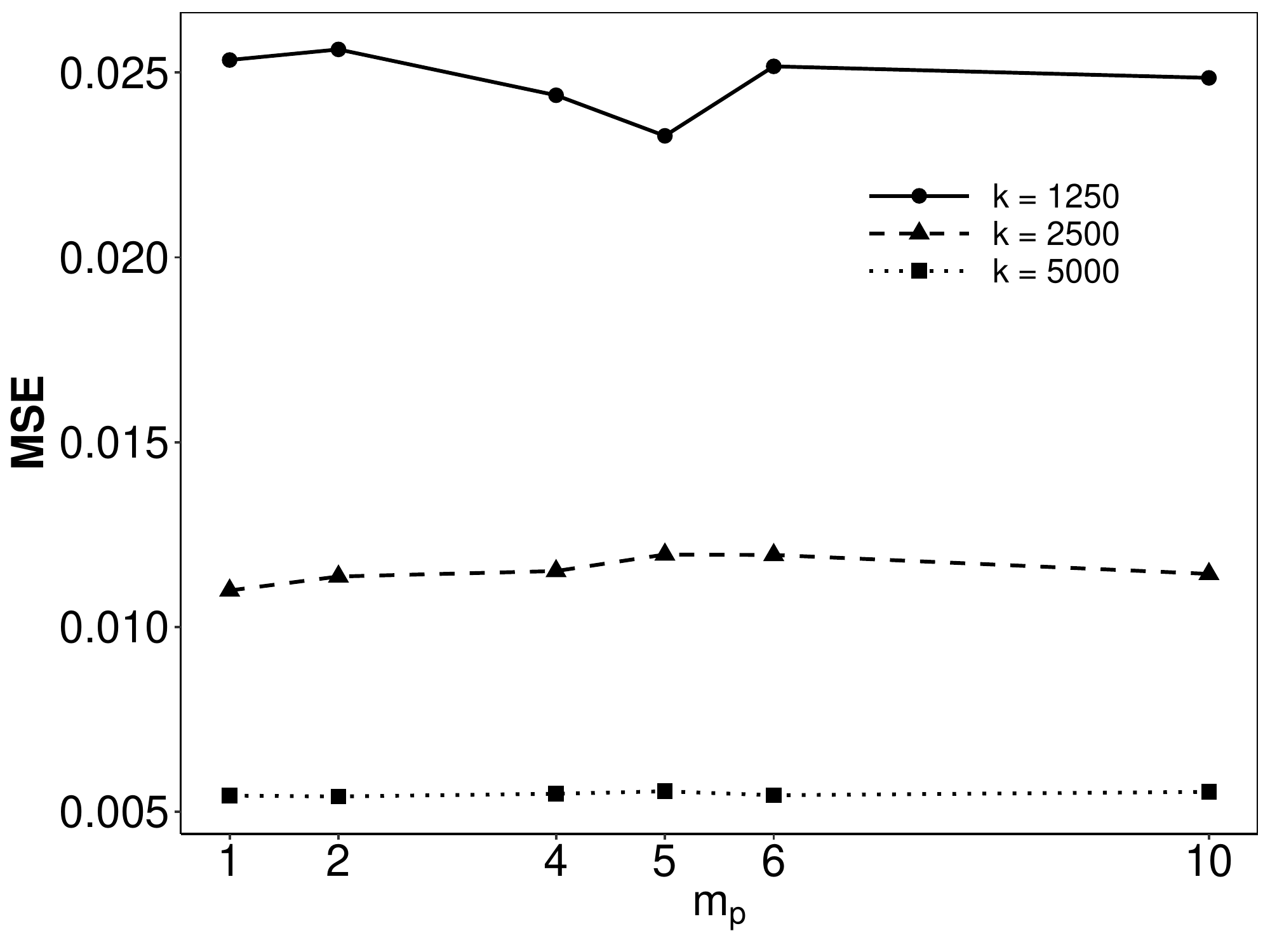}
		\caption{\scriptsize $\mathcal{T}_{p}(\pmb x;\pmb 0, \pmb \Sigma_2,3)$}
		\label{fig3_5}
	\end{subfigure}
 \hfill
  \begin{subfigure}[b]{.3\textwidth}
		\centering
		\includegraphics[width=\linewidth]{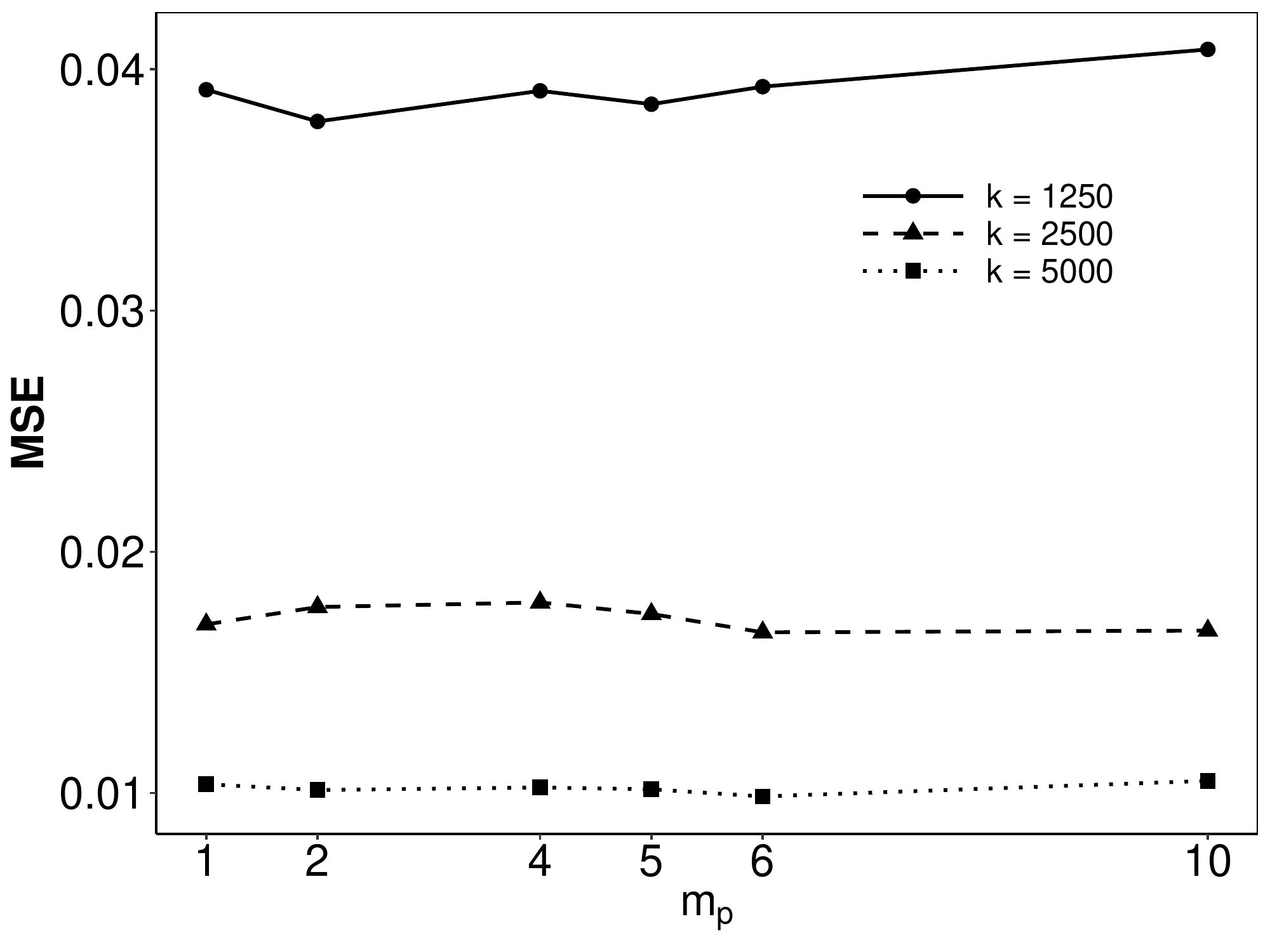}
		\caption{\scriptsize $\mathcal{T}_{p}(\pmb x;\pmb 0, \pmb \Sigma_3,3)$ }
		\label{fig3_6}
	\end{subfigure}
	\centering	\caption{ \it  MSE versus $m_p$ for the logistic regression model in equation (4.2) {with p = 7 }, $n = 10^5$, subsample of sizes $k=5000, 2500, 1250$  obtained  ODBSS (with Frobenius norm) are compared for centered normal and $t$-distribution with $3$ degrees of freedom with different covariance matrices.    
 }
	\label{Figure Supplementary 3-1} 
\end{figure}

We observe in Figure \ref{Figure Supplementary 3-1} that ODBSS is quite robust to the choice of $m_p$ when $p=7$. We obtained a very similar observation in case $p=20$ (these results are not displayed for the sake of brevity). We suggest setting values of the parameter $m_p$ to $4,5,6$. In the main paper, we have taken subsample sizes to be between 1250 to 10000, which means the initial subsample size is between 250 and 2000, and in these cases $m_p = 5$ works well. Therefore, we recommend to work with $m_p= 5$ whenever $k \geq 1250 $.  
{
\section{Time complexity of ODBSS when FIM has rank 2}
 In this section, we consider the model with a Fisher in formation matrix with rank $2$  and show that in this case the time complexity of ODBSS  remains the same as in the  rank  $1$ case. If the rank of the Fisher information matrix at the point $\pmb x$ is $2$, we have 
 \vspace{-1em}
$$
   \mathcal{I} ( \VectorBeta, \VectorX) =  \Phi_1(\VectorX, \VectorBeta) \Phi_1(\VectorX, \VectorBeta)^\top + \Phi_2(\VectorX, \VectorBeta) \Phi_2(\VectorX, \VectorBeta)^\top ,
$$
where the vectors $\Phi_1(\VectorX, \VectorBeta) $ and $ \Phi_2(\VectorX, \VectorBeta)  \in \mathbb{R}^{p+1}$ are not linearly dependent. In this case,  
\vspace{-1em}
\begin{align*}
d_F(  \VectorX , \VectorX^\prime ) & :=   \| \mathcal{I}(\VectorBeta, \pmb x) - \mathcal{I}(\VectorBeta, \pmb x^\prime)\|_F \\
  & =    
\Big \{ 
\|  \Phi_1(\VectorX, \VectorBeta)  \|^4 + 
\|  \Phi_1(\VectorX^\prime, \VectorBeta)  \|^4 +
\|  \Phi_2(\VectorX, \VectorBeta)  \|^4 + 
\|  \Phi_2(\VectorX^\prime, \VectorBeta)  \|^4 \\
& - 2 \big( \Phi_1(\VectorX  , \VectorBeta)^\top \Phi_1(\VectorX^\prime, \VectorBeta )\big)^2
- 2 \big( \Phi_1(\VectorX^\prime  , \VectorBeta)^\top \Phi_2(\VectorX, \VectorBeta )\big)^2
\\ &  - 2 \big( \Phi_1(\VectorX  , \VectorBeta)^\top \Phi_2(\VectorX^\prime, \VectorBeta )\big)^2
- 2 \big( \Phi_2(\VectorX  , \VectorBeta)^\top \Phi_2(\VectorX^\prime, \VectorBeta )\big)^2 \\
&
+ 2 \big( \Phi_1(\VectorX  , \VectorBeta)^\top \Phi_2(\VectorX, \VectorBeta )\big)^2
+ 2 \big( \Phi_1(\VectorX^\prime  , \VectorBeta)^\top \Phi_2(\VectorX^\prime, \VectorBeta )\big)^2
\Big \}^{  1/2},   
\end{align*}
where $\| \cdot \|$ is the Euclidean norm.
In particular, for the heteroskedastic model investigated in Section 4.2 of this paper the vectors $\Phi_1$ and $\Phi_2$ are given by 
$$  
\Phi_1(\VectorX  , \VectorBeta)  = \dfrac{1}{\sqrt{\sigma^2({\pmb x, \VectorBeta})}}   \begin{pmatrix}
    1 \\ \pmb x 
\end{pmatrix}   \text{ and }  \Phi_2(\VectorX  , \VectorBeta)  = \dfrac{1}{\sqrt{2}}   \begin{pmatrix}
    0 \\ \pmb x 
\end{pmatrix}.     
$$
Using the same arguments as in the Section 3.4 of the paper, we see that. the time complexity of the area estimation is  $\mathcal{O}(k^2_0 p)$ as in the rank $1$ case. The  time-complexity of the algorithm for  calculating 
the $\Psi$-optimal design on the design space ${{\pmb\chi}}_{k_0}$ is given by 
$\mathcal{O}(s p r )^3 ~ \sqrt{s p}~ log(1/\delta)$ where $ s   \leq L^p$ and  here $r=2$. 

Finally we discuss  the third component of ODBSS, the subsample allocation. If we work  with the Frobenius norm, time complexity of  calculating the distances $ \pmb{d}_i $ for $i=1, \ldots, b$ is also $\mathcal{O}(bnp) $ and then determining $\lfloor w_ik_1 \rfloor$ smallest elements $ \pmb{d}_i $ has complexity $\mathcal{O}( nb  )$. Therefore, ODBSS with Frobenius distance ($d_F(\cdot)$) has same time complexity if Fishers information matrix has rank 1 or 2. {Unfortunately, the computational complexity for subsample allocation for ODBSS with square root distance ($d_s(\cdot)$) is $\mathcal{O}(bnp^2) $ (as each term of the matrix $\mathcal{I} ( \VectorBeta, \VectorX)$ needs to be evaluated in order to obtain $\mathcal{I}^{1/2}( \VectorBeta, \VectorX)$) and even higher if we use Procrustes distance ($d_p(\cdot)$). Therefore, we recommend using the ODBSS with Frobenius norm when FIM has rank 2 or higher. }
}

\end{appendix}
\newpage


\bibliography{References} 

 \end{document}